\documentclass[twocolumn,twocolappendix]{aastex7}

\usepackage{amsmath}
\usepackage{xspace}
\newcommand {\ixpe}{{IXPE}\xspace}
\newcommand {\nustar}{\textit{NuSTAR}\xspace}

\newcommand {\nicer}{{NICER}\xspace}

\begin{document}

\title{Unchanged X-Ray Polarization During Accretion Dips in the Low Hard State of Cygnus X-1}

\author[orcid=0009-0002-4994-2528,gname=Yu-Shan,sname=Ling]{Yu-Shan Ling}
%% \altaffiliation{Kitt Peak National Observatory}
\affiliation{Guangxi Key Laboratory for Relativistic Astrophysics, School of Physical Science and Technology, Guangxi University}
\email{lingys@st.gxu.edu.cn} 

\author[orcid=0000-0002-0105-5826,gname=Fei,sname=Xie]{Fei Xie} 
\altaffiliation{Corresponding author}
\affiliation{Guangxi Key Laboratory for Relativistic Astrophysics, School of Physical Science and Technology, Guangxi University}
\email[show]{xief@gxu.edu.cn}

\author[orcid=0000-0003-0331-3259,gname=Alessandro,sname='Di Marco']{Alessandro Di Marco} 
\affiliation{INAF Istituto di Astrofisica e Planetologia Spaziali, Via del Fosso del Cavaliere 100, 00133 Roma, Italy}
\email{alessandro.dimarco@inaf.it}

\author[orcid=0000-0001-8916-4156,gname=Fabio,sname='La Monaca']{Fabio La Monaca} 
\affiliation{INAF Istituto di Astrofisica e Planetologia Spaziali, Via del Fosso del Cavaliere 100, 00133 Roma, Italy}
\email{fabio.lamonaca@inaf.it}

\author[orcid=0000-0002-3776-4536,gname=Ming-Yu,sname=Ge]{Ming-Yu Ge} 
\affiliation{Key Laboratory of Particle Astrophysics, Institute of High Energy Physics, Chinese Academy of Sciences, Beijing 100049, China}
\email{gemy@ihep.ac.cn}

\begin{abstract}
    Cygnus X-1 serves as a foundational benchmark for studying accretion mechanisms in stellar-mass black hole X-ray binaries. While traditional X-ray timing and spectroscopy have mapped its bimodal state transitions, constraining the exact geometric configuration of its accreting plasma requires additional tools. In this letter, we present a joint spectro-polarimetric analysis of Cygnus X-1 during flux-decrease events, or ``dips'', observed in its low hard state using simultaneous data from IXPE, NICER, and NuSTAR. The broadband spectral model reveals that the energy-dependent flux reduction during these dips is best described by a partial-covering absorption model, heavily impacting the softer energy bands. This absorption is consistent with obscuration by dense, cold clumps originating from the companion's stellar wind or the outer accretion disk during superior conjunction. Notably, our polarimetric analysis demonstrates that despite significant flux drops, the polarization degree and polarization angle in the 2-8 keV band remain largely consistent between dip and off-dip intervals. The stability of the polarization signature during these highly absorbed periods indicates that the obscured thermal disk emission does not significantly contribute to the total polarization. Ultimately, these results strongly support a scenario where the X-ray polarization originates from an extended, likely oblate, accretion-disk corona whose large scale renders its overall geometry unaffected by localized clumping.
\end{abstract}


\section{Introduction}\label{sec:Introduction}

X-ray binaries (XRBs) containing a black hole (BH) exhibit a characteristic bimodal spectral behavior related to their accretion mechanism \citep{Holt1976,Ichimaru1977}. Accretion is an extremely efficient mechanism for extracting energy from matter as it falls toward a compact object; this energy can be released through various channels, producing distinct X-ray emission signatures (spectral states) that correspond to different configurations/geometries of the accreting matter. In fact, the spectral variations of BH-XRBs trace a `q' in the hardness-intensity diagram \citep[HID;][]{2001ApJS..132..377H,2004MNRAS.355.1105F,2005A&A...440..207B,2016ASSL..440...61B} with each branch corresponding to a distinct spectral state. The most common states are the high soft state (HSS) and the low hard state (LHS), which are distinguished by the energies at which their emission peaks \citep{Zdziarski2004}. This state transition is driven by the behavior of the accretion flow, which switches between a standard disk model in the HSS \citep{Shakura1973} and a radiatively inefficient accretion flow in the LHS \citep{Ichimaru1977}. The soft state is characterized by dominant thermal emission from the classical geometrically thin and optically thick accretion disk. Conversely, the LHS is marked by a spectral shift to a power-law shape; this emission is believed to be produced by multiple Compton upscatterings of soft seed photons—either originating from the disk or internally produced via synchrotron emission—by hot electrons within an optically thin, rarefied medium (the ``corona'') surrounding the BH \citep{st1980, Gierlinski1997}.

Cygnus X-1 (Cyg~X-1) is one of the first discovered and most extensively studied Galactic BH-XRBs \citep{Jiang2024}; it plays a foundational role in high-energy astrophysics, serving as a unique cosmic laboratory for probing accretion physics since its discovery \citep{Bowyer1965}. It is a bright, persistent Galactic source, with a flux ranging between $\sim 0.2$ and $2\text{ Crab}$ in the $2\text{--}20\text{ keV}$ energy band. This X-ray emission is powered by the accretion onto the BH from its high-mass stellar companion, and this gives rise to different states; historically, the observed variations in the X-ray spectrum of Cyg~X-1 gave rise to the classic soft/hard classification of spectral states in BH-XRBs \citep{Tananbaum1972, Zdziarski2004, Done2007}, with Cyg~X-1 found in the LHS about two-thirds of the time \citep{2006A&A...447..245W,2013A&A...554A..88G}. Cyg~X-1 is located at a distance of $2.22^{+0.18}_{-0.17}$\,kpc with orbital inclination 27.5$^{\circ}\pm$0.8$^{\circ}$ with a near-maximal spin ($\geq$0.9985) and its BH has a mass $(21\pm2)M_{\odot}$ \citep{2021Sci...371.1046M}. The BH and its O9.7 Iab supergiant companion star are in a tightly bound system with an orbital period of ${\sim}$5.6\,days \citep{1998MNRAS.301..285L}. The Cyg~X-1 light curves show dips \citep[see, e.g.,][for a review]{Fujii2025}; these are typically observed throughout the orbital phase from 0.75 to 0.25, corresponding to the superior conjunction of the black hole \citep{Balu1995}. These dips can be due to X-ray absorption in an accretion bulge located at the accretion disk edge near the supergiant companion \citep{Poutanen2008}, or to clumps that may originate from the winds of the companion star \citep{Sund2018}. In the latter case, the duration of a dip conveys information regarding the dimensions of the X-ray generating region and clumping.

X-ray spectral and timing analysis cannot distinguish geometrical configurations of the hot, rarefied plasma responsible for the X-ray emission in BH-XRBs; however, thanks to the Imaging X-ray Polarimetry Explorer \citep[\ixpe;][]{2022JATIS...8b6002W,2021AJ....162..208S}, nowadays X-ray polarimetry is a new tool that can help in this. Polarization degree (PD), polarization angle (PA), and their spectral dependence exhibit substantial differences among different scenarios \citep{Poutanen1996,Poutanen2018,Dovciak2004,Kraw2022}. Moreover, studying the polarization across different states \citep{2024ApJ...969L..30S,2022Sci...378..650K} and orbital phases \citep{2025A&A...701A.115K} of Cyg~X-1 has improved our knowledge of this system, which has been observed by \ixpe several times. Recently, an interesting result was obtained in the case of an accreting neutron star, GX~13$+$1, observed during dips \citep{2025ApJ...979L..47D}; in this peculiar state, the source showed variability in PD and PA that can be used to investigate X-ray emission from the outer region of the disk where X-rays are scattered in an extended accretion disk corona or disk wind, allowing characterization of their geometry. Cyg~X-1, exhibiting dips on short timescales, is a compelling target for studying correlations between polarization and X-ray flux; however, its dips were not investigated in depth in previous X-ray polarimetric studies \citep{2025A&A...701A.115K,2022Sci...378..650K,2024ApJ...969L..30S}. \ixpe detected a few dips on scales ranging from minutes to hours, characterized by a rapid, distinct decrease in flux. In this work, we leverage the unique capabilities of \ixpe to characterize the polarization during such flux dips and compare the results with those obtained for dipping sources. We aim to use broadband spectral coverage from Neutron Star Interior Composition Explorer \citep[NICER;][]{2014SPIE.9144E..20A,2016SPIE.9905E..1HG} and Nuclear Spectroscopic Telescope Array \citep[\nustar;][]{2013ApJ...770..103H} to perform a joint analysis of both polarimetric and spectroscopic data. The paper is structured as follows: Section~\ref{sec:Results} describes the available observations and compares the polarimetric and spectral results between the dip and the off-dip intervals; the outcomes are discussed in Section~\ref{sec:Discussion}, and conclusions are drawn in Section~\ref{sec:Summary}.

\section{Observations and results}\label{sec:Results}

\begin{figure}[!ht]
\centering
\includegraphics[width=\linewidth]{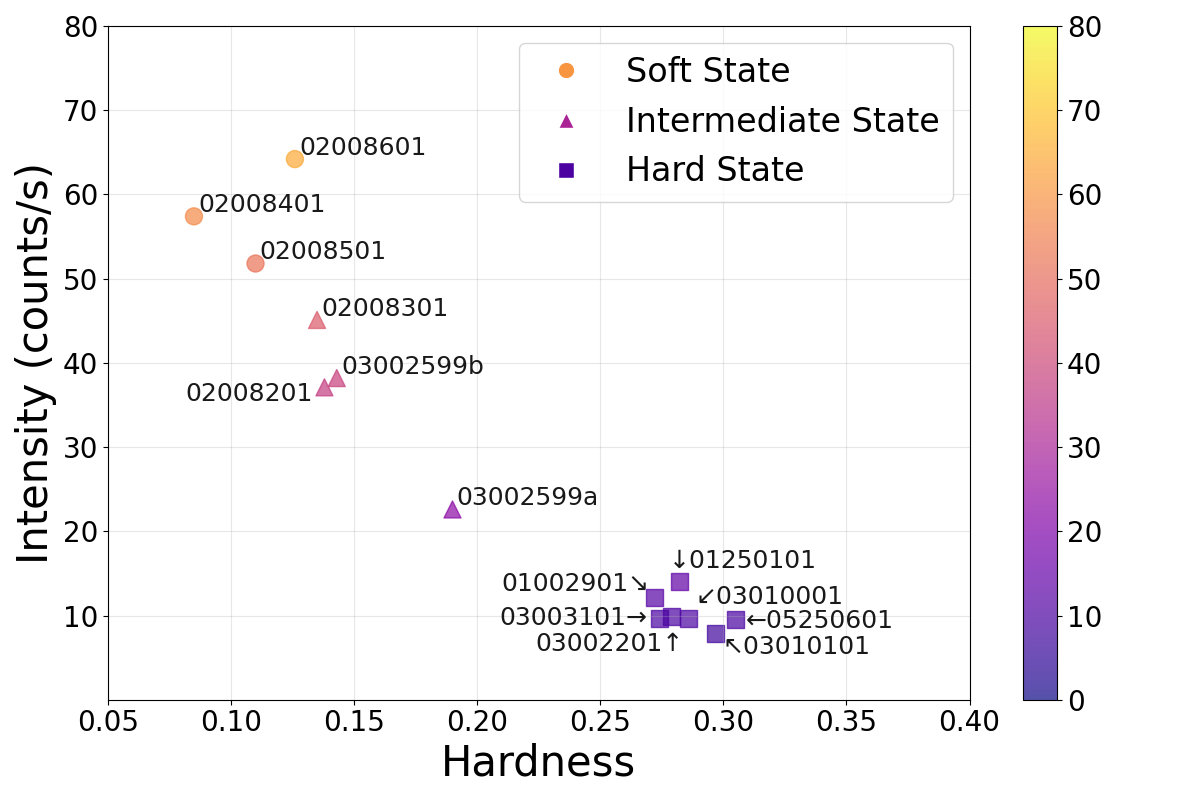}
\caption{Hardness-intensity diagram reporting the average intensity and hardness for each \ixpe observation of Cyg~X-1 performed. Hardness is defined as the ratio of photon counts in the 3.5--8\,keV range to those in 2--3.5\,keV. Circular, triangular, and square markers indicate observations in the soft, intermediate, and hard states, respectively.}
\label{fig:Cyg_X-1_HID}
\end{figure}

\begin{figure*}[!hbt]
\centering
%\epsscale{1.1}
\includegraphics[width=0.48\linewidth]{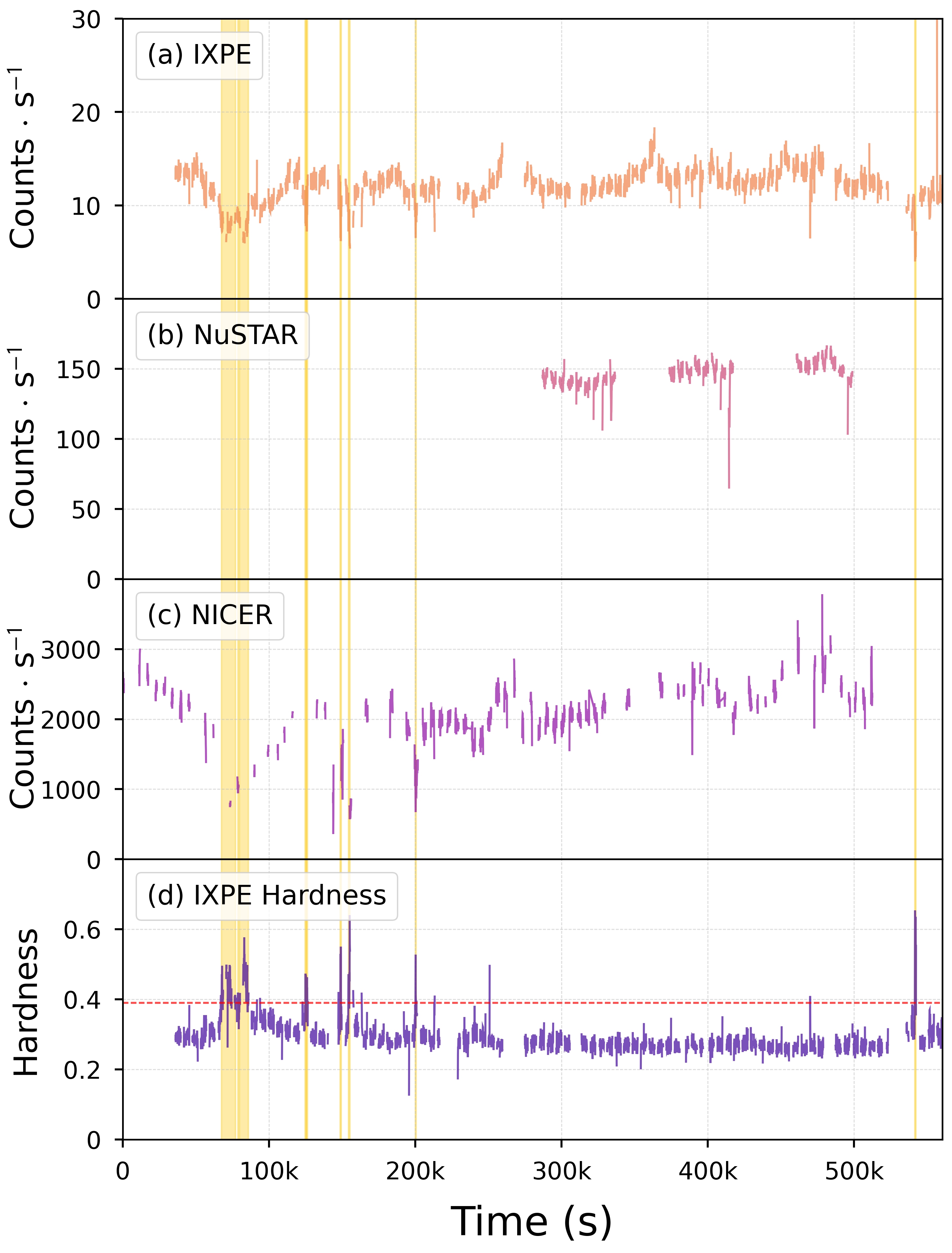}
\includegraphics[width=0.48\linewidth]{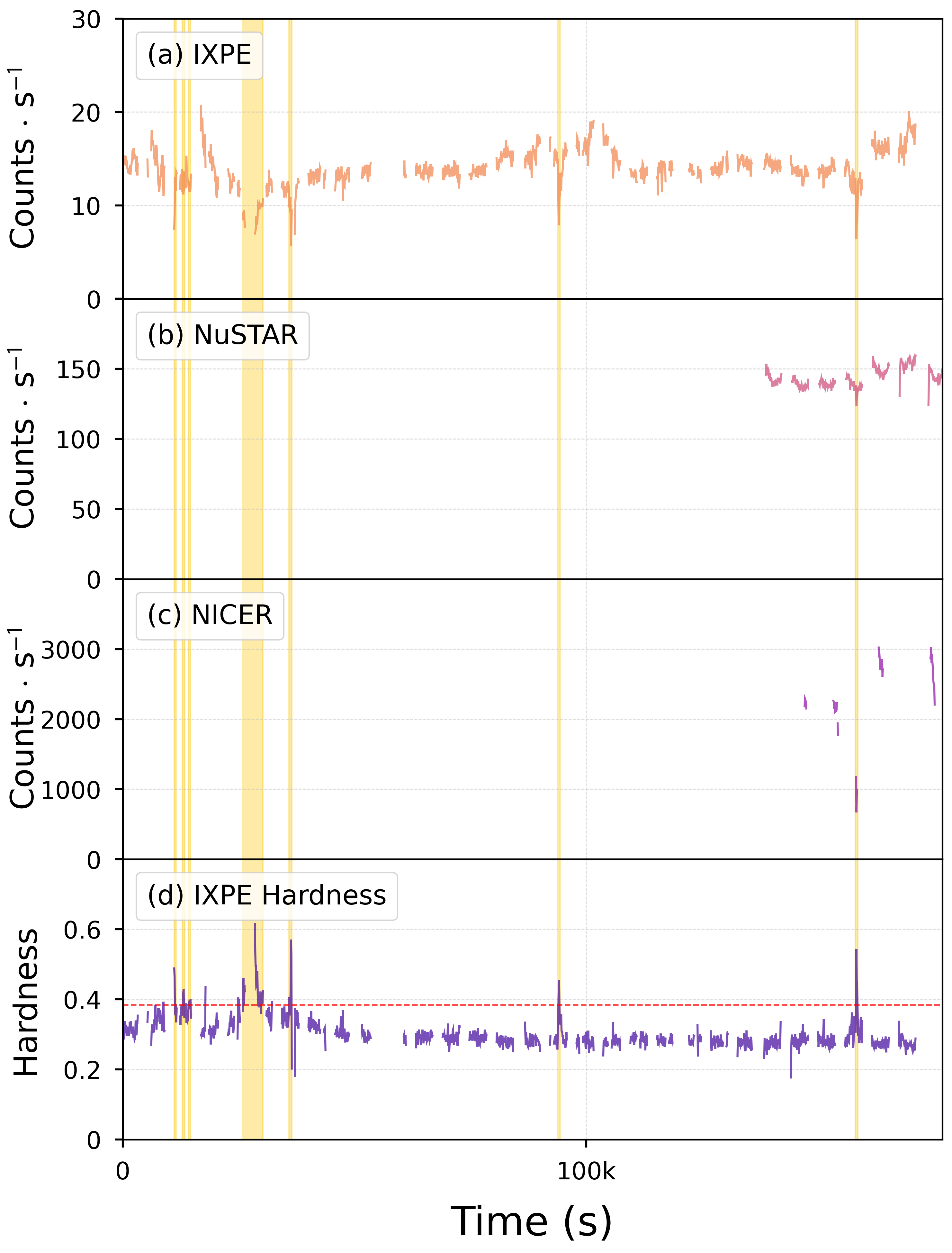}
\caption{Light curves and hardness ratio of Cyg~X-1 from observations in May (left) and June (right) 2022. Panels (a), (b), and (c) report the light curves from \ixpe, \nustar and \nicer observations, respectively; panel (d) is the hardness ratio of \ixpe. The time binning is 75 seconds for all of them. The shaded time intervals highlight periods when the \ixpe hardness ratio was persistently elevated, identified via a sliding-window analysis with a threshold of 2.0 standard deviations above the mean (see text).}
\label{fig:Light_Curves_and_IXPE_Hardness}
\end{figure*}

\ixpe has observed Cyg~X-1 14 times, as reported in Table~\ref{tab:Observation_datasets} in Appendix~\ref{data_red}, covering different spectral states, as shown in Figure~\ref{fig:Cyg_X-1_HID}. Observation ID 03002599 comprises two pointings 2 months apart that may cover two distinct spectral states; for this reason, we divided it into two segments in the following.
For each \ixpe observation, the light curve and its hardness ratio (HR) have been studied; the latter is defined as the photon counts in the 3.5--8\,keV band divided by those in the 2--3.5\,keV band. Notably, the observations in May and June 2022 show dips in the light curves, as reported in Figure~\ref{fig:Light_Curves_and_IXPE_Hardness}. It is possible to observe an increase in the \ixpe HR accompanied by simultaneous decreases in the count rate. Light curves and hardness ratios for the other observations are reported in Figure~\ref{fig:IXPE_lc_hardness_other} in Appendix~\ref{app:appendix_pol_and_lc_2023_2024}. Although observations performed in May and June 2022 are not the only ones showing dips, they are the only ones simultaneously covered by \nustar and \nicer, allowing for a broadband spectro-polarimetric analysis. Because of this, the paper focuses on these observations, while a summary of all available observations is reported in the Appendices.

Following the procedure described in Appendix~\ref{app:dips_selection}, we identified the dip and off-dip intervals for the \ixpe observations. Since Cyg~X-1 during the \ixpe observations performed in 2022 does not exhibit any significant state transition \citep{2022Sci...378..650K}, we will adopt a two-step approach to present the polarization results. The polarimetric results are first shown separately for each observation (including those not covered by \nicer and \nustar), followed by a combined analysis of 2022 datasets to provide a comprehensive assessment of dips in a spectro-polarimetric study.

As reported in Table~\ref{tab:dips_gti_2022} in Appendix~\ref{app:dips_selection}, ten dips were identified in the observation performed in May 2022 and seven in the one in June 2022, with durations spanning from 300\,s up to ${\sim}9700$\,s; all the dips are in the orbital phase 0.9--0.25 corresponding to the superior conjunction, with the exception of two short dips, both at phase ${\sim}0.34$. The maximum absorption observed in the \ixpe light curve (see Figure~\ref{fig:Light_Curves_and_IXPE_Hardness}) corresponds to a flux reduction of ${\sim}33$\% in May 2022 and ${\sim}40$\% in June 2022. 

\subsection{Model independent polarimetric results} \label{subsec:Polarimetric Analysis}

\begin{figure*}[!hbt]
\includegraphics[width=0.3\linewidth]{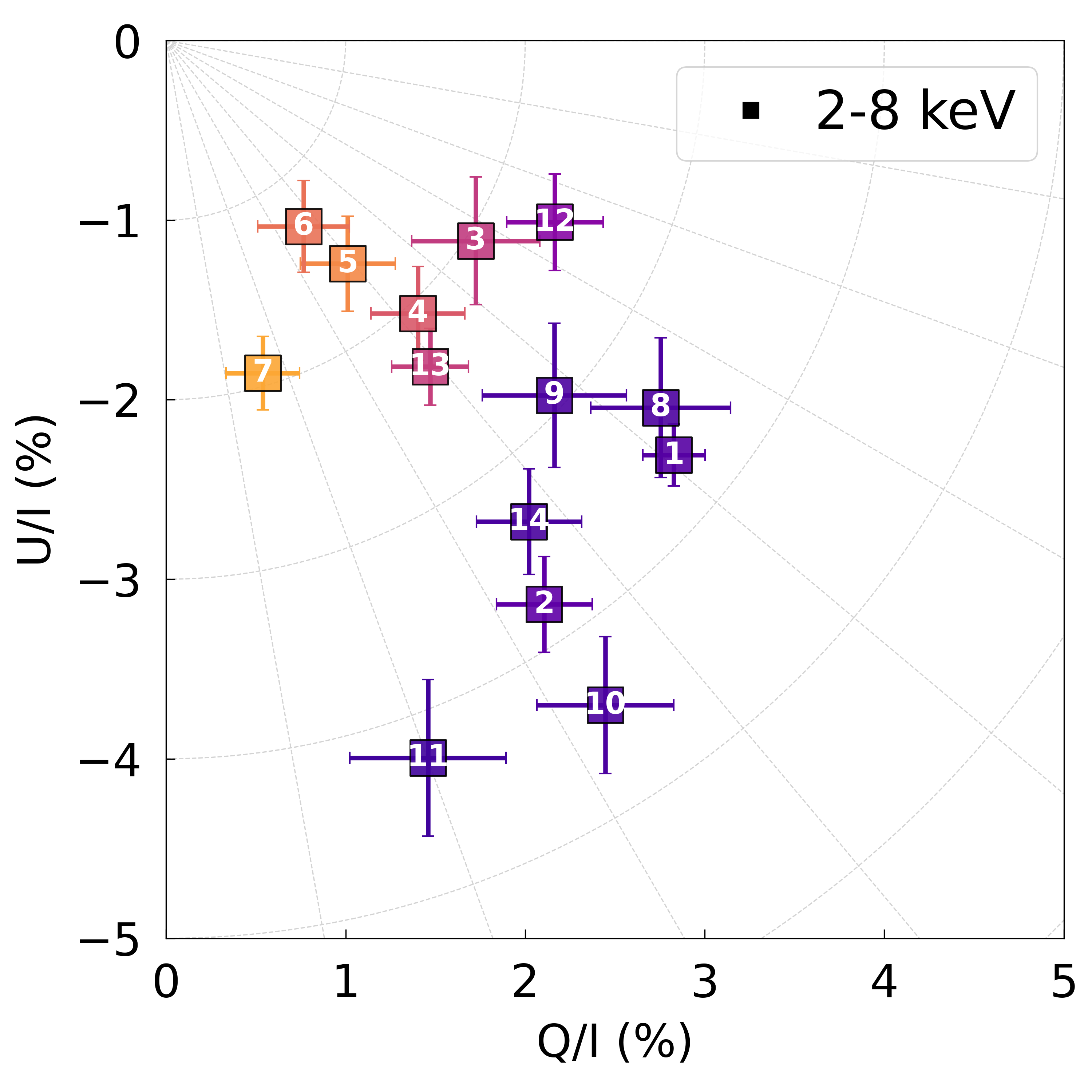}
\includegraphics[width=0.3\linewidth]{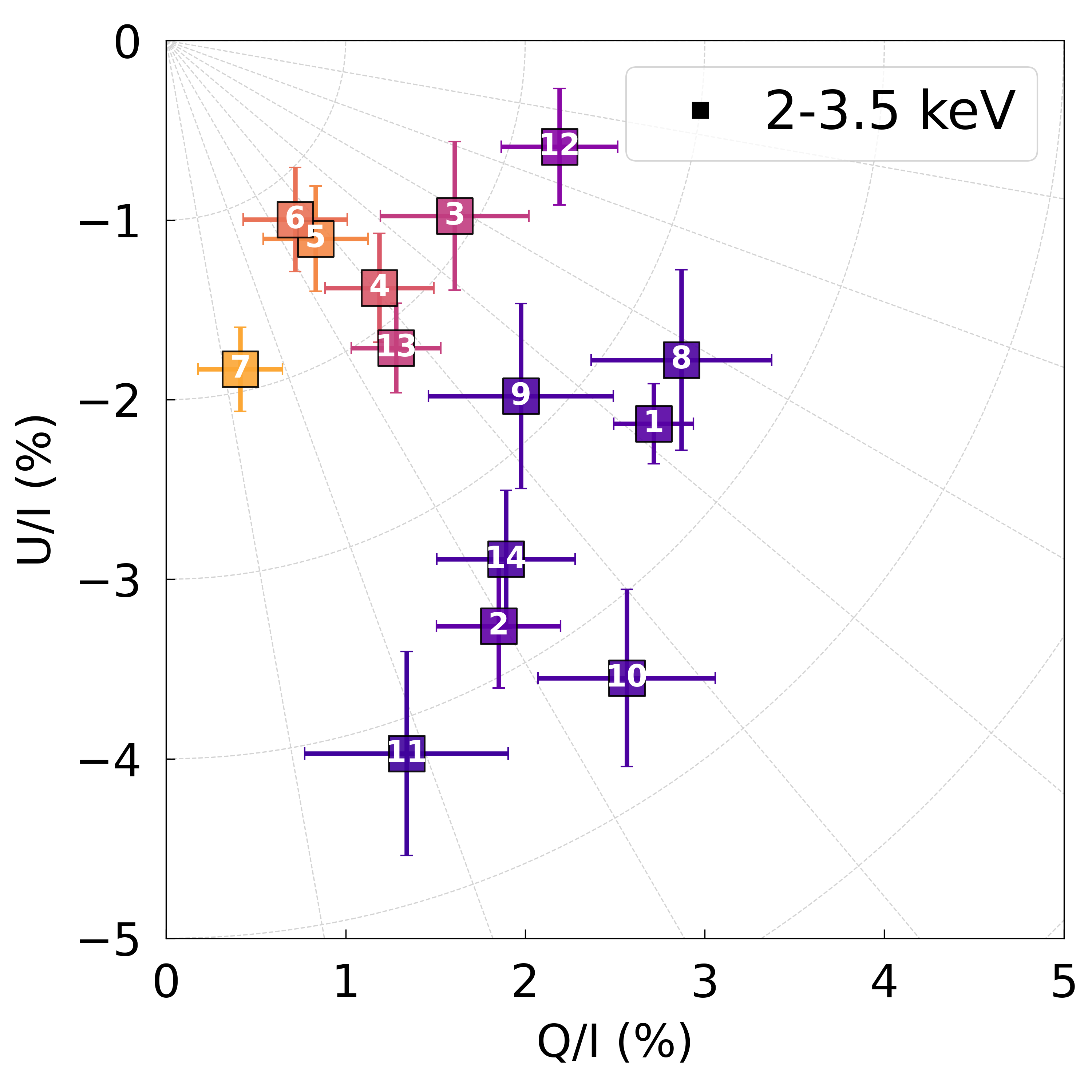}
\includegraphics[width=0.3\linewidth]{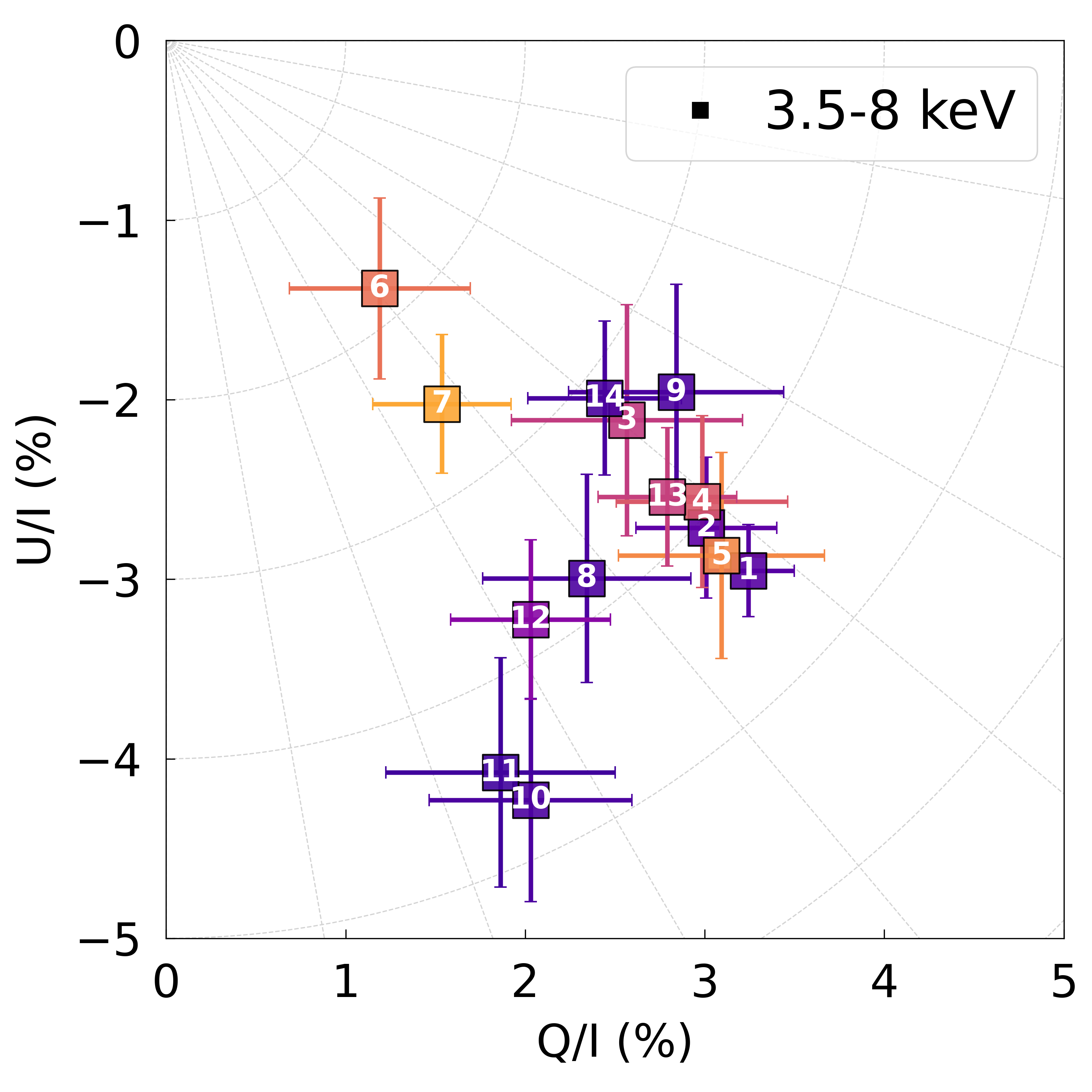}
\raisebox{0.43cm}{\includegraphics[width=0.06\linewidth]{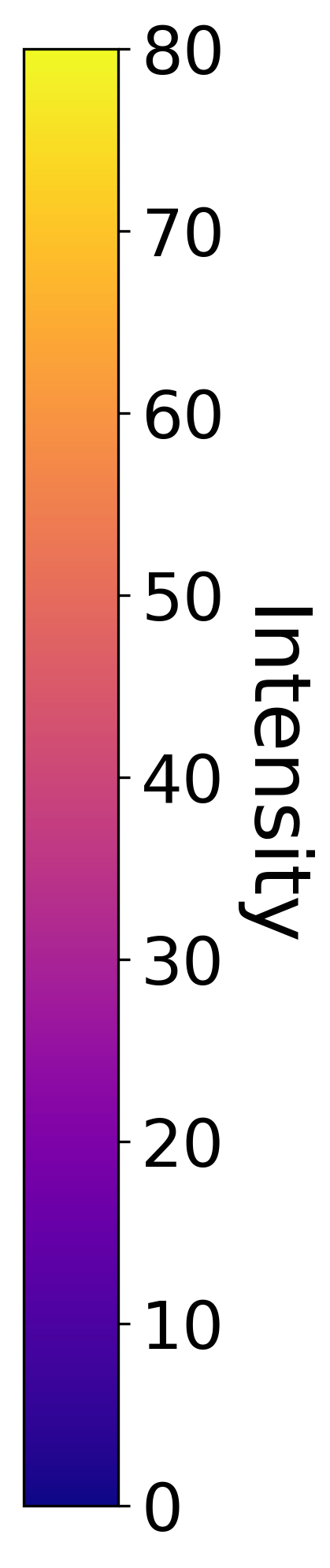}}
\caption{Stokes parameters for each \ixpe observation in 2--8\,keV (left), 2--3.5\,keV (center) and 3.5--8\,keV (right), with error bars at $68\%$. The epoch numbers refer to Table~\ref{tab:Observation_datasets}, and the intensity colorbar is consistent with Figure~\ref{fig:Cyg_X-1_HID}.}
\label{fig:Q_U_comparison}
\end{figure*}

\begin{figure}[!tb]
\plotone{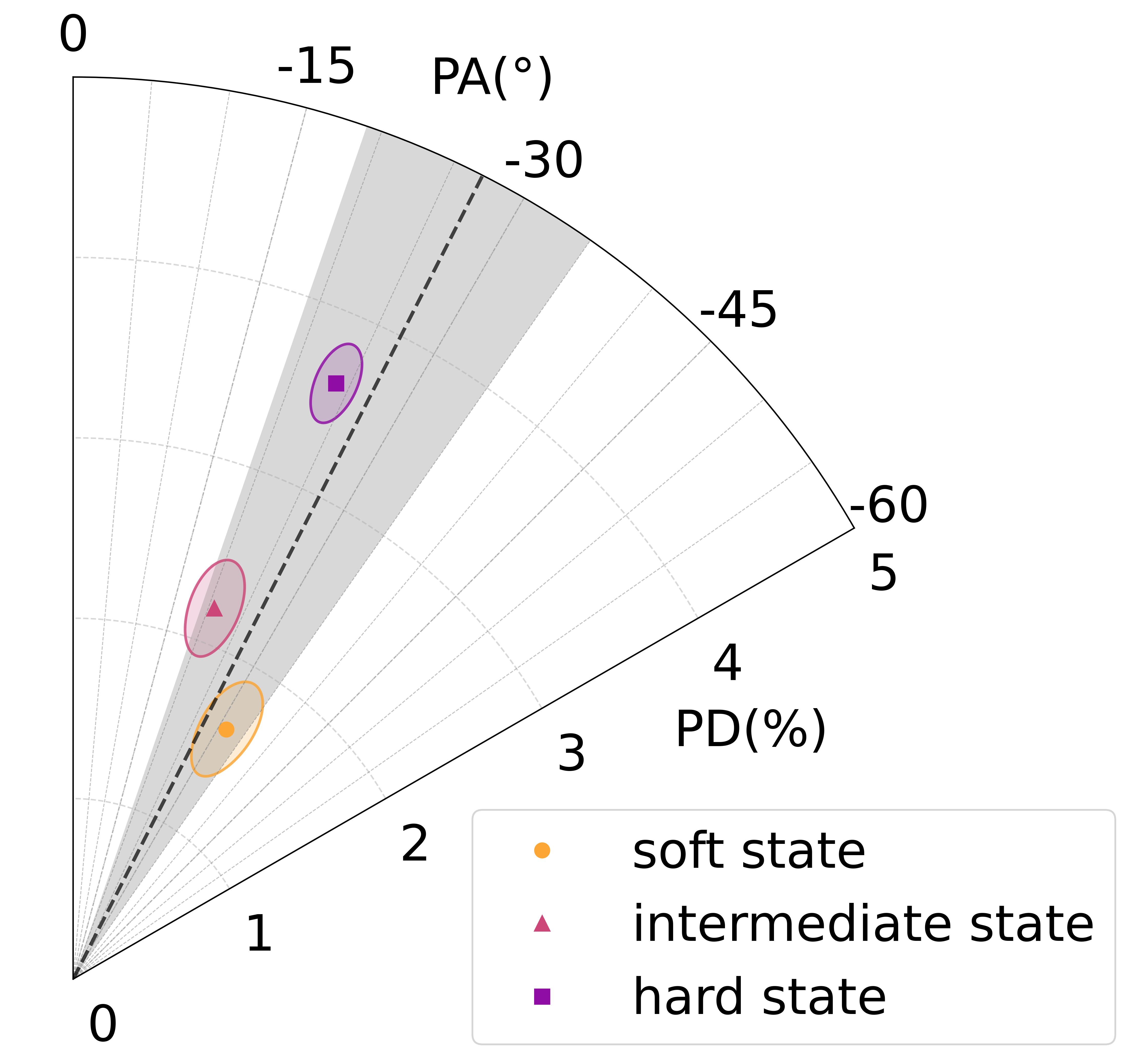}
\caption{Cyg~X-1 polarization in the soft, hard, and intermediate states. Allowed regions are at 90$\%$ confidence level (CL). The polarization angles across different spectral states align closely with the radio jet direction, which has a position angle of -27$^{\circ}\pm$8$^{\circ}$ \citep{2021Sci...371.1046M}.}
\label{fig:three_states_PD}
\end{figure}

Figure~\ref{fig:Cyg_X-1_HID} reports an intensity-hardness diagram based on all the publicly available \ixpe observations of Cyg~X$-$1. The polarization properties obtained for each observation are reported as the Stokes parameters Q and U across different energy bands in Figure~\ref{fig:Q_U_comparison}. 
The same results are also reported in Table~\ref{tab:Polarization_properties_of_all_IXPE} in Appendix~\ref{app:appendix_pol_all}.
These plots show higher PD in the 3.5--8\,keV band than in the 2--3.5\,keV band, with PD increasing with hardness ratio (and decreasing at higher intensities). In light of this, the observations can be grouped into three distinct states: the hard state (flux$\leq$20\,counts/s), the soft state (flux$\geq$50\,counts/s), and the intermediate state (20--50\,counts/s). Applying this selection and considering all observations within each flux interval, we obtained the measured polarization of Cyg~X-1, as shown in Figure~\ref{fig:three_states_PD}.

Focusing the analysis on the data from May to June 2022, we studied the polarization using the ``dip'' and ``off-dip'' selections, and the results are summarized in Table~\ref{tab:Polarization_properties_of_dip_and_off-dip_time_intervals}. We observe that, despite increasing hardness during dips, polarization in the 2--8\,keV band is consistent between states in both observations.

\begin{deluxetable*}{lcccccc}[!htb]
\tabletypesize{\normalsize}
\tablewidth{0pt}
\label{tab:Polarization_properties_of_dip_and_off-dip_time_intervals}
\tablehead{
\colhead{} & \multicolumn{3}{c}{dip} & \multicolumn{3}{c}{off-dip} \\
\cline{2-4} \cline{5-7}
\colhead{} & \colhead{2--3.5\,keV} & \colhead{3.5--8\,keV} & \colhead{2--8\,keV} & \colhead{2--3.5\,keV} & \colhead{3.5--8\,keV} & \colhead{2--8\,keV}}
\startdata
\multicolumn{7}{c}{01002901 (May 2022)} \\
\cline{1-7}
$U/I$ ($\%$) & $-2.5\pm1.2$ & $-2.8\pm1.1$ & $-2.6\pm0.8$ & $-2.1\pm0.2$ & $-3.0\pm0.3$ & $-2.3\pm0.2$ \\ 
$Q/I$ ($\%$) & $2.0\pm1.2$ & $3.5\pm1.1$ & $2.4\pm0.8$ & $2.7\pm0.2$ & $3.2\pm0.3$ & $2.8\pm0.2$ \\
\cline{1-7}
PA ($^\circ$) & $-26\pm10$ & $-19\pm7$ & $-23\pm7$ & $-19\pm2$ & $-21\pm2$ & $-19\pm1$ \\
PD ($\%$) & $3.2\pm1.2$ & $4.5\pm1.1$ & $3.5\pm0.8$ & $3.5\pm0.2$ & $4.4\pm0.3$ & $3.7\pm0.2$ \\
\cline{1-7}
MDP$_{99}$ ($\%$) & 3.5 & 3.5 & 2.5 & 0.7 & 0.8 & 0.5 \\
\cline{1-7}
\multicolumn{7}{c}{01250101 (June 2022)} \\
\cline{1-7}
$U/I$ ($\%$) & $0.4\pm1.7$ & $-5.8\pm1.7$ & $-1.3\pm1.3$ & $-3.4\pm0.4$ & $-2.5\pm0.4$ & $-3.2\pm0.3$ \\ 
$Q/I$ ($\%$) & $1.2\pm1.7$ & $4.9\pm1.7$ & $2.2\pm1.3$ & $1.9\pm0.4$ & $2.9\pm0.4$ & $2.1\pm0.3$ \\
\cline{1-7}
PA ($^\circ$) & $9\pm40$ & $-25\pm7$ & $-16\pm14$ & $-31\pm3$ & $-21\pm3$ & $-28\pm2$ \\
PD ($\%$) & $1.2\pm1.7$ & $7.6\pm1.7$ & $2.6\pm1.3$ & $3.9\pm0.4$ & $3.9\pm0.4$ & $3.8\pm0.3$ \\
\cline{1-7}
MDP$_{99}$ ($\%$) & 5.3 & 5.3 & 3.8 & 1.1 & 1.2 & 0.8 \\
\cline{1-7}
\enddata
\caption{Polarization properties of dip and off-dip time intervals for \ixpe observations of Cyg~X-1 performed in 2022. Errors are reported at 68\% CL.}
\end{deluxetable*}

Using the same energy bands of the hardness ratio and of Figure~\ref{fig:Q_U_comparison}, we obtain the results shown in Table~\ref{tab:Polarization_properties_of_dip_and_off-dip_time_intervals}; in the June 2022 observation, there is an indication of PD energy dependence only during dips.  

\begin{figure}[!htb]
\plotone{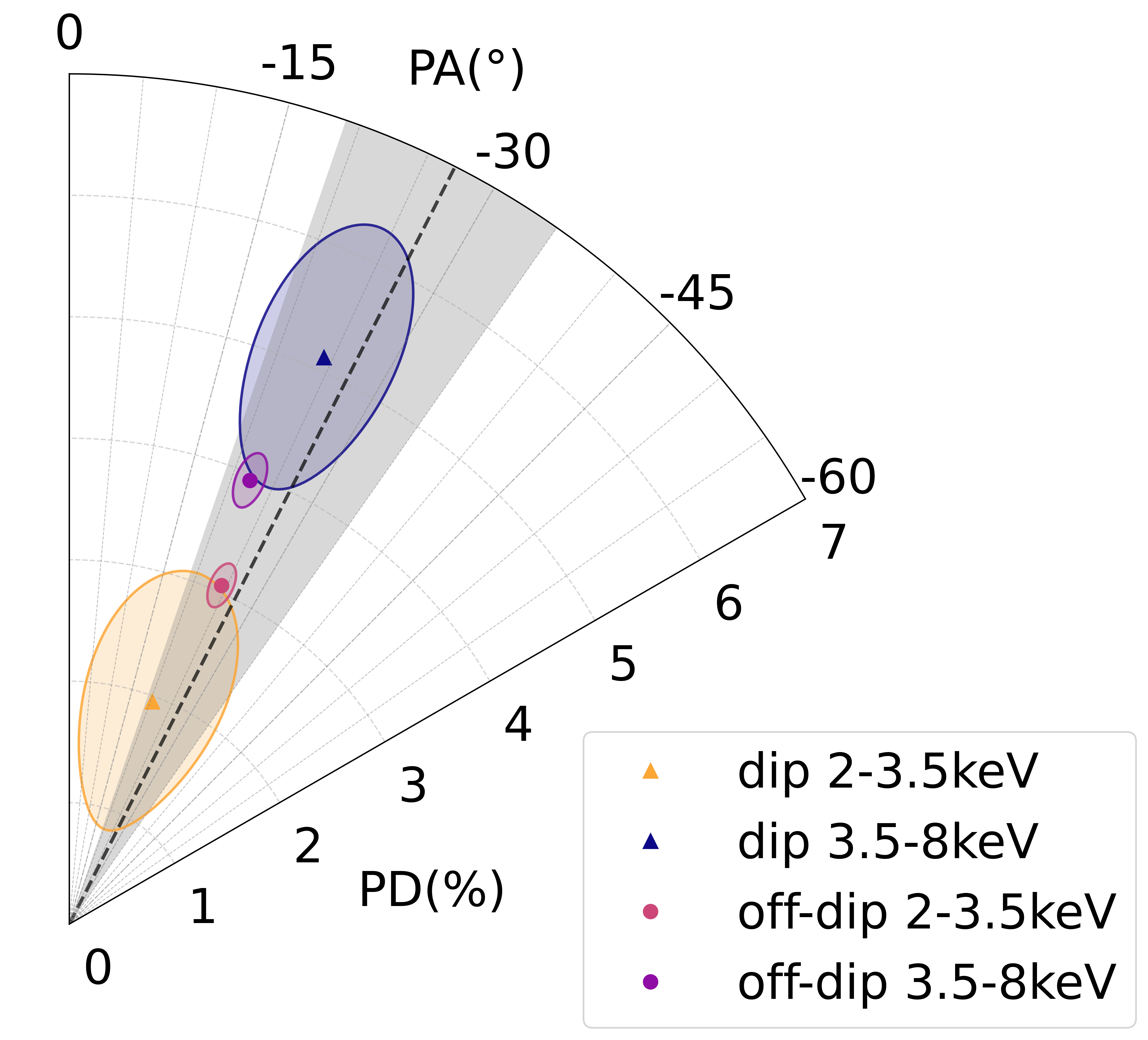}
\caption{A comparison of the polarization in the dip and off-dip states across different energy bands, employing all \ixpe datasets in which dips are detected, with a 68$\%$ CL. The polarization angles of dip and off-dip across different energies align closely with the radio jet direction, which has a position angle of -27$^{\circ}\pm$8$^{\circ}$ \citep{2021Sci...371.1046M}.}
\label{fig:dip_off-dip_comparation}
\end{figure}

In Figure~\ref{fig:dip_off-dip_comparation}, we compared the polarization properties between the dip and off-dip states across various energy bands using all datasets in which dips are detected. This study has been performed aiming to improve the significance of the energy dependence of polarization in dip and off-dip states. This is possible because the two observations are consistent in both spectral state and polarization. The result is a polarization in the two considered energy bins, compatible within 68\% CL, with a PA that is always consistent across the different states and energies. The same analysis was performed combining observations of the hard and intermediate states for each year; the results are reported in Figure~\ref{fig:dip_off-dip_other} in Appendix~\ref{app:appendix_pol_and_lc_2023_2024}.

\subsection{Spectral Analysis} \label{subsec:Spectral Analysis}

During the \ixpe Observation ID 01250101, when a dip occurs, strictly simultaneous \nicer and \nustar observations were performed (see Figure~\ref{fig:Light_Curves_and_IXPE_Hardness}). Aiming to obtain a broadband spectral model, we selected time intervals based on the hardness ratio of \ixpe to define dip and off-dip states in the \nicer and \nustar light curves. The dip selection has been performed by using \textsc{xselect} v2.5c and the GTI intervals as reported in Figure~\ref{fig:Light_Curves_and_IXPE_Hardness}. The off-dip spectrum is obtained from all remaining data from June 2022.

\begin{figure*}[!hbt]
\centering
\includegraphics[width=0.49\linewidth]{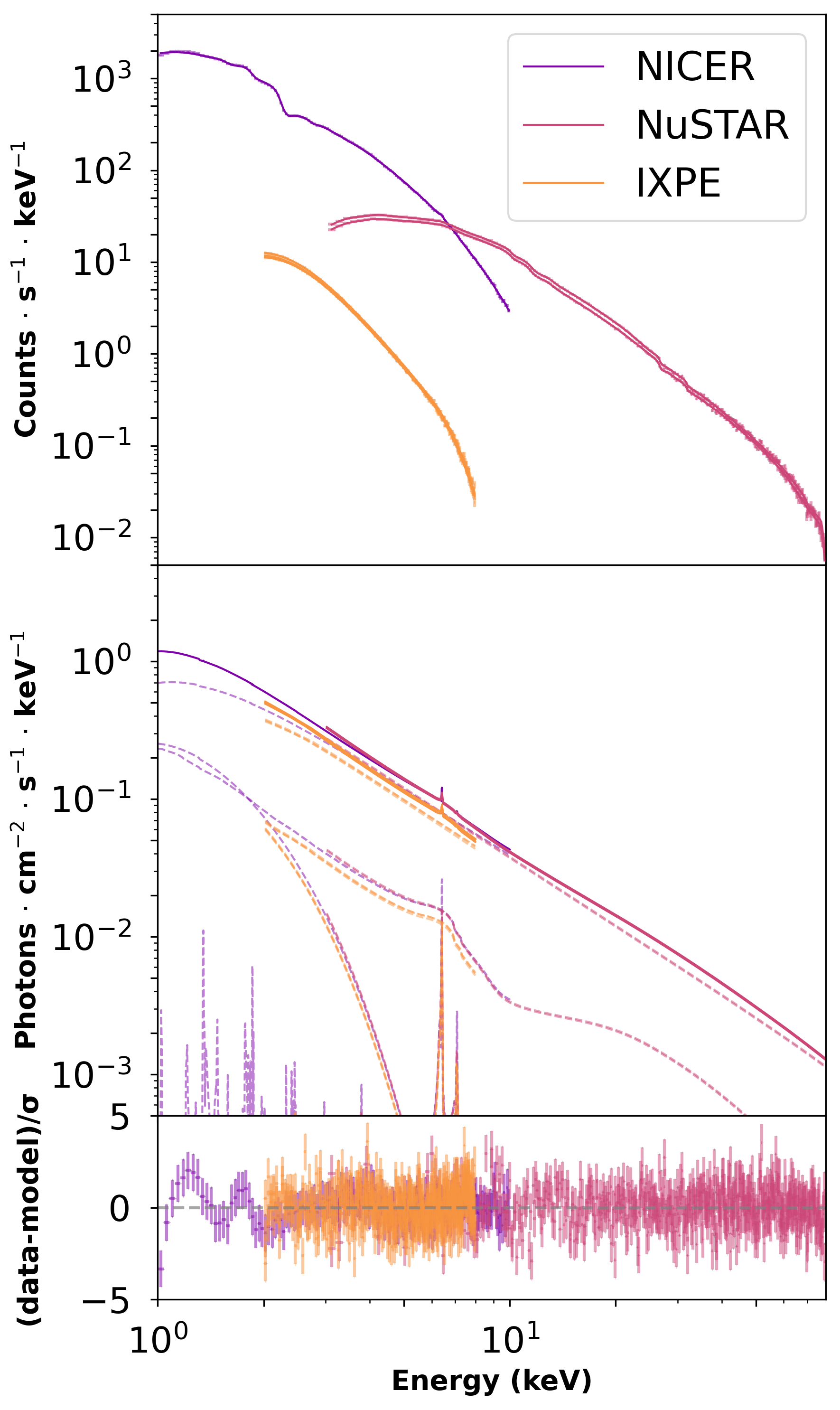}
\includegraphics[width=0.49\linewidth]{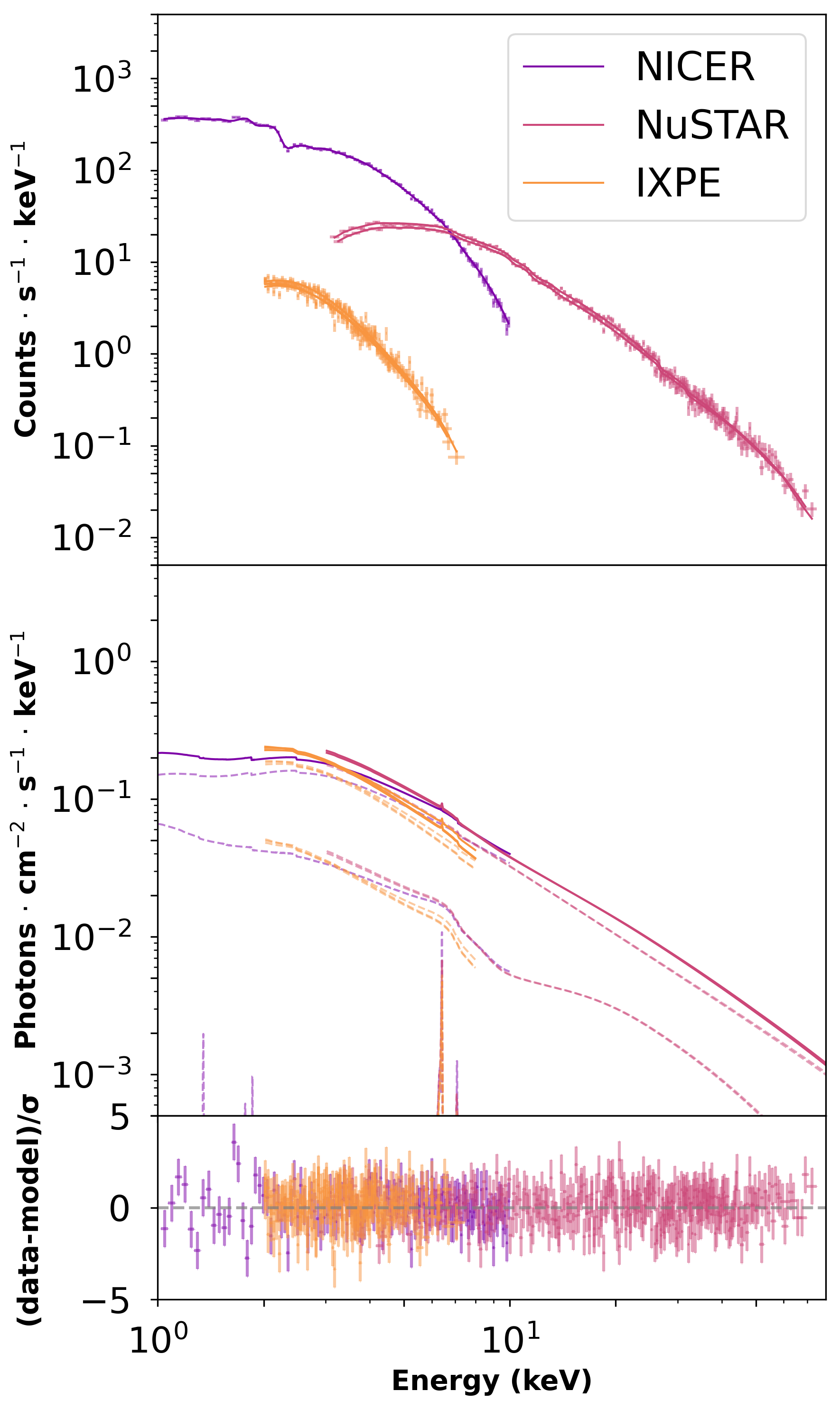}
\caption{Spectra for off-dip (left) and dip (right) time intervals.}
\label{fig:Spectrum_dip_offdip}
\end{figure*}

In this study, we apply the following spectral model in \textsc{xspec} \citep{Arnaud1996}: \texttt{MBPO * TBABS * TBPCF * (DISKBB + NTHCOMP + RELXILLCP + XILLVERCP)}. Here, following the same approach adopted by \cite{2022Sci...378..650K}, the \texttt{MBPO} model is employed to address cross-calibration inconsistencies among X-ray observatories by applying a spectral adjustment that incorporates a normalization factor and different power-law indices for energies below and above a defined break energy ($E_1{<}E_\textrm{br}{<}E_2$):
\begin{align}
\text{\texttt{MBPO}}(E_1) &= N\left(\frac{E_1}{E_{\mathrm{br}}}\right)^{d\Gamma_1}\\
\text{\texttt{MBPO}}(E_2) &= N\left(\frac{E_2}{E_{\mathrm{br}}}\right)^{d\Gamma_2}. \nonumber
\end{align}
\noindent
The \texttt{MBPO} parameters, in the case of \nicer, are \texttt{NORM = 1}, $E_{\mathrm{br}}$ = 1\,keV, and $d\Gamma_1 = d\Gamma_2 = 0$. In the case of \nustar, \texttt{NORM} is free to vary, $d\Gamma_1 = d\Gamma_2 =d\Gamma$ is free to vary, effectively reducing the model to a single power law. For \ixpe data, all \texttt{MBPO} parameters are free to vary; this is necessary to account for potential gain variations \citep{2021APh...13302628B,2022SPIE12181E..1CD}.\footnote{In Appendix~\ref{app:nicer_consistency} a cross-check using only the \nicer data is performed to verify that the \texttt{TBpcf} modeling of the dip absorption remains reliable without \texttt{MBPO}, confirming that \texttt{MBPO} does not alter the robustness of the \texttt{TBpcf} model.}
The Galactic absorption was modeled with the \texttt{TBabs} model, while \texttt{TBpcf} was included to account for absorption from the wind of the companion; both models were applied assuming element abundances at the \texttt{wilm} values \citep{2000ApJ...542..914W}. The \texttt{DISKBB} component is used to describe the direct thermal emission from the disk, with \texttt{NTHCOMP} used to model X-rays originating from Compton scattering in the corona. Both the \texttt{RELXILLCP} and \texttt{XILLVERCP} models belong to the \textsc{RELXILL} package version 2.8 \citep{2014MNRAS.444L.100D,2016A&A...590A..76D,2014ApJ...782...76G}, which is a comprehensive suite of X-ray spectral models describing the reflection from accretion disks around compact objects. The \texttt{RELXILLCP}, which accounts for blurring due to relativistic effects, is used to model X-rays reprocessed in the inner disk region, while the \texttt{XILLVERCP} models X-rays reprocessed from the outer disk and the companion star and not strongly influenced by relativistic effects. For both reflection components, we assume that the flux incident on the disk decreases with radial distance as $r^{-3}$, the black hole spin is fixed at 0.998, and the inclination is $27.5^\circ$ \citep{2021Sci...371.1046M}. Furthermore, the other common parameters in \texttt{XILLVERCP} are linked to their counterparts in \texttt{RELXILLCP}. The results obtained for this model in the dip and off-dip are reported in Table~\ref{tab:Best-fit_Parameters} and panels (a) and (b) of Figure~\ref{fig:Spectrum_dip_offdip}. Due to the distinct expected energy ranges of the thermal, Compton, and reflection components, we observe that the \texttt{DISKBB} component is heavily absorbed, with the photon index remaining nearly constant while the reflection fraction decreases.
Whereas the Comptonized emission is consistent, and the non-relativistic component is not well constrained in the dip.

\begin{deluxetable*}{lccc}[!htb]
\tabletypesize{\scriptsize}
\label{tab:Best-fit_Parameters}
\tablewidth{0pt}
\tablehead{
\colhead{Component} & 
\colhead{Parameter (unit)} & 
\colhead{Off-dip} & 
\colhead{Overlap dip}
}
\startdata
TBABS & $N_{\text{H}}$ ($10^{22}$ cm$^{-2}$) & $0.316^{+0.012}_{-0.019}$ & $[0.316]$ \\[0.5em]
TBPCF & $N_{\text{H}}$ ($10^{22}$ cm$^{-2}$) & $[5.00]$ & $5.2\pm0.2$ \\
 & $pcf$ & $<0.068$ & $0.715^{+0.012}_{-0.011}$ \\[0.5em]
DISKBB & $T_{in}$ (keV) & $0.503^{+0.013}_{-0.017}$ & $[0.503]$ \\
 & $norm$ ($10^{2}$) & $10.4^{+3.7}_{-1.1}$ & $<3.8$ \\[0.5em]
NTHCOMP & $Gamma$ & $1.620^{+0.010}_{-0.006}$ & $1.54\pm0.03$ \\
 & $kTe$ (keV) & $[94.2]$ & $[94.2]$ \\
 & $norm$ & $1.205^{+0.008}_{-0.034}$ & $0.909^{+0.079}_{-0.011}$\\[0.5em]
RELXILLCP & $Rin$ ($R_{\text{ISCO}}$) & $3.7\pm0.4$ & $4.0^{+3.1}_{-1.9}$ \\
 & $logxi$ & $3.27^{+0.03}_{-0.04}$ & $3.26^{+0.17}_{-0.14}$ \\
 & $Afe$ & $4.0\pm0.2$ & $2.2^{+1.2}_{-0.9}$ \\
 & $kTe$ (keV) & $140^{+160}_{-40}$ & $[140]$ \\
 & $norm$ ($10^{-2}$) & $4.1^{+0.2}_{-0.3}$ & $4.9^{+2.0}_{-1.7}$ \\[0.5em]
XILLVERCP & $logxi$ (cm$^{-3}$) & $1.99^{+0.05}_{-1.42}$ & $[1.99]$ \\
 & $norm$ ($10^{-3}$) & $2.5^{+0.5}_{-0.3}$ & $<6.0$ \\[0.5em]
\hline
MBPO & $d\Gamma$ & $-0.089^{+0.009}_{-0.002}$ & $-0.22\pm0.03$ \\
\nustar & $Ebr$ (keV) & $[4.00]$ & $[4.00]$ \\
(FPMA) & $norm$ & $1.053^{+0.003}_{-0.002}$ & $1.169^{+0.017}_{-0.016}$ \\[0.5em]
MBPO & $d\Gamma$ & $-0.089^{+0.009}_{-0.002}$ & $-0.20\pm0.03$ \\
\nustar & $Ebr$ (keV) & $[4.00]$ & $[4.00]$ \\
(FPMB) & $norm$ & $1.030^{+0.003}_{-0.004}$ & $1.136^{+0.017}_{-0.016}$ \\[0.5em]
MBPO & $d\Gamma_1$ & $0.11^{+0.08}_{-0.04}$ & $-0.2\pm0.4$ \\
\ixpe & $d\Gamma_2$ & $-0.098^{+0.010}_{-0.011}$ & $-0.48\pm0.13$ \\
(DU1) & $Ebr$ (keV) & $2.62^{+0.14}_{-0.13}$ & $[2.62]$ \\
 & $norm$ & $0.865\pm0.004$ & $1.13\pm0.05$ \\[0.5em]
MBPO & $d\Gamma_1$ & $0.15\pm0.04$ & $-0.4^{+0.4}_{-0.3}$ \\
\ixpe & $d\Gamma_2$ & $-0.097^{+0.012}_{-0.013}$ & $-0.32\pm0.14$ \\
(DU2) & $Ebr$ (keV) & $2.74^{+0.13}_{-0.12}$ & $[2.74]$ \\
 & $norm$ & $0.893\pm0.004$ & $1.08\pm0.05$ \\[0.5em]
MBPO & $d\Gamma_1$ & $0.21^{+0.06}_{-0.05}$ & $-0.2\pm0.4$ \\
\ixpe & $d\Gamma_2$ & $-0.061^{+0.011}_{-0.012}$ & $-0.45\pm0.14$ \\
(DU3) & $Ebr$ (keV) & $2.63^{+0.07}_{-0.09}$ & $[2.63]$ \\
 & $norm$ & $0.873\pm0.004$ & $1.09\pm0.06$ \\[0.5em]
\hline
$\chi^{2}$/d.o.f. & & 1149/1100 & 600/622 \\
\hline
$Flux_{unabsorbed}$ & ($10^{-9} erg s^{-1} cm^{-2}$) & 7.11 & 6.08 \\
$Flux_{absorbed}$ & ($10^{-9} erg s^{-1} cm^{-2}$) & 7.11 & 5.03 \\
$Flux_{diskbb}$ & ($10^{-9} erg s^{-1} cm^{-2}$) & 0.19 & 0.00 \\
$Flux_{nthcomp}$ & ($10^{-9} erg s^{-1} cm^{-2}$) & 5.93 & 4.07 \\
$Flux_{reflection}$ & ($10^{-9} erg s^{-1} cm^{-2}$) & 0.99 & 0.96 \\
\enddata
\caption{Best-fit parameters for the spectral models from the off-dip and dip in the June 2022 observation. The $kTe$ of the \texttt{NTHCOMP} model is fixed at 94.2\,keV \citep{2022Sci...378..650K}. \texttt{XILLVERCP} parameters are tied to those of \texttt{RELXILLCP}. Fluxes are evaluated in the 2.0--8.0\,keV band and cross-normalized to \nicer. Intrinsic component fluxes are reported for the disk, corona, and reflection.
Errors are reported at 68\% CL and upper limits at 90\% CL.}
\end{deluxetable*}

\subsection{Spectro-polarimetric analysis}

Based on the fitting results reported in Table~\ref{tab:Best-fit_Parameters}, we performed a spectro-polarimetric analysis by simultaneously fitting the Stokes $I$, $Q$, and $U$ spectra from \ixpe in the 2--8\,keV energy band.
To maintain the robustness of the spectral model during polarimetric fitting, all parameters of the model \texttt{MBPO * TBABS * TBPCF * (DISKBB + NTHCOMP + RELXILLCP + XILLVERCP)} were fixed at their best-fit values as obtained in Table~\ref{tab:Best-fit_Parameters}. Consequently, only the polarimetric parameters were allowed to vary during the joint fit. Hereafter, all the uncertainties are reported at the 68\% CL and upper limits at 90\% CL.

Firstly, we applied a single \texttt{polconst} component to the entire model to derive the integrated polarimetric parameters of the source, providing a direct comparison with the model-independent results obtained with the \textsc{PCUBE} algorithm. We obtained PD = $3.8\%\pm0.3\%$ and PA = $-26^{\circ}\pm2^{\circ}$ ($\chi^2/\text{d.o.f.} = 613/619$) for the off-dip and for the dip PD = $5\%\pm4\%$ and PA = $-70^{\circ}\pm30^{\circ}$ ($\chi^2/\text{d.o.f.} = 383/387$). These results are consistent with the model-independent results reported in Table~\ref{tab:Polarization_properties_of_dip_and_off-dip_time_intervals} for the off-dip, while the dip yields a low significant PD and a 1.7$\sigma$ discrepancy for the PA. Because of this, we performed a new fit using the linear polarization model \texttt{pollin} \citep{DiMarco26}. For the off-dip, the polarization parameters were constrained to $PD_\textrm{1\,keV} = 3.6\%\pm0.7\%$ and $PA_\textrm{1\,keV} = -35^{\circ}\pm5^{\circ}$, with the slopes for the PD $(0.1\pm0.2)$\,\%/keV and PA ($3\pm2$)\,$^{\circ}$/keV ($\chi^2/dof=607/617$). For the dipping phase, the fit yielded $\textrm{PD}_\textrm{1\,keV} = (22\pm10)\%$, $\textrm{PA}_\textrm{1\,keV} = -53^{+12}_{-20}{}^{\circ}$, and slopes for the PD $-6^{+9}_{-4}$\,\%/keV and PA $5^{+6}_{-8}$\,$^{\circ}$/keV ($\chi^2/\text{d.o.f.} = 379/385$). These results show no significant improvement in the fit when applying \texttt{pollin} instead of \texttt{polconst}.

To disentangle the polarization associated with the different spectral components, we performed the same spectro-polarimetric analysis, assuming different polarizations (via convolution with \texttt{polconst}) for each component. Table~\ref{tab:Best-fit_polarization} reports the results obtained for different assumptions. Leaving the polarization free to vary for each spectral component in the dip does not provide a constraint on the hard and reflection components, thereby confirming their degeneracy \citep{LaMonaca24, LaMonaca24gx340,LaMonaca25gx340NB, LaMonaca25_gx349}; similarly, in the off-dip, the polarization for the reflection component is not well determined. Considering that the reflection component is dominated by the Fe line that is expected to be unpolarized \citep{Churazov02, Veledina24} and has a low contribution to the total flux in the \ixpe energy band, as a second attempt we assumed it unpolarized as in \citealt{2022Sci...378..650K}; thus,  obtaining in the dip a polarization for the Comptonization of ${\sim}6\%$, and in the off-dip results polarization compatible with the case where the polarization of the reflection component was free to vary. On the other hand, freezing in the off-dip the polarization of the reflection at zero and the polarization for the Comptonization at the value measured in the dip, the disk component shows an unphysical polarization degree up to 89\%. An alternative scenario is one in which the polarization of the Comptonization is negligible, given the low inclination of the source \citep{st85}. It corresponds to a polarization in the dip that is due solely to the reflection component, yielding a polarization of 20--30\%. Assuming the polarization of the reflection in the off-dip at the value obtained for the dip yields a disk polarization of 72\% which is unphysical.

\begin{deluxetable*}{lccc}[!htb]
\tabletypesize{\scriptsize}
\label{tab:Best-fit_polarization}
\tablewidth{0pt}
\tablehead{
\colhead{} & 
\colhead{Parameter (unit)} & 
\colhead{Off-dip} & 
\colhead{overlap dip}
}
\startdata
POLCONST & $PD_1$ ($\%$) & $19\pm8$ & - \\
(DISKBB) & $PA_1$ ($^{\circ}$) & $-72\pm13$ & - \\[0.5em]
POLCONST & $PD_2$ ($\%$) & $5\pm5$ & $<30$ \\
(NTHCOMP) & $PA_2$ ($^{\circ}$) & $-10^{+30}_{-20}$ & - \\[0.5em]
POLCONST & $PD_3$ ($\%$) & $<40$ & $>20$ \\
(REFLECTION) & $PA_3$ ($^{\circ}$) &-& - \\[0.5em]
$\chi^{2}$/d.o.f. & & 607/615 & 380/383 \\[0.5em]
%\hline
%POLCONST & $PD_1$ ($\%$) & $[0.0]$ & $[0.0]$ \\
%(DISKBB) & $PA_1$ ($^{\circ}$) & $[0]$ & $[0]$ \\[0.5em]
%POLCONST & $PD_2$ ($\%$) & $5.5\pm4.9$ & $<25.5$ \\
%(NTHCOMP) & $PA_2$ ($^{\circ}$) & $-14^{+32}_{-26}$ & - \\[0.5em]
%POLCONST & $PD_3$ ($\%$) & $<43.9$ & - \\
%(REFLECTION) & $PA_3$ ($^{\circ}$) & - & - \\[0.5em]
%$\chi^{2}$/d.o.f. & & 612/617 & 382/385 \\[0.5em]
\hline
POLCONST & $PD_1$ ($\%$) & $20\pm8$ & - \\
(DISKBB) & $PA_1$ ($^{\circ}$) & $-72\pm12$ & - \\[0.5em]
POLCONST & $PD_2$ ($\%$) & $4.7\pm0.5$ & $6\pm5$ \\
(NTHCOMP) & $PA_2$ ($^{\circ}$) & $-20\pm3$ & $-70\pm30$ \\[0.5em]
POLCONST & $PD_3$ ($\%$) & $[0.0]$ & $[0.0]$ \\
(REFLECTION) & $PA_3$ ($^{\circ}$) & $[0]$ & $[0]$ \\[0.5em]
$\chi^{2}$/d.o.f. & & 607/617 & 383/387 \\[0.5em]
%\hline
%POLCONST & $PD_1$ ($\%$) & $[0.0]$ & $[0.0]$ \\
%(DISKBB) & $PA_1$ ($^{\circ}$) & $[0]$ & $[0]$ \\[0.5em]
%POLCONST & $PD_2$ ($\%$) & $4.6\pm0.3$ & $5.8\pm4.8$ \\
%(NTHCOMP) & $PA_2$ ($^{\circ}$) & $-26\pm2$ & $-71\pm28$ \\[0.5em]
%POLCONST & $PD_3$ ($\%$) & $[0.0]$ & $[0.0]$ \\
%(REFLECTION) & $PA_3$ ($^{\circ}$) & $[0]$ & $[0]$ \\[0.5em]
%$\chi^{2}$/d.o.f. & & 612/619 & 383/387 \\[0.5em]
%\hline
%POLCONST & $PD_1$ ($\%$) & $89\pm5$ & - \\
%(DISKBB) & $PA_1$ ($^{\circ}$) & $-2\pm2$ & - \\[0.5em]
%POLCONST & $PD_2$ ($\%$) & $[6]$ & $6\pm5$ \\
%(NTHCOMP) & $PA_2$ ($^{\circ}$) & $[-70]$ & $-70\pm30$ \\[0.5em]
%POLCONST & $PD_3$ ($\%$) & $[0.0]$ & $[0.0]$ \\
%(REFLECTION) & $PA_3$ ($^{\circ}$) & $[0]$ & $[0]$ \\[0.5em]
%$\chi^{2}$/d.o.f. & & 878/619 & 383/387 \\[0.5em]
\hline
POLCONST & $PD_1$ ($\%$) & $19.3\pm8.2$ & - \\
(DISKBB) & $PA_1$ ($^{\circ}$) & $-72\pm13$ & - \\[0.5em]
POLCONST & $PD_2$ ($\%$) & $[0.0]$ & $[0.0]$ \\
(NTHCOMP) & $PA_2$ ($^{\circ}$) & $[0]$ & $[0]$ \\[0.5em]
POLCONST & $PD_3$ ($\%$) & $29\pm3$ & $30\pm20$ \\
(REFLECTION) & $PA_3$ ($^{\circ}$) & $-21\pm3$ & $-70\pm30$ \\[0.5em]
$\chi^{2}$/d.o.f. & & 608/617 & 383/387 \\[0.5em]
%\hline
%POLCONST & $PD_1$ ($\%$) & $71.9\pm5.2$ & $[0.0]$ \\
%(DISKBB) & $PA_1$ ($^{\circ}$) & $-7\pm2$ & $[0]$ \\[0.5em]
%POLCONST & $PD_2$ ($\%$) & $[0.0]$ & $[0.0]$ \\
%(NTHCOMP) & $PA_2$ ($^{\circ}$) & $[0]$ & $[0]$ \\[0.5em]
%POLCONST & $PD_3$ ($\%$) & $[25.0]$ & $25.0\pm20.4$ \\
%(REFLECTION) & $PA_3$ ($^{\circ}$) & $[-70]$ & $-70^{+27}_{-28}$ \\[0.5em]
%$\chi^{2}$/d.o.f. & & 790/619 & 383/387 \\[0.5em]
\enddata
\caption{Spectro-polarimetric results for the off-dip and dip states using the spectral model in Table~\ref{tab:Best-fit_Parameters}. Errors are reported at 68\% CL.}
\end{deluxetable*}

\section{Discussion} \label{sec:Discussion}

\subsection{Polarization along the Cyg~X-1 states}

The PD of Cyg~X-1 shows a strong correlation with its hardness across different states, as reported in Figure~\ref{fig:Q_U_comparison}. In 2022, polarimetric measurements of Cyg~X-1 in its LHS by \ixpe revealed a PD of ${\sim}4$\% in the 2--8\,keV band with a PA aligned with the radio-jet direction \citep{2022Sci...378..650K}. In 2023, Cyg~X-1 transitioned from an intermediate state to HSS; in May and June, five \ixpe observations were performed, measuring a PD ${\sim}2$\% and a PA stable across different spectral states and consistently aligned with the direction of the radio jet \citep{2024ApJ...969L..30S,2024MNRAS.52710837J}. In 2024, \ixpe performed further observations of Cyg~X-1, four of these between April and June observed the source in the LHS with a PD varying from 3.1$\%$ to 4.6$\%$ \citep{2025A&A...701A.115K}, with an average value in agreement with that measured in the 2022 LHS observations, and the PA showed no significant variation with earlier observations. A fifth observation, which lasted from October to December in two different pointings, was split into two parts in our new study because the nearly two-month separation was sufficient for Cyg~X‑1 to undergo changes in its spectral state. Both pointings exhibited a similar PD of approximately 2.8$\%$, but showed a significant difference in hardness, indicating that the spectral state of Cyg~X‑1 continued to soften during the observation period \citep{2025A&A...701A.115K}.

Summarizing, the PD is lower in the soft state (${\sim}$1.7 $\%$), higher in the hard state (${\sim}$3.6 $\%$), and around 2.2\% in the intermediate state. This PD–hardness dependence, reported in Figure~\ref{fig:three_states_PD}, is observed in other BH-XRBs, while the PA remains relatively stable ($\sim$-25$^{\circ}$) and aligned with the radio jet across the different states. This result supports the idea that the X-ray emission is linked to the jet's origin and provides key evidence for a coronal structure extending parallel to the accretion disk, rather than being elongated along the jet axis as in the lamp-post scenarios. Given the relatively low inclination of the source, the high polarization degree also suggests that the inner accretion disk may be tilted relative to the outer disk \citep{2022Sci...378..650K}. The significant polarization observed in the soft state, higher than expected, may be attributed to the dominance of returning radiation \citep{2024ApJ...969L..30S}. 

These results align with the ones obtained for different BH-XRBs observed by \ixpe \cite{Dovciak24}. These similarities suggest remarkable consistency across different sources and luminosities in the polarimetric properties of BH-XRBs. BHs and neutron stars in low-mass XRBs (NS-LMXBs), particularly Z-sources, share a similar accretion mechanism and a jet-launching correlation with hardness, but their polarimetric behavior shows a partially distinct pattern. In fact, the hardest horizontal branch and BH hard states exhibit similar polarimetric properties, and the BH soft state resembles the normal and flaring branches in terms of polarization. The analog of the BH intermediate state seems to be missing in NS-LMXBs. A further difference is an almost constant PA across different observations of BH-XRBs, which is not clearly established for NS-LMXBs, where indications of PA variability with state and hardness have been reported \citep{Rankin24, LaMonaca24, 2025AN....34640126D,2025ApJ...979L..47D}. The observed differences can be attributed to the presence of the NS surface, which produces a different Comptonizing region in the inner region. 

\subsection{Polarization along the dips}

In particular, in this new study, we performed a polarimetric and spectro-polarimetric analysis of Cyg~X-1 during its dips. The study was driven by evidence of a correlation between flux and polarization in dipping sources, attributed to the obscuration of an extended, oblate corona \citep[see, e.g.,][]{2025ApJ...979L..47D}. The simultaneous observation of Cyg~X-1 in 2022 by \ixpe, \nicer, and \nustar provides an excellent opportunity for this study. A comparison between the light curves and the hardness ratio revealed an increase in hardness as flux decreased across data from all three observatories, enabling a clear selection of dips. Figure~\ref{fig:Light_Curves_and_IXPE_Hardness} shows the decrease in the light curves, which was more pronounced for \ixpe and \nicer due to their softer energy bands than \nustar. Between 2023 and 2026, further observations of Cyg~X-1 were conducted with \ixpe, revealing dips in the light curve (see Figure~\ref{fig:IXPE_lc_hardness_other}). However, the duration of these dips was too short to meet the statistical requirements for further analysis, and multi-observatory coverage was lacking, preventing spectro-polarimetric characterization.

Figure~\ref{fig:dip_off-dip_comparation} shows an indication of the energy dependence of the polarization in both dip and off-dip data, with no significant variation in this trend. In Table~\ref{tab:Best-fit_Parameters}, we report the best-fit for the spectral analysis in both the off-dip and dip intervals. We report, in the dip, an increase in absorption due to a covering fraction, modeled with \texttt{TBpcf}, that significantly impacts the energy bands of \ixpe and \nicer. In fact, this yields a negligible disk contribution to the X-ray emission in the dip state within the \ixpe energy band. In the dips, even if the \texttt{xillvercp} component is not well constrained, it has an upper limit on the normalization that agrees with the value measured in the off-dip; this means that both reflection and Comptonization are not strongly affected by the higher absorption. We also calculated the percentage change of low-energy photons in the off-dip and dip intervals for (1--3.5\,keV) and (3--3.5\,keV). During the dip interval, the fraction of soft photons detected in (1--3.5\,keV) decreased from ${\sim}70$\% to ${\sim}50$\%, while for (3--3.5\,keV), the drop is from ${\sim}11$\% to ${\sim}9$\%. It means that only photons below 3 keV are impacted by this effect. Given that the observed dips are due either to occultation by dense, cold clumps in the outer disk along the line of sight or to atmospheric absorption by the companion star, the spectral analysis result is consistent with this scenario. We performed the polarimetric analysis in the energy bins 2--3\,keV and 2--3.5\,keV, obtaining similar results in both cases, with compatible PD and PA between dip and off-dip. In particular, the May 2022 observation provides PD $2.5\pm1.5$\% at PA $-28^\circ\pm17^\circ$ in the dip and $3.5\pm0.3$\% at $-18^\circ\pm2^\circ$ in the off-dip for data in the 2--3\,keV energy band. Similarly, in June 2022 data, we measured PD $2\pm2$\% at PA $22^\circ\pm30^\circ$ in the dip and $3.9\pm0.4$\% at $-31^\circ\pm3^\circ$ in the off-dip for data in the energy band 2--3\,keV. These results show that the dip state is not significant enough to support any claim, but it is interesting to observe that the absence of the disk component does not seem to affect the polarization in the softer energy bins. It means that the disk component is too faint to contribute to the total polarization measured for Cyg~X-1.

The Cyg~X-1 result, although not highly significant, is very interesting to investigate in future observations of BHs, because these seem to have a completely different phenomenology with respect to NS-LMXBs, where dips sufficiently long in duration provide puzzling polarimetric variability, \citep[see, e.g.][]{2025ApJ...979L..47D,2025AN....34640126D} which is strongly related to the presence of dips \citep{2025ApJ...979L..47D}. Moreover, a recent study of the high-mass XRB \mbox{4U~1700–37} displayed a similar polarimetric pattern during eclipses \citep{2026MNRAS.tmp.1047W}. The fact that Cyg~X-1 shows no polarimetric variability could mean that the corona in BH-XRBs is completely different from that of NS-LMXBs and that the polarization is dominated by the corona. Also, the unchanged PA in the dip and off-dip states indicates that the geometry of the X-ray-emitting regions that give rise to the measured polarization in the hard state is not affected by the dips, confirming the idea of an extended slab-like corona as reported in \cite{Kraw2022}.

\section{Conclusions} \label{sec:Summary}

Dips in the Cyg~X-1 light curve are associated with the appearance of clumps, which are episodic and strongly absorb low-energy photons, producing dips in the light curve and hardening in the spectra \citep{2025ApJ...979L..47D}. Cyg~X-1 and its companion star undergo periodic orbital motion. Although the companion star does not directly block photons to produce eclipses, its atmosphere can strongly absorb photons during the superior conjunction phase, and both the \ixpe observations show dips starting around this phase: the orbital phase in May 2022 ranges from -0.10 to 0.98, while in June 2022 it ranges from 0.01 to 0.36. Beyond Cyg~X-1, the same phenomenon has been identified in other black hole X-ray binaries observed by \ixpe. An example is also provided by the observations of the source IGR J17091-3624 conducted in March 2025 \citep{2025MNRAS.541.1774E,2025ApJ...989..165D}. Despite a high PD, the source was not sufficiently bright, and the limited exposure time resulted in a polarization measurement that was not statistically significant. For dip polarization studies, the first limitation is the duration, which limits statistical significance. Therefore, the 2022 observations of Cyg~X-1, which provide adequate data, make it an exceptionally valuable case. Observations of Cyg~X-1 during May and June 2022 revealed a distinct anti-correlation between source flux and hardness ratio. Spectral analysis confirmed that the changes during dips are predominantly concentrated in lower-energy bands. The joint spectral analysis indicates that the dip spectrum is best fitted by a model incorporating additional partial-covering absorption (the \texttt{TBPCF} model). The results show that during dips the polarization remains unchanged, which means that the disk is not affecting the polarimetric result and possibly that the emitting region where polarization originates is an extended accretion-disk corona, possibly oblate, with a size sufficiently large to make obscuration from the observed clumps negligible in changing its geometry. Future observations, possibly with higher statistical significance, will help to dig into the geometry of the outer region of the accretion disk; considering the limited effective area of \ixpe, eXTP should be able in the future to better investigate polarimetric properties of dipping sources, also for narrow dips lasting ${\sim}1$\,h.

%% Please use the acknowledgment and contribution environments. This will 
%% be anonomyized when the "anonymous" style option is used. 
\begin{acknowledgments}
This work is supported by National Natural Science Foundation of China (grant No. 12422306 and No. 12373041), and Natural Science Foundation of Guangxi (grant Nos. 2025GXNSFDA02850001), and Bagui Scholars Program (XF). 
This work is also supported by the Guangxi Science and Technology Innovation Platform Program (Leitai Action Plan, Grant No. Guike LT2600640026), Guangxi Key R$\&$D Program (Guangxi Funeng Action Plan, Grant No. Guike FN2504240040), and the ``Guangxi Highland of Innovation Talents'' Program.

\end{acknowledgments}

\bibliography{References}{}

@ARTICLE{Holt1976,
       author = {{Holt}, S.~S. and {Boldt}, E.~A. and {Serlemitsos}, P.~J. and {Kaluzienski}, L.~J.},
        title = "{New results from long-term observations of Cygnus X-1.}",
      journal = {\apjl},
         year = 1976,
        month = jan,
       volume = {203},
        pages = {L63-L66},
          doi = {10.1086/182020},
       adsurl = {https://ui.adsabs.harvard.edu/abs/1976ApJ...203L..63H}
}

@ARTICLE{Ichimaru1977,
       author = {{Ichimaru}, S.},
        title = "{Bimodal behavior of accretion disks: theory and application to Cygnus X-1 transitions.}",
      journal = {\apj},
         year = 1977,
        month = jun,
       volume = {214},
        pages = {840-855},
          doi = {10.1086/155314},
       adsurl = {https://ui.adsabs.harvard.edu/abs/1977ApJ...214..840I}
}

@ARTICLE{2001ApJS..132..377H,
       author = {{Homan}, Jeroen and {Wijnands}, Rudy and {van der Klis}, Michiel and {Belloni}, Tomaso and {van Paradijs}, Jan and {Klein-Wolt}, Marc and {Fender}, Rob and {M{\'e}ndez}, Mariano},
        title = "{Correlated X-Ray Spectral and Timing Behavior of the Black Hole Candidate XTE J1550-564: A New Interpretation of Black Hole States}",
      journal = {\apjs},
         year = 2001,
        month = feb,
       volume = {132},
       number = {2},
        pages = {377-402},
          doi = {10.1086/318954},
archivePrefix = {arXiv},
       eprint = {astro-ph/0001163},
 primaryClass = {astro-ph},
       adsurl = {https://ui.adsabs.harvard.edu/abs/2001ApJS..132..377H}
}

@ARTICLE{2004MNRAS.355.1105F,
       author = {{Fender}, R.~P. and {Belloni}, T.~M. and {Gallo}, E.},
        title = "{Towards a unified model for black hole X-ray binary jets}",
      journal = {\mnras},
         year = 2004,
        month = dec,
       volume = {355},
       number = {4},
        pages = {1105-1118},
          doi = {10.1111/j.1365-2966.2004.08384.x},
archivePrefix = {arXiv},
       eprint = {astro-ph/0409360},
 primaryClass = {astro-ph},
       adsurl = {https://ui.adsabs.harvard.edu/abs/2004MNRAS.355.1105F}
}

@ARTICLE{2005A&A...440..207B,
       author = {{Belloni}, T. and {Homan}, J. and {Casella}, P. and {van der Klis}, M. and {Nespoli}, E. and {Lewin}, W.~H.~G. and {Miller}, J.~M. and {M{\'e}ndez}, M.},
        title = "{The evolution of the timing properties of the black-hole transient GX 339-4 during its 2002/2003 outburst}",
      journal = {\aap},
         year = 2005,
        month = sep,
       volume = {440},
       number = {1},
        pages = {207-222},
          doi = {10.1051/0004-6361:20042457},
archivePrefix = {arXiv},
       eprint = {astro-ph/0504577},
 primaryClass = {astro-ph},
       adsurl = {https://ui.adsabs.harvard.edu/abs/2005A&A...440..207B}
}

@INPROCEEDINGS{2016ASSL..440...61B,
       author = {{Belloni}, Tomaso M. and {Motta}, Sara E.},
        title = "{Transient Black Hole Binaries}",
    booktitle = {Astrophysics of Black Holes: From Fundamental Aspects to Latest Developments},
         year = 2016,
       editor = {{Bambi}, Cosimo},
       series = {Astrophysics and Space Science Library},
       volume = {440},
        month = jan,
        pages = {61},
          doi = {10.1007/978-3-662-52859-4_2},
archivePrefix = {arXiv},
       eprint = {1603.07872},
 primaryClass = {astro-ph.HE},
       adsurl = {https://ui.adsabs.harvard.edu/abs/2016ASSL..440...61B}
}

@ARTICLE{Zdziarski2004,
       author = {{Zdziarski}, A.~A. and {Gierli{\'n}ski}, M.},
        title = "{Radiative Processes, Spectral States and Variability of Black-Hole Binaries}",
      journal = {Progress of Theoretical Physics Supplement},
         year = 2004,
        month = jan,
       volume = {155},
        pages = {99-119},
          doi = {10.1143/PTPS.155.99},
archivePrefix = {arXiv},
       eprint = {astro-ph/0403683},
 primaryClass = {astro-ph},
       adsurl = {https://ui.adsabs.harvard.edu/abs/2004PThPS.155...99Z}
}

@INPROCEEDINGS{Shakura1973,
       author = {{Shakura}, N.~I. and {Sunyaev}, R.~A.},
        title = "{Black Holes in Binary Systems: Observational Appearances}",
    booktitle = {X- and Gamma-Ray Astronomy},
         year = 1973,
       editor = {{Bradt}, H. and {Giacconi}, Riccardo},
       series = {IAU Symposium},
       volume = {55},
        month = jan,
        pages = {155},
       adsurl = {https://ui.adsabs.harvard.edu/abs/1973IAUS...55..155S}
}

@ARTICLE{st1980,
       author = {{Sunyaev}, R.~A. and {Titarchuk}, L.~G.},
        title = "{Comptonization of X-Rays in Plasma Clouds - Typical Radiation Spectra}",
      journal = {\aap},
         year = 1980,
        month = jun,
       volume = {86},
        pages = {121},
       adsurl = {https://ui.adsabs.harvard.edu/abs/1980A&A....86..121S}
}

@ARTICLE{Gierlinski1997,
       author = {{Gierlinski}, Marek and {Zdziarski}, Andrzej A. and {Done}, Chris and {Johnson}, W. Neil and {Ebisawa}, Ken and {Ueda}, Yoshihiro and {Haardt}, Francesco and {Phlips}, Bernard F.},
        title = "{Simultaneous X-ray and gamma-ray observations of CYG X-1 in the hard state by GINGA and OSSE}",
      journal = {\mnras},
         year = 1997,
        month = jul,
       volume = {288},
       number = {4},
        pages = {958-964},
          doi = {10.1093/mnras/288.4.958},
archivePrefix = {arXiv},
       eprint = {astro-ph/9610156},
 primaryClass = {astro-ph},
       adsurl = {https://ui.adsabs.harvard.edu/abs/1997MNRAS.288..958G}
}

@ARTICLE{Jiang2024,
       author = {{Jiang}, Jiachen},
        title = "{Fifty Years After the Discovery of the First Stellar-Mass Black Hole: A Review of Cyg X-1}",
      journal = {Galaxies},
         year = 2024,
        month = nov,
       volume = {12},
       number = {6},
          eid = {80},
        pages = {80},
          doi = {10.3390/galaxies12060080},
archivePrefix = {arXiv},
       eprint = {2411.12507},
 primaryClass = {astro-ph.HE},
       adsurl = {https://ui.adsabs.harvard.edu/abs/2024Galax..12...80J}
}

@ARTICLE{Bowyer1965,
       author = {{Bowyer}, S. and {Byram}, E.~T. and {Chubb}, T.~A. and {Friedman}, H.},
        title = "{Cosmic X-ray Sources}",
      journal = {Science},
         year = 1965,
        month = jan,
       volume = {147},
       number = {3656},
        pages = {394-398},
          doi = {10.1126/science.147.3656.394},
       adsurl = {https://ui.adsabs.harvard.edu/abs/1965Sci...147..394B}
}

@ARTICLE{Tananbaum1972,
       author = {{Tananbaum}, H. and {Gursky}, H. and {Kellogg}, E. and {Giacconi}, R. and {Jones}, C.},
        title = "{Observation of a Correlated X-Ray Transition in Cygnus X-1}",
      journal = {\apjl},
         year = 1972,
        month = oct,
       volume = {177},
        pages = {L5},
          doi = {10.1086/181042},
       adsurl = {https://ui.adsabs.harvard.edu/abs/1972ApJ...177L...5T}
}

@ARTICLE{Done2007,
       author = {{Done}, Chris and {Gierli{\'n}ski}, Marek and {Kubota}, Aya},
        title = "{Modelling the behaviour of accretion flows in X-ray binaries. Everything you always wanted to know about accretion but were afraid to ask}",
      journal = {\aapr},
         year = 2007,
        month = dec,
       volume = {15},
       number = {1},
        pages = {1-66},
          doi = {10.1007/s00159-007-0006-1},
archivePrefix = {arXiv},
       eprint = {0708.0148},
 primaryClass = {astro-ph},
       adsurl = {https://ui.adsabs.harvard.edu/abs/2007A&ARv..15....1D}
}

@ARTICLE{2006A&A...447..245W,
       author = {{Wilms}, J. and {Nowak}, M.~A. and {Pottschmidt}, K. and {Pooley}, G.~G. and {Fritz}, S.},
        title = "{Long term variability of Cygnus X-1. IV. Spectral evolution 1999-2004}",
      journal = {\aap},
         year = 2006,
        month = feb,
       volume = {447},
       number = {1},
        pages = {245-261},
          doi = {10.1051/0004-6361:20053938},
archivePrefix = {arXiv},
       eprint = {astro-ph/0510193},
 primaryClass = {astro-ph},
       adsurl = {https://ui.adsabs.harvard.edu/abs/2006A&A...447..245W}
}

@ARTICLE{2013A&A...554A..88G,
       author = {{Grinberg}, V. and {Hell}, N. and {Pottschmidt}, K. and {B{\"o}ck}, M. and {Nowak}, M.~A. and {Rodriguez}, J. and {Bodaghee}, A. and {Cadolle Bel}, M. and {Case}, G.~L. and {Hanke}, M. and {K{\"u}hnel}, M. and {Markoff}, S.~B. and {Pooley}, G.~G. and {Rothschild}, R.~E. and {Tomsick}, J.~A. and {Wilson-Hodge}, C.~A. and {Wilms}, J.},
        title = "{Long term variability of Cygnus X-1. V. State definitions with all sky monitors}",
      journal = {\aap},
         year = 2013,
        month = jun,
       volume = {554},
          eid = {A88},
        pages = {A88},
          doi = {10.1051/0004-6361/201321128},
archivePrefix = {arXiv},
       eprint = {1303.1198},
 primaryClass = {astro-ph.HE},
       adsurl = {https://ui.adsabs.harvard.edu/abs/2013A&A...554A..88G}
}

@ARTICLE{Fujii2025,
       author = {{Fujii}, Taisei and {Kitamoto}, Shunji and {Fukuichi}, Makoto and {Ohwada}, Soma and {Sawada}, Makoto},
        title = "{X-ray emitting region of Cygnus X-1 constrained by short absorption dips}",
      journal = {\pasj},
         year = 2025,
        month = dec,
       volume = {77},
       number = {6},
        pages = {1290-1300},
          doi = {10.1093/pasj/psaf107},
       adsurl = {https://ui.adsabs.harvard.edu/abs/2025PASJ...77.1290F}
}

@ARTICLE{Balu1995,
       author = {{Balucinska-Church}, M. and {Belloni}, T. and {Church}, M.~J. and {Hasinger}, G.},
        title = "{Identification of the soft X-ray excess in Cygnus X-1 with disc emission.}",
      journal = {\aap},
         year = 1995,
        month = oct,
       volume = {302},
        pages = {L5},
          doi = {10.48550/arXiv.astro-ph/9509020},
archivePrefix = {arXiv},
       eprint = {astro-ph/9509020},
 primaryClass = {astro-ph},
       adsurl = {https://ui.adsabs.harvard.edu/abs/1995A&A...302L...5B}
}

@ARTICLE{Poutanen2008,
       author = {{Poutanen}, Juri and {Zdziarski}, Andrzej A. and {Ibragimov}, Askar},
        title = "{Superorbital variability of X-ray and radio emission of Cyg X-1 - II. Dependence of the orbital modulation and spectral hardness on the superorbital phase}",
      journal = {\mnras},
         year = 2008,
        month = sep,
       volume = {389},
       number = {3},
        pages = {1427-1438},
          doi = {10.1111/j.1365-2966.2008.13666.x},
archivePrefix = {arXiv},
       eprint = {0802.1391},
 primaryClass = {astro-ph},
       adsurl = {https://ui.adsabs.harvard.edu/abs/2008MNRAS.389.1427P}
}

@ARTICLE{Sund2018,
       author = {{Sundqvist}, J.~O. and {Owocki}, S.~P. and {Puls}, J.},
        title = "{2D wind clumping in hot, massive stars from hydrodynamical line-driven instability simulations using a pseudo-planar approach}",
      journal = {\aap},
         year = 2018,
        month = mar,
       volume = {611},
          eid = {A17},
        pages = {A17},
          doi = {10.1051/0004-6361/201731718},
archivePrefix = {arXiv},
       eprint = {1710.07780},
 primaryClass = {astro-ph.SR},
       adsurl = {https://ui.adsabs.harvard.edu/abs/2018A&A...611A..17S}
}

@ARTICLE{2022JATIS...8b6002W,
       author = {{Weisskopf}, Martin C. and {Soffitta}, Paolo and {Baldini}, Luca and {Ramsey}, Brian D. and {O'Dell}, Stephen L. and {Romani}, Roger W. and {Matt}, Giorgio and {Deininger}, William D. and {Baumgartner}, Wayne H. and {Bellazzini}, Ronaldo and {Costa}, Enrico and {Kolodziejczak}, Jeffery J. and {Latronico}, Luca and {Marshall}, Herman L. and {Muleri}, Fabio and {Bongiorno}, Stephen D. and {Tennant}, Allyn and {Bucciantini}, Niccolo and {Dovciak}, Michal and {Marin}, Frederic and {Marscher}, Alan and {Poutanen}, Juri and {Slane}, Pat and {Turolla}, Roberto and {Kalinowski}, William and {Di Marco}, Alessandro and {Fabiani}, Sergio and {Minuti}, Massimo and {La Monaca}, Fabio and {Pinchera}, Michele and {Rankin}, John and {Sgro'}, Carmelo and {Trois}, Alessio and {Xie}, Fei and {Alexander}, Cheryl and {Allen}, D. Zachery and {Amici}, Fabrizio and {Andersen}, Jason and {Antonelli}, Angelo and {Antoniak}, Spencer and {Attin{\`a}}, Primo and {Barbanera}, Mattia and {Bachetti}, Matteo and {Baggett}, Randy M. and {Bladt}, Jeff and {Brez}, Alessandro and {Bonino}, Raffaella and {Boree}, Christopher and {Borotto}, Fabio and {Breeding}, Shawn and {Brienza}, Daniele and {Bygott}, H. Kyle and {Caporale}, Ciro and {Cardelli}, Claudia and {Carpentiero}, Rita and {Castellano}, Simone and {Castronuovo}, Marco and {Cavalli}, Luca and {Cavazzuti}, Elisabetta and {Ceccanti}, Marco and {Centrone}, Mauro and {Citraro}, Saverio and {D'Amico}, Fabio and {D'Alba}, Elisa and {Di Gesu}, Laura and {Del Monte}, Ettore and {Dietz}, Kurtis L. and {Di Lalla}, Niccolo' and {Persio}, Giuseppe Di and {Dolan}, David and {Donnarumma}, Immacolata and {Evangelista}, Yuri and {Ferrant}, Kevin and {Ferrazzoli}, Riccardo and {Ferrie}, MacKenzie and {Footdale}, Joseph and {Forsyth}, Brent and {Foster}, Michelle and {Garelick}, Benjamin and {Gunji}, Shuichi and {Gurnee}, Eli and {Head}, Michael and {Hibbard}, Grant and {Johnson}, Samantha and {Kelly}, Erik and {Kilaru}, Kiranmayee and {Lefevre}, Carlo and {Roy}, Shelley Le and {Loffredo}, Pasqualino and {Lorenzi}, Paolo and {Lucchesi}, Leonardo and {Maddox}, Tyler and {Magazzu}, Guido and {Maldera}, Simone and {Manfreda}, Alberto and {Mangraviti}, Elio and {Marengo}, Marco and {Marrocchesi}, Alessandra and {Massaro}, Francesco and {Mauger}, David and {McCracken}, Jeffrey and {McEachen}, Michael and {Mize}, Rondal and {Mereu}, Paolo and {Mitchell}, Scott and {Mitsuishi}, Ikuyuki and {Morbidini}, Alfredo and {Mosti}, Federico and {Nasimi}, Hikmat and {Negri}, Barbara and {Negro}, Michela and {Nguyen}, Toan and {Nitschke}, Isaac and {Nuti}, Alessio and {Onizuka}, Mitch and {Oppedisano}, Chiara and {Orsini}, Leonardo and {Osborne}, Darren and {Pacheco}, Richard and {Paggi}, Alessandro and {Painter}, Will and {Pavelitz}, Steven D. and {Pentz}, Christina and {Piazzolla}, Raffaele and {Perri}, Matteo and {Pesce-Rollins}, Melissa and {Peterson}, Colin and {Pilia}, Maura and {Profeti}, Alessandro and {Puccetti}, Simonetta and {Ranganathan}, Jaganathan and {Ratheesh}, Ajay and {Reedy}, Lee and {Root}, Noah and {Rubini}, Alda and {Ruswick}, Stephanie and {Sanchez}, Javier and {Sarra}, Paolo and {Santoli}, Francesco and {Scalise}, Emanuele and {Sciortino}, Andrea and {Schroeder}, Christopher and {Seek}, Tim and {Sosdian}, Kalie and {Spandre}, Gloria and {Speegle}, Chet O. and {Tamagawa}, Toru and {Tardiola}, Marcello and {Tobia}, Antonino and {Thomas}, Nicholas E. and {Valerie}, Robert and {Vimercati}, Marco and {Walden}, Amy L. and {Weddendorf}, Bruce and {Wedmore}, Jeffrey and {Welch}, David and {Zanetti}, Davide and {Zanetti}, Francesco},
        title = "{The Imaging X-Ray Polarimetry Explorer (IXPE): Pre-Launch}",
      journal = {Journal of Astronomical Telescopes, Instruments, and Systems},
         year = 2022,
        month = apr,
       volume = {8},
       number = {2},
          eid = {026002},
        pages = {026002},
          doi = {10.1117/1.JATIS.8.2.026002},
archivePrefix = {arXiv},
       eprint = {2112.01269},
 primaryClass = {astro-ph.IM},
       adsurl = {https://ui.adsabs.harvard.edu/abs/2022JATIS...8b6002W}
}

@ARTICLE{Poutanen1996,
       author = {{Poutanen}, Juri and {Svensson}, Roland},
        title = "{The Two-Phase Pair Corona Model for Active Galactic Nuclei and X-Ray Binaries: How to Obtain Exact Solutions}",
      journal = {\apj},
         year = 1996,
        month = oct,
       volume = {470},
        pages = {249},
          doi = {10.1086/177865},
archivePrefix = {arXiv},
       eprint = {astro-ph/9605073},
 primaryClass = {astro-ph},
       adsurl = {https://ui.adsabs.harvard.edu/abs/1996ApJ...470..249P}
}

@ARTICLE{2021AJ....162..208S,
       author = {{Soffitta}, Paolo and {Baldini}, Luca and {Bellazzini}, Ronaldo and {Costa}, Enrico and {Latronico}, Luca and {Muleri}, Fabio and {Del Monte}, Ettore and {Fabiani}, Sergio and {Minuti}, Massimo and {Pinchera}, Michele and {Sgro'}, Carmelo and {Spandre}, Gloria and {Trois}, Alessio and {Amici}, Fabrizio and {Andersson}, Hans and {Attina'}, Primo and {Bachetti}, Matteo and {Barbanera}, Mattia and {Borotto}, Fabio and {Brez}, Alessandro and {Brienza}, Daniele and {Caporale}, Ciro and {Cardelli}, Claudia and {Carpentiero}, Rita and {Castellano}, Simone and {Castronuovo}, Marco and {Cavalli}, Luca and {Cavazzuti}, Elisabetta and {Ceccanti}, Marco and {Centrone}, Mauro and {Ciprini}, Stefano and {Citraro}, Saverio and {D'Amico}, Fabio and {D'Alba}, Elisa and {Di Cosimo}, Sergio and {Di Lalla}, Niccolo' and {Di Marco}, Alessandro and {Di Persio}, Giuseppe and {Donnarumma}, Immacolata and {Evangelista}, Yuri and {Ferrazzoli}, Riccardo and {Hayato}, Asami and {Kitaguchi}, Takao and {La Monaca}, Fabio and {Lefevre}, Carlo and {Loffredo}, Pasqualino and {Lorenzi}, Paolo and {Lucchesi}, Leonardo and {Magazzu}, Carlo and {Maldera}, Simone and {Manfreda}, Alberto and {Mangraviti}, Elio and {Marengo}, Marco and {Matt}, Giorgio and {Mereu}, Paolo and {Morbidini}, Alfredo and {Mosti}, Federico and {Nakano}, Toshio and {Nasimi}, Hikmat and {Negri}, Barbara and {Nenonen}, Seppo and {Nuti}, Alessio and {Orsini}, Leonardo and {Perri}, Matteo and {Pesce-Rollins}, Melissa and {Piazzolla}, Raffaele and {Pilia}, Maura and {Profeti}, Alessandro and {Puccetti}, Simonetta and {Rankin}, John and {Ratheesh}, Ajay and {Rubini}, Alda and {Santoli}, Francesco and {Sarra}, Paolo and {Scalise}, Emanuele and {Sciortino}, Andrea and {Tamagawa}, Toru and {Tardiola}, Marcello and {Tobia}, Antonino and {Vimercati}, Marco and {Xie}, Fei},
        title = "{The Instrument of the Imaging X-Ray Polarimetry Explorer}",
      journal = {\aj},
         year = 2021,
        month = nov,
       volume = {162},
       number = {5},
          eid = {208},
        pages = {208},
          doi = {10.3847/1538-3881/ac19b0},
archivePrefix = {arXiv},
       eprint = {2108.00284},
 primaryClass = {astro-ph.IM},
       adsurl = {https://ui.adsabs.harvard.edu/abs/2021AJ....162..208S}
}

@ARTICLE{Poutanen2018,
       author = {{Poutanen}, Juri and {Veledina}, Alexandra and {Zdziarski}, Andrzej A.},
        title = "{Doughnut strikes sandwich: the geometry of hot medium in accreting black hole X-ray binaries}",
      journal = {\aap},
         year = 2018,
        month = jun,
       volume = {614},
          eid = {A79},
        pages = {A79},
          doi = {10.1051/0004-6361/201732345},
archivePrefix = {arXiv},
       eprint = {1711.08509},
 primaryClass = {astro-ph.HE},
       adsurl = {https://ui.adsabs.harvard.edu/abs/2018A&A...614A..79P}
}

@ARTICLE{Dovciak2004,
       author = {{Dov{\v{c}}iak}, M. and {Karas}, V. and {Matt}, G.},
        title = "{Polarization signatures of strong gravity in active galactic nuclei accretion discs}",
      journal = {\mnras},
         year = 2004,
        month = dec,
       volume = {355},
       number = {3},
        pages = {1005-1009},
          doi = {10.1111/j.1365-2966.2004.08396.x},
archivePrefix = {arXiv},
       eprint = {astro-ph/0409356},
 primaryClass = {astro-ph},
       adsurl = {https://ui.adsabs.harvard.edu/abs/2004MNRAS.355.1005D}
}

@ARTICLE{Kraw2022,
       author = {{Krawczynski}, H. and {Beheshtipour}, B.},
        title = "{New Constraints on the Spin of the Black Hole Cygnus X-1 and the Physical Properties of its Accretion Disk Corona}",
      journal = {\apj},
         year = 2022,
        month = jul,
       volume = {934},
       number = {1},
          eid = {4},
        pages = {4},
          doi = {10.3847/1538-4357/ac7725},
archivePrefix = {arXiv},
       eprint = {2201.07360},
 primaryClass = {astro-ph.HE},
       adsurl = {https://ui.adsabs.harvard.edu/abs/2022ApJ...934....4K}
}

@ARTICLE{2022Sci...378..650K,
       author = {{Krawczynski}, Henric and {Muleri}, Fabio and {Dov{\v{c}}iak}, Michal and {Veledina}, Alexandra and {Rodriguez Cavero}, Nicole and {Svoboda}, Jiri and {Ingram}, Adam and {Matt}, Giorgio and {Garcia}, Javier A. and {Loktev}, Vladislav and {Negro}, Michela and {Poutanen}, Juri and {Kitaguchi}, Takao and {Podgorn{\'y}}, Jakub and {Rankin}, John and {Zhang}, Wenda and {Berdyugin}, Andrei and {Berdyugina}, Svetlana V. and {Bianchi}, Stefano and {Blinov}, Dmitry and {Capitanio}, Fiamma and {Di Lalla}, Niccol{\`o} and {Draghis}, Paul and {Fabiani}, Sergio and {Kagitani}, Masato and {Kravtsov}, Vadim and {Kiehlmann}, Sebastian and {Latronico}, Luca and {Lutovinov}, Alexander A. and {Mandarakas}, Nikos and {Marin}, Fr{\'e}d{\'e}ric and {Marinucci}, Andrea and {Miller}, Jon M. and {Mizuno}, Tsunefumi and {Molkov}, Sergey V. and {Omodei}, Nicola and {Petrucci}, Pierre-Olivier and {Ratheesh}, Ajay and {Sakanoi}, Takeshi and {Semena}, Andrei N. and {Skalidis}, Raphael and {Soffitta}, Paolo and {Tennant}, Allyn F. and {Thalhammer}, Phillipp and {Tombesi}, Francesco and {Weisskopf}, Martin C. and {Wilms}, Joern and {Zhang}, Sixuan and {Agudo}, Iv{\'a}n and {Antonelli}, Lucio A. and {Bachetti}, Matteo and {Baldini}, Luca and {Baumgartner}, Wayne H. and {Bellazzini}, Ronaldo and {Bongiorno}, Stephen D. and {Bonino}, Raffaella and {Brez}, Alessandro and {Bucciantini}, Niccol{\`o} and {Castellano}, Simone and {Cavazzuti}, Elisabetta and {Ciprini}, Stefano and {Costa}, Enrico and {De Rosa}, Alessandra and {Del Monte}, Ettore and {Di Gesu}, Laura and {Di Marco}, Alessandro and {Donnarumma}, Immacolata and {Doroshenko}, Victor and {Ehlert}, Steven R. and {Enoto}, Teruaki and {Evangelista}, Yuri and {Ferrazzoli}, Riccardo and {Gunji}, Shuichi and {Hayashida}, Kiyoshi and {Heyl}, Jeremy and {Iwakiri}, Wataru and {Jorstad}, Svetlana G. and {Karas}, Vladimir and {Kolodziejczak}, Jeffery J. and {La Monaca}, Fabio and {Liodakis}, Ioannis and {Maldera}, Simone and {Manfreda}, Alberto and {Marscher}, Alan P. and {Marshall}, Herman L. and {Mitsuishi}, Ikuyuki and {Ng}, Chi-Yung and {O{\textquoteright}Dell}, Stephen L. and {Oppedisano}, Chiara and {Papitto}, Alessandro and {Pavlov}, George G. and {Peirson}, Abel L. and {Perri}, Matteo and {Pesce-Rollins}, Melissa and {Pilia}, Maura and {Possenti}, Andrea and {Puccetti}, Simonetta and {Ramsey}, Brian D. and {Romani}, Roger W. and {Sgr{\`o}}, Carmelo and {Slane}, Patrick and {Spandre}, Gloria and {Tamagawa}, Toru and {Tavecchio}, Fabrizio and {Taverna}, Roberto and {Tawara}, Yuzuru and {Thomas}, Nicholas E. and {Trois}, Alessio and {Tsygankov}, Sergey and {Turolla}, Roberto and {Vink}, Jacco and {Wu}, Kinwah and {Xie}, Fei and {Zane}, Silvia},
        title = "{Polarized x-rays constrain the disk-jet geometry in the black hole x-ray binary Cygnus X-1}",
      journal = {Science},
         year = 2022,
        month = nov,
       volume = {378},
       number = {6620},
        pages = {650-654},
          doi = {10.1126/science.add5399},
archivePrefix = {arXiv},
       eprint = {2206.09972},
 primaryClass = {astro-ph.HE},
       adsurl = {https://ui.adsabs.harvard.edu/abs/2022Sci...378..650K}
}

@ARTICLE{2025ApJ...979L..47D,
       author = {{Di Marco}, Alessandro and {La Monaca}, Fabio and {Bobrikova}, Anna and {Stella}, Luigi and {Papitto}, Alessandro and {Poutanen}, Juri and {Baglio}, Maria Cristina and {Bachetti}, Matteo and {Loktev}, Vladislav and {Pilia}, Maura and {Rogantini}, Daniele},
        title = "{X-Ray Dips and Polarization Angle Swings in GX 13+1}",
      journal = {\apjl},
         year = 2025,
        month = feb,
       volume = {979},
       number = {2},
          eid = {L47},
        pages = {L47},
          doi = {10.3847/2041-8213/ada7f8},
archivePrefix = {arXiv},
       eprint = {2501.05511},
 primaryClass = {astro-ph.HE},
       adsurl = {https://ui.adsabs.harvard.edu/abs/2025ApJ...979L..47D}
}

@ARTICLE{2025A&A...701A.115K,
       author = {{Kravtsov}, Vadim and {Bocharova}, Anastasiia and {Veledina}, Alexandra and {Poutanen}, Juri and {Hughes}, Andrew K. and {Dov{\v{c}}iak}, Michal and {Egron}, Elise and {Muleri}, Fabio and {Podgorny}, Jakub and {Svoboda}, Ji{\v{r}}i and {Forsblom}, Sofia V. and {Berdyugin}, Andrei V. and {Blinov}, Dmitry and {Bright}, Joe S. and {Carotenuto}, Francesco and {Green}, David A. and {Ingram}, Adam and {Liodakis}, Ioannis and {Mandarakas}, Nikos and {Nitindala}, Anagha P. and {Rhodes}, Lauren and {Trushkin}, Sergei A. and {Tsygankov}, Sergey S. and {Brigitte}, Ma{\"\i}mouna and {Di Marco}, Alessandro and {Iacolina}, Noemi and {Krawczynski}, Henric and {La Monaca}, Fabio and {Loktev}, Vladislav and {Mastroserio}, Guglielmo and {Petrucci}, Pierre-Olivier and {Pilia}, Maura and {Tombesi}, Francesco and {Zdziarski}, Andrzej A.},
        title = "{Variability of X-ray polarization of Cyg X-1}",
      journal = {\aap},
         year = 2025,
        month = sep,
       volume = {701},
          eid = {A115},
        pages = {A115},
          doi = {10.1051/0004-6361/202555411},
archivePrefix = {arXiv},
       eprint = {2505.03942},
 primaryClass = {astro-ph.HE},
       adsurl = {https://ui.adsabs.harvard.edu/abs/2025A&A...701A.115K}
}

@ARTICLE{2024ApJ...969L..30S,
       author = {{Steiner}, James F. and {Nathan}, Edward and {Hu}, Kun and {Krawczynski}, Henric and {Dov{\v{c}}iak}, Michal and {Veledina}, Alexandra and {Muleri}, Fabio and {Svoboda}, Jiri and {Alabarta}, Kevin and {Parra}, Maxime and {Bhargava}, Yash and {Matt}, Giorgio and {Poutanen}, Juri and {Petrucci}, Pierre-Olivier and {Tennant}, Allyn F. and {Baglio}, M. Cristina and {Baldini}, Luca and {Barnier}, Samuel and {Bhattacharyya}, Sudip and {Bianchi}, Stefano and {Brigitte}, Maimouna and {Cabezas}, Mauricio and {Cangemi}, Floriane and {Capitanio}, Fiamma and {Casey}, Jacob and {Rodriguez Cavero}, Nicole and {Castellano}, Simone and {Cavazzuti}, Elisabetta and {Chun}, Sohee and {Churazov}, Eugene and {Costa}, Enrico and {Di Lalla}, Niccol{\`o} and {Di Marco}, Alessandro and {Egron}, Elise and {Ewing}, Melissa and {Fabiani}, Sergio and {Garc{\'\i}a}, Javier A. and {Green}, David A. and {Grinberg}, Victoria and {Hadrava}, Petr and {Ingram}, Adam and {Kaaret}, Philip and {Kislat}, Fabian and {Kitaguchi}, Takao and {Kravtsov}, Vadim and {Kub{\'a}tov{\'a}}, Brankica and {La Monaca}, Fabio and {Latronico}, Luca and {Loktev}, Vladislav and {Malacaria}, Christian and {Marin}, Fr{\'e}d{\'e}ric and {Marinucci}, Andrea and {Maryeva}, Olga and {Mastroserio}, Guglielmo and {Mizuno}, Tsunefumi and {Negro}, Michela and {Omodei}, Nicola and {Podgorn{\'y}}, Jakub and {Rankin}, John and {Ratheesh}, Ajay and {Rhodes}, Lauren and {Russell}, David M. and {{\v{S}}lechta}, Miroslav and {Soffitta}, Paolo and {Spooner}, Sean and {Suleimanov}, Valery and {Tombesi}, Francesco and {Trushkin}, Sergei A. and {Weisskopf}, Martin C. and {Zane}, Silvia and {Zdziarski}, Andrzej A. and {Zhang}, Sixuan and {Zhang}, Wenda and {Zhou}, Menglei and {Agudo}, Iv{\'a}n and {Antonelli}, Lucio A. and {Bachetti}, Matteo and {Baumgartner}, Wayne H. and {Bellazzini}, Ronaldo and {Bongiorno}, Stephen D. and {Bonino}, Raffaella and {Brez}, Alessandro and {Bucciantini}, Niccol{\`o} and {Chen}, Chien-Ting and {Ciprini}, Stefano and {De Rosa}, Alessandra and {Del Monte}, Ettore and {Di Gesu}, Laura and {Donnarumma}, Immacolata and {Doroshenko}, Victor and {Ehlert}, Steven R. and {Enoto}, Teruaki and {Evangelista}, Yuri and {Ferrazzoli}, Riccardo and {Gunji}, Shuichi and {Hayashida}, Kiyoshi and {Heyl}, Jeremy and {Iwakiri}, Wataru and {Jorstad}, Svetlana G. and {Karas}, Vladimir and {Kolodziejczak}, Jeffery J. and {Liodakis}, Ioannis and {Maldera}, Simone and {Manfreda}, Alberto and {Marscher}, Alan P. and {Marshall}, Herman L. and {Massaro}, Francesco and {Mitsuishi}, Ikuyuki and {Ng}, Chi-Yung and {O'Dell}, Stephen L. and {Oppedisano}, Chiara and {Papitto}, Alessandro and {Pavlov}, George G. and {Peirson}, Abel L. and {Perri}, Matteo and {Pesce-Rollins}, Melissa and {Pilia}, Maura and {Possenti}, Andrea and {Puccetti}, Simonetta and {Ramsey}, Brian D. and {Roberts}, Oliver J. and {Romani}, Roger W. and {Sgr{\`o}}, Carmelo and {Slane}, Patrick and {Spandre}, Gloria and {Swartz}, Douglas A. and {Tamagawa}, Toru and {Tavecchio}, Fabrizio and {Taverna}, Roberto and {Tawara}, Yuzuru and {Thomas}, Nicholas E. and {Trois}, Alessio and {Tsygankov}, Sergey S. and {Turolla}, Roberto and {Vink}, Jacco and {Wu}, Kinwah and {Xie}, Fei},
        title = "{An IXPE-led X-Ray Spectropolarimetric Campaign on the Soft State of Cygnus X-1: X-Ray Polarimetric Evidence for Strong Gravitational Lensing}",
      journal = {\apjl},
         year = 2024,
        month = jul,
       volume = {969},
       number = {2},
          eid = {L30},
        pages = {L30},
          doi = {10.3847/2041-8213/ad58e4},
archivePrefix = {arXiv},
       eprint = {2406.12014},
 primaryClass = {astro-ph.HE},
       adsurl = {https://ui.adsabs.harvard.edu/abs/2024ApJ...969L..30S}
}

@INPROCEEDINGS{2014SPIE.9144E..20A,
       author = {{Arzoumanian}, Z. and {Gendreau}, K.~C. and {Baker}, C.~L. and {Cazeau}, T. and {Hestnes}, P. and {Kellogg}, J.~W. and {Kenyon}, S.~J. and {Kozon}, R.~P. and {Liu}, K.-C. and {Manthripragada}, S.~S. and {Markwardt}, C.~B. and {Mitchell}, A.~L. and {Mitchell}, J.~W. and {Monroe}, C.~A. and {Okajima}, T. and {Pollard}, S.~E. and {Powers}, D.~F. and {Savadkin}, B.~J. and {Winternitz}, L.~B. and {Chen}, P.~T. and {Wright}, M.~R. and {Foster}, R. and {Prigozhin}, G. and {Remillard}, R. and {Doty}, J.},
        title = "{The neutron star interior composition explorer (NICER): mission definition}",
    booktitle = {Space Telescopes and Instrumentation 2014: Ultraviolet to Gamma Ray},
         year = 2014,
       editor = {{Takahashi}, Tadayuki and {den Herder}, Jan-Willem A. and {Bautz}, Mark},
       series = {Society of Photo-Optical Instrumentation Engineers (SPIE) Conference Series},
       volume = {9144},
        month = jul,
          eid = {914420},
        pages = {914420},
          doi = {10.1117/12.2056811},
       adsurl = {https://ui.adsabs.harvard.edu/abs/2014SPIE.9144E..20A}
}

@INPROCEEDINGS{2016SPIE.9905E..1HG,
       author = {{Gendreau}, Keith C. and {Arzoumanian}, Zaven and {Adkins}, Phillip W. and {Albert}, Cheryl L. and {Anders}, John F. and {Aylward}, Andrew T. and {Baker}, Charles L. and {Balsamo}, Erin R. and {Bamford}, William A. and {Benegalrao}, Suyog S. and {Berry}, Daniel L. and {Bhalwani}, Shiraz and {Black}, J. Kevin and {Blaurock}, Carl and {Bronke}, Ginger M. and {Brown}, Gary L. and {Budinoff}, Jason G. and {Cantwell}, Jeffrey D. and {Cazeau}, Thoniel and {Chen}, Philip T. and {Clement}, Thomas G. and {Colangelo}, Andrew T. and {Coleman}, Jerry S. and {Coopersmith}, Jonathan D. and {Dehaven}, William E. and {Doty}, John P. and {Egan}, Mark D. and {Enoto}, Teruaki and {Fan}, Terry W. and {Ferro}, Deneen M. and {Foster}, Richard and {Galassi}, Nicholas M. and {Gallo}, Luis D. and {Green}, Chris M. and {Grosh}, Dave and {Ha}, Kong Q. and {Hasouneh}, Monther A. and {Heefner}, Kristofer B. and {Hestnes}, Phyllis and {Hoge}, Lisa J. and {Jacobs}, Tawanda M. and {J{\o}rgensen}, John L. and {Kaiser}, Michael A. and {Kellogg}, James W. and {Kenyon}, Steven J. and {Koenecke}, Richard G. and {Kozon}, Robert P. and {LaMarr}, Beverly and {Lambertson}, Mike D. and {Larson}, Anne M. and {Lentine}, Steven and {Lewis}, Jesse H. and {Lilly}, Michael G. and {Liu}, Kuochia Alice and {Malonis}, Andrew and {Manthripragada}, Sridhar S. and {Markwardt}, Craig B. and {Matonak}, Bryan D. and {Mcginnis}, Isaac E. and {Miller}, Roger L. and {Mitchell}, Alissa L. and {Mitchell}, Jason W. and {Mohammed}, Jelila S. and {Monroe}, Charles A. and {Montt de Garcia}, Kristina M. and {Mul{\'e}}, Peter D. and {Nagao}, Louis T. and {Ngo}, Son N. and {Norris}, Eric D. and {Norwood}, Dwight A. and {Novotka}, Joseph and {Okajima}, Takashi and {Olsen}, Lawrence G. and {Onyeachu}, Chimaobi O. and {Orosco}, Henry Y. and {Peterson}, Jacqualine R. and {Pevear}, Kristina N. and {Pham}, Karen K. and {Pollard}, Sue E. and {Pope}, John S. and {Powers}, Daniel F. and {Powers}, Charles E. and {Price}, Samuel R. and {Prigozhin}, Gregory Y. and {Ramirez}, Julian B. and {Reid}, Winston J. and {Remillard}, Ronald A. and {Rogstad}, Eric M. and {Rosecrans}, Glenn P. and {Rowe}, John N. and {Sager}, Jennifer A. and {Sanders}, Claude A. and {Savadkin}, Bruce and {Saylor}, Maxine R. and {Schaeffer}, Alexander F. and {Schweiss}, Nancy S. and {Semper}, Sean R. and {Serlemitsos}, Peter J. and {Shackelford}, Larry V. and {Soong}, Yang and {Struebel}, Jonathan and {Vezie}, Michael L. and {Villasenor}, Joel S. and {Winternitz}, Luke B. and {Wofford}, George I. and {Wright}, Michael R. and {Yang}, Mike Y. and {Yu}, Wayne H.},
        title = "{The Neutron star Interior Composition Explorer (NICER): design and development}",
    booktitle = {Space Telescopes and Instrumentation 2016: Ultraviolet to Gamma Ray},
         year = 2016,
       editor = {{den Herder}, Jan-Willem A. and {Takahashi}, Tadayuki and {Bautz}, Marshall},
       series = {Society of Photo-Optical Instrumentation Engineers (SPIE) Conference Series},
       volume = {9905},
        month = jul,
          eid = {99051H},
        pages = {99051H},
          doi = {10.1117/12.2231304},
       adsurl = {https://ui.adsabs.harvard.edu/abs/2016SPIE.9905E..1HG}
}

@ARTICLE{2021APh...13302628B,
       author = {{Baldini}, L. and {Barbanera}, M. and {Bellazzini}, R. and {Bonino}, R. and {Borotto}, F. and {Brez}, A. and {Caporale}, C. and {Cardelli}, C. and {Castellano}, S. and {Ceccanti}, M. and {Citraro}, S. and {Di Lalla}, N. and {Latronico}, L. and {Lucchesi}, L. and {Magazz{\`u}}, C. and {Magazz{\`u}}, G. and {Maldera}, S. and {Manfreda}, A. and {Marengo}, M. and {Marrocchesi}, A. and {Mereu}, P. and {Minuti}, M. and {Mosti}, F. and {Nasimi}, H. and {Nuti}, A. and {Oppedisano}, C. and {Orsini}, L. and {Pesce-Rollins}, M. and {Pinchera}, M. and {Profeti}, A. and {Sgr{\`o}}, C. and {Spandre}, G. and {Tardiola}, M. and {Zanetti}, D. and {Amici}, F. and {Andersson}, H. and {Attin{\`a}}, P. and {Bachetti}, M. and {Baumgartner}, W. and {Brienza}, D. and {Carpentiero}, R. and {Castronuovo}, M. and {Cavalli}, L. and {Cavazzuti}, E. and {Centrone}, M. and {Costa}, E. and {D'Alba}, E. and {D'Amico}, F. and {Del Monte}, E. and {Di Cosimo}, S. and {Di Marco}, A. and {Di Persio}, G. and {Donnarumma}, I. and {Evangelista}, Y. and {Fabiani}, S. and {Ferrazzoli}, R. and {Kitaguchi}, T. and {La Monaca}, F. and {Lefevre}, C. and {Loffredo}, P. and {Lorenzi}, P. and {Mangraviti}, E. and {Matt}, G. and {Meilahti}, T. and {Morbidini}, A. and {Muleri}, F. and {Nakano}, T. and {Negri}, B. and {Nenonen}, S. and {O'Dell}, S.~L. and {Perri}, M. and {Piazzolla}, R. and {Pieraccini}, S. and {Pilia}, M. and {Puccetti}, S. and {Ramsey}, B.~D. and {Rankin}, J. and {Ratheesh}, A. and {Rubini}, A. and {Santoli}, F. and {Sarra}, P. and {Scalise}, E. and {Sciortino}, A. and {Soffitta}, P. and {Tamagawa}, T. and {Tennant}, A.~F. and {Tobia}, A. and {Trois}, A. and {Uchiyama}, K. and {Vimercati}, M. and {Weisskopf}, M.~C. and {Xie}, F. and {Zanetti}, F. and {Zhou}, Y.},
        title = "{Design, construction, and test of the Gas Pixel Detectors for the IXPE mission}",
      journal = {Astroparticle Physics},
         year = 2021,
        month = dec,
       volume = {133},
          eid = {102628},
        pages = {102628},
          doi = {10.1016/j.astropartphys.2021.102628},
archivePrefix = {arXiv},
       eprint = {2107.05496},
 primaryClass = {astro-ph.IM},
       adsurl = {https://ui.adsabs.harvard.edu/abs/2021APh...13302628B}
}

@INPROCEEDINGS{2022SPIE12181E..1CD,
       author = {{Di Marco}, A. and {Muleri}, F. and {Fabiani}, S. and {La Monaca}, F. and {Rankin}, J. and {Soffitta}, P. and {Baldini}, L. and {Costa}, E. and {Del Monte}, E. and {Ferrazzoli}, R. and {Lefevre}, C. and {Maiolo}, L. and {Maita}, F. and {Manfreda}, A. and {Morbidini}, A. and {O'Dell}, S.~L. and {Ramsey}, B.~D. and {Ratheesh}, A. and {Sgro'}, C. and {Trois}, A. and {Tennant}, A.~F. and {Weisskopf}, M.~C.},
        title = "{In-orbit monitoring of the imaging x-ray polarimeters on-board IXPE}",
    booktitle = {Space Telescopes and Instrumentation 2022: Ultraviolet to Gamma Ray},
         year = 2022,
       editor = {{den Herder}, Jan-Willem A. and {Nikzad}, Shouleh and {Nakazawa}, Kazuhiro},
       series = {Society of Photo-Optical Instrumentation Engineers (SPIE) Conference Series},
       volume = {12181},
        month = aug,
          eid = {121811C},
        pages = {121811C},
          doi = {10.1117/12.2629413},
       adsurl = {https://ui.adsabs.harvard.edu/abs/2022SPIE12181E..1CD}
}

@article{Arnaud1996,
  title={XSPEC: The First Ten Years},
  author={Arnaud, K. A.},
  journal={Astronomical Data Analysis Software and Systems V, A.S.P. Conference Series},
  year={1996},
  volume={101}
}

@ARTICLE{2000ApJ...542..914W,
       author = {{Wilms}, J. and {Allen}, A. and {McCray}, R.},
        title = "{On the Absorption of X-Rays in the Interstellar Medium}",
      journal = {\apj},
         year = 2000,
        month = oct,
       volume = {542},
       number = {2},
        pages = {914-924},
          doi = {10.1086/317016},
archivePrefix = {arXiv},
       eprint = {astro-ph/0008425},
 primaryClass = {astro-ph},
       adsurl = {https://ui.adsabs.harvard.edu/abs/2000ApJ...542..914W}
}

@ARTICLE{2014MNRAS.444L.100D,
       author = {{Dauser}, T. and {Garcia}, J. and {Parker}, M.~L. and {Fabian}, A.~C. and {Wilms}, J.},
        title = "{The role of the reflection fraction in constraining black hole spin.}",
      journal = {\mnras},
         year = 2014,
        month = oct,
       volume = {444},
        pages = {L100-L104},
          doi = {10.1093/mnrasl/slu125},
archivePrefix = {arXiv},
       eprint = {1408.2347},
 primaryClass = {astro-ph.HE},
       adsurl = {https://ui.adsabs.harvard.edu/abs/2014MNRAS.444L.100D}
}

@ARTICLE{2016A&A...590A..76D,
       author = {{Dauser}, T. and {Garc{\'\i}a}, J. and {Walton}, D.~J. and {Eikmann}, W. and {Kallman}, T. and {McClintock}, J. and {Wilms}, J.},
        title = "{Normalizing a relativistic model of X-ray reflection. Definition of the reflection fraction and its implementation in relxill}",
      journal = {\aap},
         year = 2016,
        month = may,
       volume = {590},
          eid = {A76},
        pages = {A76},
          doi = {10.1051/0004-6361/201628135},
archivePrefix = {arXiv},
       eprint = {1601.03771},
 primaryClass = {astro-ph.HE},
       adsurl = {https://ui.adsabs.harvard.edu/abs/2016A&A...590A..76D}
}

@ARTICLE{2014ApJ...782...76G,
       author = {{Garc{\'\i}a}, J. and {Dauser}, T. and {Lohfink}, A. and {Kallman}, T.~R. and {Steiner}, J.~F. and {McClintock}, J.~E. and {Brenneman}, L. and {Wilms}, J. and {Eikmann}, W. and {Reynolds}, C.~S. and {Tombesi}, F.},
        title = "{Improved Reflection Models of Black Hole Accretion Disks: Treating the Angular Distribution of X-Rays}",
      journal = {\apj},
         year = 2014,
        month = feb,
       volume = {782},
       number = {2},
          eid = {76},
        pages = {76},
          doi = {10.1088/0004-637X/782/2/76},
archivePrefix = {arXiv},
       eprint = {1312.3231},
 primaryClass = {astro-ph.HE},
       adsurl = {https://ui.adsabs.harvard.edu/abs/2014ApJ...782...76G}
}

@ARTICLE{2021Sci...371.1046M,
       author = {{Miller-Jones}, James C.~A. and {Bahramian}, Arash and {Orosz}, Jerome A. and {Mandel}, Ilya and {Gou}, Lijun and {Maccarone}, Thomas J. and {Neijssel}, Coenraad J. and {Zhao}, Xueshan and {Zi{\'o}{\l}kowski}, Janusz and {Reid}, Mark J. and {Uttley}, Phil and {Zheng}, Xueying and {Byun}, Do-Young and {Dodson}, Richard and {Grinberg}, Victoria and {Jung}, Taehyun and {Kim}, Jeong-Sook and {Marcote}, Benito and {Markoff}, Sera and {Rioja}, Mar{\'\i}a J. and {Rushton}, Anthony P. and {Russell}, David M. and {Sivakoff}, Gregory R. and {Tetarenko}, Alexandra J. and {Tudose}, Valeriu and {Wilms}, Joern},
        title = "{Cygnus X-1 contains a 21-solar mass black hole{\textemdash}Implications for massive star winds}",
      journal = {Science},
         year = 2021,
        month = mar,
       volume = {371},
       number = {6533},
        pages = {1046-1049},
          doi = {10.1126/science.abb3363},
archivePrefix = {arXiv},
       eprint = {2102.09091},
 primaryClass = {astro-ph.HE},
       adsurl = {https://ui.adsabs.harvard.edu/abs/2021Sci...371.1046M}
}

@ARTICLE{DiMarco26,
       author = {{Di Marco}, Alessandro},
        title = "{The hitchhiker's guide to the IXPE data analysis}",
      journal = {arXiv e-prints},
         year = 2026,
        month = apr,
          eid = {arXiv:2604.03366},
        pages = {arXiv:2604.03366},
          doi = {10.48550/arXiv.2604.03366},
archivePrefix = {arXiv},
       eprint = {2604.03366},
 primaryClass = {astro-ph.HE},
       adsurl = {https://ui.adsabs.harvard.edu/abs/2026arXiv260403366D}
}

@ARTICLE{2026MNRAS.tmp.1047W,
       author = {{Wu}, WanYun and {Xie}, Fei and {La Monaca}, Fabio},
        title = "{High Polarization and Complex Accretion Wind Structure in 4U 1700-37}",
      journal = {\mnras},
         year = 2026,
        month = jun,
          doi = {10.1093/mnras/stag1109},
       adsurl = {https://ui.adsabs.harvard.edu/abs/2026MNRAS.tmp.1047W}
}

@ARTICLE{2025AN....34640126D,
       author = {{Di Marco}, Alessandro},
        title = "{Weakly Magnetized Accreting Neutron Stars as Seen by IXPE}",
      journal = {Astronomische Nachrichten},
         year = 2025,
        month = jan,
       volume = {346},
       number = {1},
          eid = {e20240126},
        pages = {e20240126},
          doi = {10.1002/asna.20240126},
       adsurl = {https://ui.adsabs.harvard.edu/abs/2025AN....34640126D}
}

@ARTICLE{2025MNRAS.541.1774E,
       author = {{Ewing}, Melissa and {Parra}, Maxime and {Mastroserio}, Guglielmo and {Veledina}, Alexandra and {Ingram}, Adam and {Dov{\v{c}}iak}, Michal and {Garc{\'\i}a}, Javier A. and {Russell}, Thomas D. and {Baglio}, Maria C. and {Poutanen}, Juri and {Adegoke}, Oluwashina and {Bianchi}, Stefano and {Capitanio}, Fiamma and {Connors}, Riley and {Del Santo}, Melania and {De Marco}, Barbara and {Trigo}, Mar{\'\i}a D{\'\i}az and {Gandhi}, Poshak and {Gupta}, Maitrayee and {Kang}, Chulsoo and {Kammoun}, Elias and {Loktev}, Vladislav and {Marra}, Lorenzo and {Matt}, Giorgio and {Nathan}, Edward and {Petrucci}, Pierre-Olivier and {Shidatsu}, Megumi and {Steiner}, James F. and {Tombesi}, Francesco and {Vincentelli}, Federico M.},
        title = "{The very high X-ray polarization of accreting black hole IGR J17091‑3624 in the hard state}",
      journal = {\mnras},
         year = 2025,
        month = aug,
       volume = {541},
       number = {2},
        pages = {1774-1781},
          doi = {10.1093/mnras/staf859},
archivePrefix = {arXiv},
       eprint = {2503.22665},
 primaryClass = {astro-ph.HE},
       adsurl = {https://ui.adsabs.harvard.edu/abs/2025MNRAS.541.1774E}
}

@ARTICLE{2025ApJ...989..165D,
       author = {{Debnath}, Dipak and {Srimani}, Subham and {Chang}, Hsiang-Kuang},
        title = "{Detection of X-Ray Polarization in the Hard State of IGR J17091-3624: Spectropolarimetric Study with IXPE and NuSTAR Data}",
      journal = {\apj},
         year = 2025,
        month = aug,
       volume = {989},
       number = {2},
          eid = {165},
        pages = {165},
          doi = {10.3847/1538-4357/adf1e6},
archivePrefix = {arXiv},
       eprint = {2504.16402},
 primaryClass = {astro-ph.HE},
       adsurl = {https://ui.adsabs.harvard.edu/abs/2025ApJ...989..165D}
}

@ARTICLE{2023AJ....165..143D,
       author = {{Di Marco}, Alessandro and {Soffitta}, Paolo and {Costa}, Enrico and {Ferrazzoli}, Riccardo and {La Monaca}, Fabio and {Rankin}, John and {Ratheesh}, Ajay and {Xie}, Fei and {Baldini}, Luca and {Del Monte}, Ettore and {Ehlert}, Steven R. and {Fabiani}, Sergio and {Kim}, Dawoon E. and {Muleri}, Fabio and {O'Dell}, Stephen L. and {Ramsey}, Brian D. and {Rubini}, Alda and {Sgr{\`o}}, Carmelo and {Silvestri}, Stefano and {Tennant}, Allyn F. and {Weisskopf}, Martin C.},
        title = "{Handling the Background in IXPE Polarimetric Data}",
      journal = {\aj},
         year = 2023,
        month = apr,
       volume = {165},
       number = {4},
          eid = {143},
        pages = {143},
          doi = {10.3847/1538-3881/acba0f},
archivePrefix = {arXiv},
       eprint = {2302.02927},
 primaryClass = {astro-ph.IM},
       adsurl = {https://ui.adsabs.harvard.edu/abs/2023AJ....165..143D}
}

@ARTICLE{2013ApJ...770..103H,
       author = {{Harrison}, Fiona A. and {Craig}, William W. and {Christensen}, Finn E. and {Hailey}, Charles J. and {Zhang}, William W. and {Boggs}, Steven E. and {Stern}, Daniel and {Cook}, W. Rick and {Forster}, Karl and {Giommi}, Paolo and {Grefenstette}, Brian W. and {Kim}, Yunjin and {Kitaguchi}, Takao and {Koglin}, Jason E. and {Madsen}, Kristin K. and {Mao}, Peter H. and {Miyasaka}, Hiromasa and {Mori}, Kaya and {Perri}, Matteo and {Pivovaroff}, Michael J. and {Puccetti}, Simonetta and {Rana}, Vikram R. and {Westergaard}, Niels J. and {Willis}, Jason and {Zoglauer}, Andreas and {An}, Hongjun and {Bachetti}, Matteo and {Barri{\`e}re}, Nicolas M. and {Bellm}, Eric C. and {Bhalerao}, Varun and {Brejnholt}, Nicolai F. and {Fuerst}, Felix and {Liebe}, Carl C. and {Markwardt}, Craig B. and {Nynka}, Melania and {Vogel}, Julia K. and {Walton}, Dominic J. and {Wik}, Daniel R. and {Alexander}, David M. and {Cominsky}, Lynn R. and {Hornschemeier}, Ann E. and {Hornstrup}, Allan and {Kaspi}, Victoria M. and {Madejski}, Greg M. and {Matt}, Giorgio and {Molendi}, Silvano and {Smith}, David M. and {Tomsick}, John A. and {Ajello}, Marco and {Ballantyne}, David R. and {Balokovi{\'c}}, Mislav and {Barret}, Didier and {Bauer}, Franz E. and {Blandford}, Roger D. and {Brandt}, W. Niel and {Brenneman}, Laura W. and {Chiang}, James and {Chakrabarty}, Deepto and {Chenevez}, Jerome and {Comastri}, Andrea and {Dufour}, Francois and {Elvis}, Martin and {Fabian}, Andrew C. and {Farrah}, Duncan and {Fryer}, Chris L. and {Gotthelf}, Eric V. and {Grindlay}, Jonathan E. and {Helfand}, David J. and {Krivonos}, Roman and {Meier}, David L. and {Miller}, Jon M. and {Natalucci}, Lorenzo and {Ogle}, Patrick and {Ofek}, Eran O. and {Ptak}, Andrew and {Reynolds}, Stephen P. and {Rigby}, Jane R. and {Tagliaferri}, Gianpiero and {Thorsett}, Stephen E. and {Treister}, Ezequiel and {Urry}, C. Megan},
        title = "{The Nuclear Spectroscopic Telescope Array (NuSTAR) High-energy X-Ray Mission}",
      journal = {\apj},
         year = 2013,
        month = jun,
       volume = {770},
       number = {2},
          eid = {103},
        pages = {103},
          doi = {10.1088/0004-637X/770/2/103},
archivePrefix = {arXiv},
       eprint = {1301.7307},
 primaryClass = {astro-ph.IM},
       adsurl = {https://ui.adsabs.harvard.edu/abs/2013ApJ...770..103H}
}

@INPROCEEDINGS{2003ASPC..295..489J,
       author = {{Joye}, W.~A. and {Mandel}, E.},
        title = "{New Features of SAOImage DS9}",
    booktitle = {Astronomical Data Analysis Software and Systems XII},
         year = 2003,
       editor = {{Payne}, H.~E. and {Jedrzejewski}, R.~I. and {Hook}, R.~N.},
       series = {Astronomical Society of the Pacific Conference Series},
       volume = {295},
        month = jan,
        pages = {489},
       adsurl = {https://ui.adsabs.harvard.edu/abs/2003ASPC..295..489J}
}

@ARTICLE{2022SoftX..1901194B,
       author = {{Baldini}, Luca and {Bucciantini}, Niccol{\`o} and {Lalla}, Niccol{\`o} Di and {Ehlert}, Steven and {Manfreda}, Alberto and {Negro}, Michela and {Omodei}, Nicola and {Pesce-Rollins}, Melissa and {Sgr{\`o}}, Carmelo and {Silvestri}, Stefano},
        title = "{ixpeobssim: A simulation and analysis framework for the imaging X-ray polarimetry explorer}",
      journal = {SoftwareX},
         year = 2022,
        month = jul,
       volume = {19},
          eid = {101194},
        pages = {101194},
          doi = {10.1016/j.softx.2022.101194},
archivePrefix = {arXiv},
       eprint = {2203.06384},
 primaryClass = {astro-ph.IM},
       adsurl = {https://ui.adsabs.harvard.edu/abs/2022SoftX..1901194B}
}

@ARTICLE{2016A&A...587A.151K,
       author = {{Kaastra}, J.~S. and {Bleeker}, J.~A.~M.},
        title = "{Optimal binning of X-ray spectra and response matrix design}",
      journal = {\aap},
         year = 2016,
        month = mar,
       volume = {587},
          eid = {A151},
        pages = {A151},
          doi = {10.1051/0004-6361/201527395},
archivePrefix = {arXiv},
       eprint = {1601.05309},
 primaryClass = {astro-ph.IM},
       adsurl = {https://ui.adsabs.harvard.edu/abs/2016A&A...587A.151K}
}

@ARTICLE{2024MNRAS.52710837J,
       author = {{Jana}, Arghajit and {Chang}, Hsiang-Kuang},
        title = "{X-ray polarization changes with the state transition in Cygnus X-1}",
      journal = {\mnras},
         year = 2024,
        month = feb,
       volume = {527},
       number = {4},
        pages = {10837-10843},
          doi = {10.1093/mnras/stad3961},
archivePrefix = {arXiv},
       eprint = {2307.14604},
 primaryClass = {astro-ph.HE},
       adsurl = {https://ui.adsabs.harvard.edu/abs/2024MNRAS.52710837J}
}

@ARTICLE{2022AJ....164..103D,
       author = {{Di Marco}, Alessandro and {Fabiani}, Sergio and {La Monaca}, Fabio and {Muleri}, Fabio and {Rankin}, John and {Soffitta}, Paolo and {Xie}, Fei and {Amici}, Fabrizio and {attin{\`a}}, Primo and {Bachetti}, Matteo and {Baldini}, Luca and {Barbanera}, Mattia and {Baumgartner}, Wayne and {Bellazzini}, Ronaldo and {Borotto}, Fabio and {Brez}, Alessandro and {Brienza}, Daniele and {Caporale}, Ciro and {Cardelli}, Claudia and {Carpentiero}, Rita and {Castellano}, Simone and {Castronuovo}, Marco and {Cavalli}, Luca and {Cavazzuti}, Elisabetta and {Ceccanti}, Marco and {Centrone}, Mauro and {Citraro}, Saverio and {Costa}, Enrico and {D'Alba}, Elisa and {D'Amico}, Fabio and {Del Monte}, Ettore and {Di Cosimo}, Sergio and {Di Lalla}, Niccol{\`o} and {Di Persio}, Giuseppe and {Donnarumma}, Immacolata and {Evangelista}, Yuri and {Ferrazzoli}, Riccardo and {Latronico}, Luca and {Lefevre}, Carlo and {Loffredo}, Pasqualino and {Lorenzi}, Paolo and {Lucchesi}, Leonardo and {Magazz{\`u}}, Carlo and {Magazz{\`u}}, Guido and {Maldera}, Simone and {Manfreda}, Alberto and {Mangraviti}, Elio and {Marengo}, Marco and {Matt}, Giorgio and {Mereu}, Paolo and {Minuti}, Massimo and {Morbidini}, Alfredo and {Mosti}, Federico and {Nasimi}, Hikmat and {Negri}, Barbara and {Nuti}, Alessio and {O'Dell}, Stephen L. and {Orsini}, Leonardo and {Perri}, Matteo and {Pesce-Rollins}, Melissa and {Piazzolla}, Raffaele and {Pieraccini}, Stefano and {Pilia}, Maura and {Pinchera}, Michele and {Profeti}, Alessandro and {Puccetti}, Simonetta and {Ramsey}, Brian D. and {Ratheesh}, Ajay and {Rubini}, Alda and {Santoli}, Francesco and {Sarra}, Paolo and {Scalise}, Emanuele and {Sciortino}, Andrea and {Sgr{\`o}}, Carmelo and {Spandre}, Gloria and {Tardiola}, Marcello and {Tennant}, Allyn F. and {Tobia}, Antonino and {Trois}, Alessio and {Vimercati}, Marco and {Weisskopf}, Martin C. and {Zanetti}, Davide and {Zanetti}, Francesco},
        title = "{Calibration of the IXPE Focal Plane X-Ray Polarimeters to Polarized Radiation}",
      journal = {\aj},
         year = 2022,
        month = sep,
       volume = {164},
       number = {3},
          eid = {103},
        pages = {103},
          doi = {10.3847/1538-3881/ac7719},
archivePrefix = {arXiv},
       eprint = {2206.07582},
 primaryClass = {astro-ph.IM},
       adsurl = {https://ui.adsabs.harvard.edu/abs/2022AJ....164..103D}
}

@ARTICLE{2022AJ....163..170D,
       author = {{Di Marco}, Alessandro and {Costa}, Enrico and {Muleri}, Fabio and {Soffitta}, Paolo and {Fabiani}, Sergio and {La Monaca}, Fabio and {Rankin}, John and {Xie}, Fei and {Bachetti}, Matteo and {Baldini}, Luca and {Baumgartner}, Wayne and {Bellazzini}, Ronaldo and {Brez}, Alessandro and {Castellano}, Simone and {Del Monte}, Ettore and {Di Lalla}, Niccol{\`o} and {Ferrazzoli}, Riccardo and {Latronico}, Luca and {Maldera}, Simone and {Manfreda}, Alberto and {O'Dell}, Stephen L. and {Perri}, Matteo and {Pesce-Rollins}, Melissa and {Puccetti}, Simonetta and {Ramsey}, Brian D. and {Ratheesh}, Ajay and {Sgr{\`o}}, Carmelo and {Spandre}, Gloria and {Tennant}, Allyn F. and {Tobia}, Antonino and {Trois}, Alessio and {Weisskopf}, Martin C.},
        title = "{A Weighted Analysis to Improve the X-Ray Polarization Sensitivity of the Imaging X-ray Polarimetry Explorer}",
      journal = {\aj},
         year = 2022,
        month = apr,
       volume = {163},
       number = {4},
          eid = {170},
        pages = {170},
          doi = {10.3847/1538-3881/ac51c9},
archivePrefix = {arXiv},
       eprint = {2202.01093},
 primaryClass = {astro-ph.IM},
       adsurl = {https://ui.adsabs.harvard.edu/abs/2022AJ....163..170D}
}

@INPROCEEDINGS{1996ASPC..101...17A,
       author = {{Arnaud}, K.~A.},
        title = "{XSPEC: The First Ten Years}",
    booktitle = {Astronomical Data Analysis Software and Systems V},
         year = 1996,
       editor = {{Jacoby}, George H. and {Barnes}, Jeannette},
       series = {Astronomical Society of the Pacific Conference Series},
       volume = {101},
        month = jan,
        pages = {17},
       adsurl = {https://ui.adsabs.harvard.edu/abs/1996ASPC..101...17A}
}

@ARTICLE{1998MNRAS.301..285L,
       author = {{LaSala}, J. and {Charles}, P.~A. and {Smith}, R.~A.~D. and {Balucinska-Church}, M. and {Church}, M.~J.},
        title = "{The orbital period of HDE 226868/Cyg X-1}",
      journal = {\mnras},
         year = 1998,
        month = nov,
       volume = {301},
       number = {1},
        pages = {285-288},
          doi = {10.1046/j.1365-8711.1998.02064.x},
archivePrefix = {arXiv},
       eprint = {astro-ph/9812101},
 primaryClass = {astro-ph},
       adsurl = {https://ui.adsabs.harvard.edu/abs/1998MNRAS.301..285L}
}

@ARTICLE{LaMonaca24,
       author = {{La Monaca}, Fabio and {Di Marco}, Alessandro and {Poutanen}, Juri and {Bachetti}, Matteo and {Motta}, Sara Elisa and {Papitto}, Alessandro and {Pilia}, Maura and {Xie}, Fei and {Bianchi}, Stefano and {Bobrikova}, Anna and {Costa}, Enrico and {Deng}, Wei and {Ge}, Ming-Yu and {Illiano}, Giulia and {Jia}, Shu-Mei and {Krawczynski}, Henric and {Lai}, Eleonora Veronica and {Liu}, Kuan and {Mastroserio}, Guglielmo and {Muleri}, Fabio and {Rankin}, John and {Soffitta}, Paolo and {Veledina}, Alexandra and {Ambrosino}, Filippo and {Del Santo}, Melania and {Chen}, Wei and {Garcia}, Javier A. and {Kaaret}, Philip and {Russell}, Thomas D. and {Wei}, Wen-Hao and {Zhang}, Shuang-Nan and {Zuo}, Chao and {Arzoumanian}, Zaven and {Cocchi}, Massimo and {Gnarini}, Andrea and {Farinelli}, Ruben and {Gendreau}, Keith and {Ursini}, Francesco and {Weisskopf}, Martin C. and {Zane}, Silvia and {Agudo}, Iv{\'a}n and {Antonelli}, Lucio A. and {Baldini}, Luca and {Baumgartner}, Wayne H. and {Bellazzini}, Ronaldo and {Bongiorno}, Stephen D. and {Bonino}, Raffaella and {Brez}, Alessandro and {Bucciantini}, Niccol{\`o} and {Capitanio}, Fiamma and {Castellano}, Simone and {Cavazzuti}, Elisabetta and {Chen}, Chien-Ting and {Ciprini}, Stefano and {De Rosa}, Alessandra and {Del Monte}, Ettore and {Di Gesu}, Laura and {Di Lalla}, Niccol{\`o} and {Donnarumma}, Immacolata and {Doroshenko}, Victor and {Dov{\v{c}}iak}, Michal and {Ehlert}, Steven R. and {Enoto}, Teruaki and {Evangelista}, Yuri and {Fabiani}, Sergio and {Ferrazzoli}, Riccardo and {Gunji}, Shuichi and {Hayashida}, Kiyoshi and {Heyl}, Jeremy and {Iwakiri}, Wataru and {Jorstad}, Svetlana G. and {Karas}, Vladimir and {Kislat}, Fabian and {Kitaguchi}, Takao and {Kolodziejczak}, Jeffery J. and {Latronico}, Luca and {Liodakis}, Ioannis and {Maldera}, Simone and {Manfreda}, Alberto and {Marin}, Fr{\'e}d{\'e}ric and {Marinucci}, Andrea and {Marscher}, Alan P. and {Marshall}, Herman L. and {Massaro}, Francesco and {Matt}, Giorgio and {Mitsuishi}, Ikuyuki and {Mizuno}, Tsunefumi and {Negro}, Michela and {Ng}, Chi-Yung and {O'Dell}, Stephen L. and {Omodei}, Nicola and {Oppedisano}, Chiara and {Pavlov}, George G. and {Peirson}, Abel L. and {Perri}, Matteo and {Pesce-Rollins}, Melissa and {Petrucci}, Pierre-Olivier and {Possenti}, Andrea and {Puccetti}, Simonetta and {Ramsey}, Brian D. and {Ratheesh}, Ajay and {Roberts}, Oliver J. and {Romani}, Roger W. and {Sgr{\`o}}, Carmelo and {Slane}, Patrick and {Spandre}, Gloria and {Swartz}, Douglas A. and {Tamagawa}, Toru and {Tavecchio}, Fabrizio and {Taverna}, Roberto and {Tawara}, Yuzuru and {Tennant}, Allyn F. and {Thomas}, Nicholas E. and {Tombesi}, Francesco and {Trois}, Alessio and {Tsygankov}, Sergey S. and {Turolla}, Roberto and {Vink}, Jacco and {Wu}, Kinwah and {IXPE Collaboration}},
        title = "{Highly Significant Detection of X-Ray Polarization from the Brightest Accreting Neutron Star Sco X-1}",
      journal = {\apjl},
         year = 2024,
        month = jan,
       volume = {960},
       number = {2},
          eid = {L11},
        pages = {L11},
          doi = {10.3847/2041-8213/ad132d},
archivePrefix = {arXiv},
       eprint = {2311.06359},
 primaryClass = {astro-ph.HE},
       adsurl = {https://ui.adsabs.harvard.edu/abs/2024ApJ...960L..11L}
}

@ARTICLE{LaMonaca24gx340,
       author = {{La Monaca}, Fabio and {Di Marco}, Alessandro and {Ludlam}, Renee M. and {Bobrikova}, Anna and {Poutanen}, Juri and {Li}, Songwei and {Xie}, Fei},
        title = "{X-ray spectropolarimetric characterization of GX 340+0 in the horizontal branch: A highly inclined source?}",
      journal = {\aap},
         year = 2024,
        month = nov,
       volume = {691},
          eid = {A253},
        pages = {A253},
          doi = {10.1051/0004-6361/202451966},
archivePrefix = {arXiv},
       eprint = {2410.00972},
 primaryClass = {astro-ph.HE},
       adsurl = {https://ui.adsabs.harvard.edu/abs/2024A&A...691A.253L}
}

@ARTICLE{LaMonaca25gx340NB,
       author = {{La Monaca}, Fabio and {Di Marco}, Alessandro and {Coti Zelati}, Francesco and {Bobrikova}, Anna and {Ludlam}, Renee M. and {Poutanen}, Juri and {Marino}, Alessio and {Li}, Songwei and {Xie}, Fei and {Feng}, Hua and {Jin}, Chichuan and {Rea}, Nanda and {Tao}, Lian and {Yuan}, Weimin},
        title = "{X-ray spectropolarimetric characterization of the Z-source GX 340+0 in the normal branch}",
      journal = {\aap\ in press},
         year = 2025,
        month = aug,
          eid = {arXiv:2508.13278},
        pages = {arXiv:2508.13278},
          doi = {https://doi.org/10.1051/0004-6361/202555134},
archivePrefix = {arXiv},
       eprint = {2508.13278},
 primaryClass = {astro-ph.HE},
       adsurl = {https://ui.adsabs.harvard.edu/abs/2025arXiv250813278L}
}

@ARTICLE{LaMonaca25_gx349,
       author = {{La Monaca}, Fabio and {Bobrikova}, Anna and {Poutanen}, Juri and {Coti Zelati}, Francesco and {Pilia}, Maura and {Veledina}, Alexandra and {Bachetti}, Matteo and {Loktev}, Vladislav and {Xie}, Fei},
        title = "{IXPE view of the Sco-like source GX 349+2 in the normal branch}",
      journal = {\aap},
         year = 2025,
        month = oct,
       volume = {702},
          eid = {A40},
        pages = {A40},
          doi = {10.1051/0004-6361/202555647},
archivePrefix = {arXiv},
       eprint = {2507.07163},
 primaryClass = {astro-ph.HE},
       adsurl = {https://ui.adsabs.harvard.edu/abs/2025A&A...702A..40L}
}

@ARTICLE{Churazov02,
       author = {{Churazov}, E. and {Sunyaev}, R. and {Sazonov}, S.},
        title = "{Polarization of X-ray emission from the Sgr B2 cloud}",
      journal = {\mnras},
         year = 2002,
        month = mar,
       volume = {330},
       number = {4},
        pages = {817-820},
          doi = {10.1046/j.1365-8711.2002.05113.x},
archivePrefix = {arXiv},
       eprint = {astro-ph/0111065},
 primaryClass = {astro-ph},
       adsurl = {https://ui.adsabs.harvard.edu/abs/2002MNRAS.330..817C}
}

@ARTICLE{Veledina24,
       author = {{Veledina}, Alexandra and {Muleri}, Fabio and {Poutanen}, Juri and {Podgorn{\'y}}, Jakub and {Dov{\v{c}}iak}, Michal and {Capitanio}, Fiamma and {Churazov}, Eugene and {De Rosa}, Alessandra and {Di Marco}, Alessandro and {Forsblom}, Sofia V. and {Kaaret}, Philip and {Krawczynski}, Henric and {La Monaca}, Fabio and {Loktev}, Vladislav and {Lutovinov}, Alexander A. and {Molkov}, Sergey V. and {Mushtukov}, Alexander A. and {Ratheesh}, Ajay and {Rodriguez Cavero}, Nicole and {Steiner}, James F. and {Sunyaev}, Rashid A. and {Tsygankov}, Sergey S. and {Weisskopf}, Martin C. and {Zdziarski}, Andrzej A. and {Bianchi}, Stefano and {Bright}, Joe S. and {Bursov}, Nikolaj and {Costa}, Enrico and {Egron}, Elise and {Garcia}, Javier A. and {Green}, David A. and {Gurwell}, Mark and {Ingram}, Adam and {Kajava}, Jari J.~E. and {Kale}, Ruta and {Kraus}, Alex and {Malyshev}, Denys and {Marin}, Fr{\'e}d{\'e}ric and {Matt}, Giorgio and {McCollough}, Michael and {Mereminskiy}, Ilya A. and {Nizhelsky}, Nikolaj and {Piano}, Giovanni and {Pilia}, Maura and {Pittori}, Carlotta and {Rao}, Ramprasad and {Righini}, Simona and {Soffitta}, Paolo and {Shevchenko}, Anton and {Svoboda}, Jiri and {Tombesi}, Francesco and {Trushkin}, Sergei A. and {Tsybulev}, Peter and {Ursini}, Francesco and {Wu}, Kinwah and {Agudo}, Iv{\'a}n and {Antonelli}, Lucio A. and {Bachetti}, Matteo and {Baldini}, Luca and {Baumgartner}, Wayne H. and {Bellazzini}, Ronaldo and {Bongiorno}, Stephen D. and {Bonino}, Raffaella and {Brez}, Alessandro and {Bucciantini}, Niccol{\`o} and {Castellano}, Simone and {Cavazzuti}, Elisabetta and {Chen}, Chien-Ting and {Ciprini}, Stefano and {Del Monte}, Ettore and {Di Gesu}, Laura and {Di Lalla}, Niccol{\`o} and {Donnarumma}, Immacolata and {Doroshenko}, Victor and {Ehlert}, Steven R. and {Enoto}, Teruaki and {Evangelista}, Yuri and {Fabiani}, Sergio and {Ferrazzoli}, Riccardo and {Gunji}, Shuichi and {Hayashida}, Kiyoshi and {Heyl}, Jeremy and {Iwakiri}, Wataru and {Jorstad}, Svetlana G. and {Karas}, Vladimir and {Kislat}, Fabian and {Kitaguchi}, Takao and {Kolodziejczak}, Jeffery J. and {Latronico}, Luca and {Liodakis}, Ioannis and {Maldera}, Simone and {Manfreda}, Alberto and {Marinucci}, Andrea and {Marscher}, Alan P. and {Marshall}, Herman L. and {Massaro}, Francesco and {Mitsuishi}, Ikuyuki and {Mizuno}, Tsunefumi and {Negro}, Michela and {Ng}, Chi-Yung and {O'Dell}, Stephen L. and {Omodei}, Nicola and {Oppedisano}, Chiara and {Papitto}, Alessandro and {Pavlov}, George G. and {Peirson}, Abel L. and {Perri}, Matteo and {Pesce-Rollins}, Melissa and {Petrucci}, Pierre-Olivier and {Possenti}, Andrea and {Puccetti}, Simonetta and {Ramsey}, Brian D. and {Rankin}, John and {Roberts}, Oliver and {Romani}, Roger W. and {Sgr{\`o}}, Carmelo and {Slane}, Patrick and {Spandre}, Gloria and {Swartz}, Doug and {Tamagawa}, Toru and {Tavecchio}, Fabrizio and {Taverna}, Roberto and {Tawara}, Yuzuru and {Tennant}, Allyn F. and {Thomas}, Nicholas E. and {Trois}, Alessio and {Turolla}, Roberto and {Vink}, Jacco and {Xie}, Fei and {Zane}, Silvia},
        title = "{Cygnus X-3 revealed as a Galactic ultraluminous X-ray source by IXPE}",
      journal = {Nature Astronomy},
         year = 2024,
        month = aug,
       volume = {8},
        pages = {1031-1046},
          doi = {10.1038/s41550-024-02294-9},
archivePrefix = {arXiv},
       eprint = {2303.01174},
 primaryClass = {astro-ph.HE},
       adsurl = {https://ui.adsabs.harvard.edu/abs/2024NatAs...8.1031V}
}

@ARTICLE{st85,
       author = {{Sunyaev}, R.~A. and {Titarchuk}, L.~G.},
        title = "{Comptonization of low-frequency radiation in accretion disks Angular distribution and polarization of hard radiation}",
      journal = {\aap},
         year = 1985,
        month = feb,
       volume = {143},
       number = {2},
        pages = {374-388},
       adsurl = {https://ui.adsabs.harvard.edu/abs/1985A&A...143..374S}
}

@ARTICLE{Rankin24,
       author = {{Rankin}, John and {La Monaca}, Fabio and {Di Marco}, Alessandro and {Poutanen}, Juri and {Bobrikova}, Anna and {Kravtsov}, Vadim and {Muleri}, Fabio and {Pilia}, Maura and {Veledina}, Alexandra and {Fender}, Rob and {Kaaret}, Philip and {Kim}, Dawoon E. and {Marinucci}, Andrea and {Marshall}, Herman L. and {Papitto}, Alessandro and {Tennant}, Allyn F. and {Tsygankov}, Sergey S. and {Weisskopf}, Martin C. and {Wu}, Kinwah and {Zane}, Silvia and {Ambrosino}, Filippo and {Farinelli}, Ruben and {Gnarini}, Andrea and {Agudo}, Iv{\'a}n and {Antonelli}, Lucio A. and {Bachetti}, Matteo and {Baldini}, Luca and {Baumgartner}, Wayne H. and {Bellazzini}, Ronaldo and {Bianchi}, Stefano and {Bongiorno}, Stephen D. and {Bonino}, Raffaella and {Brez}, Alessandro and {Bucciantini}, Niccol{\`o} and {Capitanio}, Fiamma and {Castellano}, Simone and {Cavazzuti}, Elisabetta and {Chen}, Chien-Ting and {Ciprini}, Stefano and {Costa}, Enrico and {De Rosa}, Alessandra and {Del Monte}, Ettore and {Di Gesu}, Laura and {Di Lalla}, Niccol{\`o} and {Donnarumma}, Immacolata and {Doroshenko}, Victor and {Dov{\v{c}}iak}, Michal and {Ehlert}, Steven R. and {Enoto}, Teruaki and {Evangelista}, Yuri and {Fabiani}, Sergio and {Ferrazzoli}, Riccardo and {Garcia}, Javier A. and {Gunji}, Shuichi and {Hayashida}, Kiyoshi and {Heyl}, Jeremy and {Iwakiri}, Wataru and {Jorstad}, Svetlana G. and {Karas}, Vladimir and {Kislat}, Fabian and {Kitaguchi}, Takao and {Kolodziejczak}, Jeffery J. and {Krawczynski}, Henric and {Latronico}, Luca and {Liodakis}, Ioannis and {Maldera}, Simone and {Manfreda}, Alberto and {Marin}, Fr{\'e}d{\'e}ric and {Marscher}, Alan P. and {Massaro}, Francesco and {Matt}, Giorgio and {Mitsuishi}, Ikuyuki and {Mizuno}, Tsunefumi and {Negro}, Michela and {Ng}, Chi-Yung and {O'Dell}, Stephen L. and {Omodei}, Nicola and {Oppedisano}, Chiara and {Pavlov}, George G. and {Peirson}, Abel L. and {Perri}, Matteo and {Pesce-Rollins}, Melissa and {Petrucci}, Pierre-Olivier and {Possenti}, Andrea and {Puccetti}, Simonetta and {Ramsey}, Brian D. and {Ratheesh}, Ajay and {Roberts}, Oliver J. and {Romani}, Roger W. and {Sgr{\`o}}, Carmelo and {Slane}, Patrick and {Soffitta}, Paolo and {Spandre}, Gloria and {Swartz}, Douglas A. and {Tamagawa}, Toru and {Tavecchio}, Fabrizio and {Taverna}, Roberto and {Tawara}, Yuzuru and {Thomas}, Nicholas E. and {Tombesi}, Francesco and {Trois}, Alessio and {Turolla}, Roberto and {Vink}, Jacco and {Xie}, Fei},
        title = "{X-Ray Polarized View of the Accretion Geometry in the X-Ray Binary Circinus X-1}",
      journal = {\apjl},
         year = 2024,
        month = jan,
       volume = {961},
       number = {1},
          eid = {L8},
        pages = {L8},
          doi = {10.3847/2041-8213/ad1832},
archivePrefix = {arXiv},
       eprint = {2311.04632},
 primaryClass = {astro-ph.HE},
       adsurl = {https://ui.adsabs.harvard.edu/abs/2024ApJ...961L...8R}
}

@ARTICLE{Dovciak24,
       author = {{Dov{\v{c}}iak}, Michal and {Podgorn{\'y}}, Jakub and {Svoboda}, Ji{\v{r}}{\'\i} and {Steiner}, James F. and {Kaaret}, Philip and {Krawczynski}, Henric and {Ingram}, Adam and {Kravtsov}, Vadim and {Marra}, Lorenzo and {Muleri}, Fabio and {Garc{\'\i}a}, Javier A. and {Mastroserio}, Guglielmo and {Miku{\v{s}}incov{\'a}}, Romana and {Ratheesh}, Ajay and {Cavero}, Nicole Rodriguez},
        title = "{IXPE View of BH XRBs during the First 2.5 Years of the Mission}",
      journal = {Galaxies},
         year = 2024,
        month = sep,
       volume = {12},
       number = {5},
          eid = {54},
        pages = {54},
          doi = {10.3390/galaxies12050054},
       adsurl = {https://ui.adsabs.harvard.edu/abs/2024Galax..12...54D}
}
\bibliographystyle{aasjournalv7}

\appendix

\section{Observatories and data reduction}\label{data_red}

In this section, the methods applied to extract and analyze the data from the different observatories are reported; the complete dataset is summarized in Table~\ref{tab:Observation_datasets}. For the timing analysis, we used custom code based on the \textsc{astropy} package. The spectral analysis was conducted within the \textsc{xspec} version 12.15.0d \citep{1996ASPC..101...17A}, utilizing the \texttt{wilm} abundance \citep{2000ApJ...542..914W}.

\begin{deluxetable}{lcccc}[!htb]
\tabletypesize{\scriptsize}
\tablewidth{0pt}
\label{tab:Observation_datasets}
\tablehead{
\colhead{Observatory} & \colhead{Epoch} & \colhead{Obs ID} & \colhead{Date} & \colhead{Exposure (ks)}}
\startdata
\ixpe &1& 01002901 & 2022-05-15 & 242 \\
     &2& 01250101 & 2022-06-18 & 86 \\
     &3& 02008201 & 2023-05-02 & 21 \\
     &4& 02008301 & 2023-05-09 & 31 \\
     &5& 02008401 & 2023-05-24 & 25 \\
     &6& 02008501 & 2023-06-13 & 29 \\
     &7& 02008601 & 2023-06-20 & 34 \\
     &8& 03002201 & 2024-04-12 & 56 \\
     &9& 03003101 & 2024-05-06 & 54 \\
     &10& 03010001 & 2024-05-26 & 58 \\
     &11& 03010101 & 2024-06-14 & 56 \\
     &12& 03002599a & 2024-10-10 & 55 \\
     &13& 03002599b & 2024-12-12 & 56 \\
     &14& 05250601 & 2026-06-06 & 99 \\
\hline
\nicer && 5100320101 & 2022-05-15 & 8 \\
      && 5100320102 & 2022-05-16 & 4 \\
      && 5100320103 & 2022-05-16 & 24 \\
      && 5100320104 & 2022-05-18 & 27 \\
      && 5100320105 & 2022-05-18 & 14 \\
      && 5100320106 & 2022-05-20 & 11 \\
      && 5100320107 & 2022-05-21 & 3 \\
      && 5100320108 & 2022-06-20 & 3 \\
\hline
\nustar && 30702017002 & 2022-05-18 & 16 \\
      && 30702017004 & 2022-05-19 & 14 \\
      && 30702017006 & 2022-05-20 & 12 \\
      && 90802013002 & 2022-06-20 & 1 \\
\enddata
\caption{List of the Cyg~X-1 \ixpe observations and simultaneous observations by \nicer and \nustar used in this work. \ixpe Observation ID 03002599 was divided into two parts.}
\end{deluxetable}

\subsection{\ixpe}

\ixpe \citep{2022JATIS...8b6002W,2021AJ....162..208S}, a mission led by NASA and the Italian Space Agency (ASI), was launched on December 9, 2021. It carries three co-aligned telescopes, which provide data in the 2--8\,keV energy range \citep{2021APh...13302628B,2021AJ....162..208S,2022AJ....164..103D,2022AJ....163..170D}.

The \ixpe level 2 datasets are provided by the HEASARC archive; the data have been extracted by selecting the source in an 80" circular region by SAOImage~DS9 version 7.5 \citep{2003ASPC..295..489J}. Given the high source flux, the background need not be rejected or subtracted \citep{2023AJ....165..143D}. A model‑independent, unweighted polarization analysis \citep{2022AJ....163..170D} was performed using the \texttt{PCUBE} algorithm in the \textsc{ixpeobssim} package version 31.0.1 \citep{2022SoftX..1901194B} and applying the response matrices appropriate for each observation: \texttt{20220702\_v013} for the 2022 data, \texttt{20230702\_v013} for the 2023 data, and \texttt{20240701\_v013} for the 2024 and 2026 data. We generated the unweighted \ixpe spectra using the same response matrix. The spectra were binned with a minimum of 30 counts per bin for Stokes $I$ and a constant binning of 200\,eV for Stokes $U$ and $Q$. We performed an additional PI correction for Observation ID 01002901 to properly correct the PI column following instructions reported in a dedicated README file.

\subsection{\nicer}
\nicer is a non-imaging observatory hosted on board the International Space Station, operating in the 0.2--12\,keV energy range, and consists of 56 silicon drift detectors placed at the focus of co-aligned concentrator X-ray optics \citep{2016SPIE.9905E..1HG}. The \nicer observations reported in Table~\ref{tab:Observation_datasets} were selected for their overlap with the \ixpe ones; the data were processed using the standard \nicer pipeline tasks \texttt{nicerl2} and \texttt{nicerl3-spec} in the HEASoft version 6.35.2 using the SCORPEON model for the background. GTIs were selected by excluding periods exhibiting anomalously high count rates in 12--15\,keV. The spectral analysis was performed in the 1--10\,keV energy band using Kaastra \& Bleeker binning \citep{2016A&A...587A.151K} with a minimum of 30 counts/bin to ensure statistical quality of the data.

\subsection{\nustar}
\nustar is an X-ray observatory operating in the energy range 3--79\,keV, and providing broadband X-ray imaging, spectroscopy, and timing \citep{2013ApJ...770..103H}. \nustar data were spatially selected by defining a circular region with a radius of 120'' centered on the source in SAOImage DS9, while the background was selected in a region of the same size away from the source. The spectra were extracted using the standard \nustar pipeline tool \texttt{nuproducts}, considering the 3--79\,keV energy band. The considered \nustar observations also correspond to periods that overlap with the \ixpe ones, and the binning method is the same as that adopted for \nicer.

\section{Dips selection}\label{app:dips_selection}

To select these dips for the analysis, we first defined the hardness ratio (HR) using \ixpe data, as the photon counts in the 3.5--8\,keV band divided by those in the 2--3.5\,keV band.
The HR curves were binned with a time bin of 75\,seconds, and we calculated the average HR and its standard deviation $\sigma_{HR}$ every five consecutive time bins. A selection for time bins having HR${>}2\sigma_{HR}+\textrm{HR}_{\textrm{average}}$ is applied to identify ``dip'' intervals. The resulting good time intervals (GTIs) for the dip state are listed in Table~\ref{tab:dips_gti_2022}.

\begin{figure}[!hbt]
\centering
\includegraphics[width=\linewidth]{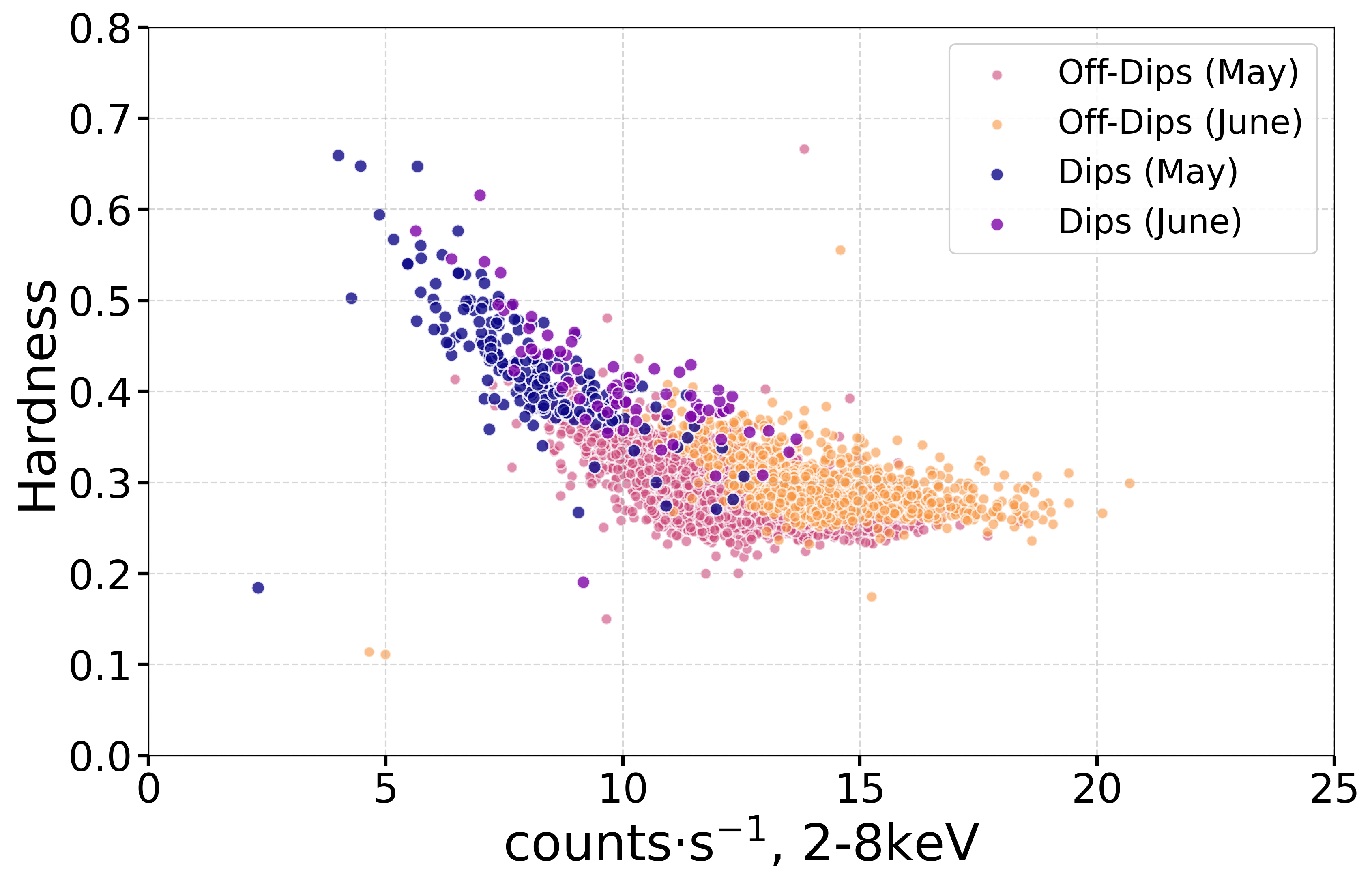}
\caption{\ixpe HID obtained by applying 75\,seconds time bins. The darker points represent the ``dip'' time intervals, while the lighter ones are the ``off-dip'' intervals. The states of Cyg~X-1 show no significant changes between May and June 2022.}
\label{hardness_and_count_rate_for_time_bins}
\end{figure}

\begin{deluxetable*}{lccccc}[!htb]
\tabletypesize{\normalsize}
\tablewidth{0pt}
\label{tab:dips_gti_2022}
\tablehead{
\colhead{Dip \#} & \colhead{START (s)} & \colhead{STOP (s)} & \colhead{Start Phase} & \colhead{End Phase} & \colhead{Duration (s)}}
\startdata
\multicolumn{6}{c}{Obs ID 01002901 (2022 May)} \\
1 & 169433650.50 & 169443325.50 & 0.9620 & 0.9820 & 9675.0 \\
2 & 169444825.50 & 169445350.50 & 0.9851 & 0.9861 & 525.0 \\
3 & 169446100.50 & 169452250.50 & 0.9877 & 0.0004 & 6150.0 \\
4 & 169490875.50 & 169491700.50 & 0.0802 & 0.0819 & 825.0 \\
5 & 169491775.50 & 169492525.50 & 0.0821 & 0.0837 & 750.0 \\
6 & 169514875.50 & 169515700.50 & 0.1298 & 0.1316 & 825.0 \\
7 & 169520425.50 & 169521625.50 & 0.1413 & 0.1438 & 1200.0 \\
8 & 169566175.50 & 169566850.50 & 0.2359 & 0.2373 & 675.0 \\
9 & 169616800.50 & 169617100.50 & 0.3405 & 0.3411 & 300.0 \\
10 & 169907650.50 & 169908625.50 & 0.9417 & 0.9437 & 975.0 \\
\multicolumn{6}{c}{Obs ID 01250101 (2022 June)} \\
1 & 172370240.72 & 172370615.72 & 0.0315 & 0.0323 & 375.0 \\
2 & 172371965.72 & 172372415.72 & 0.0351 & 0.0360 & 450.0 \\
3 & 172373315.72 & 172373765.72 & 0.0379 & 0.0388 & 450.0 \\
4 & 172385015.72 & 172389365.72 & 0.0621 & 0.0711 & 4350.0 \\
5 & 172395065.72 & 172395590.72 & 0.0828 & 0.0839 & 525.0 \\
6 & 172453040.72 & 172453490.72 & 0.2027 & 0.2036 & 450.0 \\
7 & 172517165.72 & 172517690.72 & 0.3352 & 0.3363 & 525.0 \\
\enddata
\caption{The selected GTI dip intervals for \ixpe observations 01002901 (May 2022) and 01250101 (June 2022).}
\end{deluxetable*}
Figure~\ref{hardness_and_count_rate_for_time_bins} shows the HID where ``dip'' (darker points) and ``off-dip'' (lighter points) intervals for the considered \ixpe observations are reported.

\section{Polarization Properties of All \ixpe Observations} \label{app:appendix_pol_all}

Table~\ref{tab:Polarization_properties_of_all_IXPE} presents the polarization properties of all \ixpe observations in the 2--3.5\,keV, 3.5--8\,keV, and 2--8\,keV energy bands.

\begin{deluxetable*}{lccccccccc}[!htb]
\tabletypesize{\scriptsize}
\tablewidth{0pt}
\label{tab:Polarization_properties_of_all_IXPE}
\tablehead{
\colhead{} & \multicolumn{3}{c}{01002901} & \multicolumn{3}{c}{01250101} & \multicolumn{3}{c}{02008201} \\
\cline{2-4} \cline{5-7} \cline{8-10}
\colhead{} & \colhead{2--3.5\,keV} & \colhead{3.5--8\,keV} & \colhead{2--8\,keV} & \colhead{2--3.5\,keV} & \colhead{3.5--8\,keV} & \colhead{2--8\,keV} &
\colhead{2--3.5\,keV} & \colhead{3.5--8\,keV} & \colhead{2--8\,keV}} 
\startdata
$U/I$ ($\%$) & $-2.1\pm0.2$ & $-3.0\pm0.3$ & $-2.3\pm0.2$ 
& $-3.3\pm0.3$ & $-2.7\pm0.4$ & $-3.1\pm0.3$ 
& $-1.0\pm0.4$ & $-2.1\pm0.6$ & $-1.1\pm0.4$ \\ 
$Q/I$ ($\%$) & $2.7\pm0.2$ & $3.2\pm0.3$ & $2.8\pm0.2$ 
& $1.9\pm0.3$ & $3.0\pm0.4$ & $2.1\pm0.3$ 
& $1.6\pm0.4$ & $2.6\pm0.6$ & $1.7\pm0.4$ \\
\cline{1-10}
PA ($^\circ$) & $-19\pm2$ & $-21\pm2$ & $-20\pm1$ 
& $-30\pm3$ & $-21\pm3$ & $-28\pm2$ 
& $-16\pm6$ & $-20\pm6$ & $-16\pm5$ \\
PD ($\%$) & $3.5\pm0.2$ & $4.4\pm0.3$ & $3.7\pm0.2$ 
& $3.8\pm0.3$ & $4.1\pm0.4$ & $3.8\pm0.3$ 
& $1.9\pm0.4$ & $3.3\pm0.6$ & $2.1\pm0.4$ \\
\cline{1-10}
MDP$_{99}$ ($\%$) & 0.7 & 0.8 & 0.5 
& 1.1 & 1.2 & 0.8 
& 1.3 & 2.0 & 1.1 \\
\cline{1-10}
\colhead{} & \multicolumn{3}{c}{02008301} & \multicolumn{3}{c}{02008401} & \multicolumn{3}{c}{02008501} \\
\cline{1-10}
$U/I$ ($\%$) & $-1.4\pm0.3$ & $-2.6\pm0.5$ & $-1.5\pm0.3$ 
& $-1.1\pm0.3$ & $-2.9\pm0.6$ & $-1.2\pm0.3$ 
& $-1.0\pm0.3$ & $-1.4\pm0.5$ & $-1.0\pm0.3$ \\ 
$Q/I$ ($\%$) & $1.2\pm0.3$ & $3.0\pm0.5$ & $1.4\pm0.3$ 
& $0.8\pm0.3$ & $3.1\pm0.6$ & $1.0\pm0.3$ 
& $0.7\pm0.3$ & $1.2\pm0.5$ & $0.8\pm0.3$ \\
\cline{1-10}
PA ($^\circ$) & $-25\pm5$ & $-20\pm3$ & $-24\pm4$ 
& $-26\pm6$ & $-21\pm4$ & $-25\pm5$ 
& $-27\pm7$ & $-25\pm8$ & $-27\pm6$ \\
PD ($\%$) & $1.8\pm0.3$ & $3.9\pm0.5$ & $2.1\pm0.3$ 
& $1.4\pm0.3$ & $4.2\pm0.6$ & $1.6\pm0.3$ 
& $1.2\pm0.3$ & $1.8\pm0.5$ & $1.3\pm0.3$ \\
\cline{1-10}
MDP$_{99}$ ($\%$) & 0.9 & 1.5 & 0.8 
& 0.9 & 1.7 & 0.8 
& 0.9 & 1.5 & 0.8 \\
\cline{1-10}
\colhead{} & \multicolumn{3}{c}{02008601} & \multicolumn{3}{c}{03002201} & \multicolumn{3}{c}{03003101} \\
\cline{1-10}
$U/I$ ($\%$) & $-1.8\pm0.2$ & $-2.0\pm0.4$ & $-1.9\pm0.2$ 
& $-1.8\pm0.5$ & $-3.0\pm0.6$ & $-2.0\pm0.4$ 
& $-2.0\pm0.5$ & $-2.0\pm0.6$ & $-2.0\pm0.4$ \\ 
$Q/I$ ($\%$) & $0.4\pm0.2$ & $1.5\pm0.4$ & $0.5\pm0.2$ 
& $2.9\pm0.5$ & $2.3\pm0.6$ & $2.8\pm0.4$ 
& $2.0\pm0.5$ & $2.8\pm0.6$ & $2.2\pm0.4$ \\
\cline{1-10}
PA ($^\circ$) & $-39\pm4$ & $-26\pm4$ & $-37\pm3$ 
& $-16\pm4$ & $-26\pm4$ & $-18\pm3$ 
& $-23\pm5$ & $-17\pm5$ & $-21\pm4$ \\
PD ($\%$) & $1.9\pm0.2$ & $2.5\pm0.4$ & $1.9\pm0.2$ 
& $3.4\pm0.5$ & $3.8\pm0.6$ & $3.4\pm0.4$ 
& $2.8\pm0.5$ & $3.5\pm0.6$ & $2.9\pm0.4$ \\
\cline{1-10}
MDP$_{99}$ ($\%$) & 0.7 & 1.2 & 0.6 
& 1.5 & 1.8 & 1.2 
& 1.6 & 1.8 & 1.2 \\
\cline{1-10}
\colhead{} & \multicolumn{3}{c}{03010001} & \multicolumn{3}{c}{03010101} & \multicolumn{3}{c}{03002599a} \\
\cline{1-10}
$U/I$ ($\%$) & $-3.6\pm0.5$ & $-4.2\pm0.6$ & $-3.7\pm0.4$ 
& $-4.0\pm0.6$ & $-4.1\pm0.6$ & $-4.0\pm0.4$ 
& $-0.6\pm0.3$ & $-3.2\pm0.4$ & $-1.0\pm0.3$ \\ 
$Q/I$ ($\%$) & $2.6\pm0.5$ & $2.0\pm0.6$ & $2.5\pm0.4$ 
& $1.3\pm0.6$ & $1.9\pm0.6$ & $1.5\pm0.4$ 
& $2.2\pm0.3$ & $2.0\pm0.4$ & $2.2\pm0.3$ \\
\cline{1-10}
PA ($^\circ$) & $-27\pm3$ & $-32\pm3$ & $-28\pm2$ 
& $-36\pm4$ & $-33\pm4$ & $-35\pm3$ 
& $-8\pm4$ & $-29\pm3$ & $-13\pm3$ \\
PD ($\%$) & $4.4\pm0.5$ & $4.7\pm0.6$ & $4.4\pm0.4$ 
& $4.2\pm0.6$ & $4.5\pm0.6$ & $4.3\pm0.4$ 
& $2.3\pm0.3$ & $3.8\pm0.4$ & $2.4\pm0.3$ \\
\cline{1-10}
MDP$_{99}$ ($\%$) & 1.5 & 1.7 & 1.2 
& 1.7 & 1.9 & 1.3 
& 1.0 & 1.4 & 0.8 \\
\cline{1-10}
\colhead{} & \multicolumn{3}{c}{03002599b} & \multicolumn{3}{c}{05250601} & \multicolumn{3}{c}{} \\
\cline{1-7}
$U/I$ ($\%$) & $-1.7\pm0.2$ & $-2.5\pm0.4$ & $-1.8\pm0.2$
& $-2.9\pm0.4$ & $-2.0\pm0.4$ & $-2.7\pm0.3$ &&&\\ 
$Q/I$ ($\%$) & $1.3\pm0.2$ & $2.8\pm0.4$ & $1.5\pm0.2$
& $1.9\pm0.4$ & $2.4\pm0.4$ & $2.0\pm0.3$ &&&\\ 
\cline{1-7}
PA ($^\circ$) & $-27\pm3$ & $-21\pm3$ & $-25\pm3$ 
& $-28\pm3$ & $-20\pm4$ & $-26\pm3$ &&&\\ 
PD ($\%$) & $2.1\pm0.2$ & $3.8\pm0.4$ & $2.3\pm0.2$ 
& $3.5\pm0.4$ & $3.2\pm0.4$ & $3.4\pm0.3$ &&&\\ 
\cline{1-7}
MDP$_{99}$ ($\%$) & 0.7 & 1.2 & 0.6 
& 1.2 & 1.3 & 0.9 &&&\\ 
\enddata
\caption{Polarization properties of all \ixpe observations. \ixpe observation ID 03002599 is divided into two parts. Errors are reported at 68\% CL.}
\end{deluxetable*}

\begin{deluxetable*}{lccc}[!htb]
\tabletypesize{\normalsize}
\tablewidth{0pt}
\label{tab:Polarization_properties_of_dip_time_intervals_2}
\tablehead{
\colhead{} & \multicolumn{3}{c}{dip} \\
\cline{2-4}
\colhead{} & \colhead{2--3.5\,keV} & \colhead{3.5--8\,keV} & \colhead{2--8\,keV}}
\startdata
\colhead{} & \multicolumn{3}{c}{02008201} \\
\cline{1-4}
$U/I$ ($\%$) & $8.1\pm4.4$ & $-12.0\pm6.5$ & $5.4\pm3.8$ \\ 
$Q/I$ ($\%$) & $4.9\pm4.4$ & $3.3\pm6.5$ & $4.7\pm3.8$ \\
\cline{1-4}
PA ($^\circ$) & $29\pm13$ & $-37\pm15$ & $24\pm15$ \\
PD ($\%$) & $9.4\pm4.4$ & $12.5\pm6.5$ & $7.1\pm3.8$ \\
\cline{1-4}
MDP$_{99}$ ($\%$) & 13.4 & 19.8 & 11.4 \\
\cline{1-4}
\colhead{} & \multicolumn{3}{c}{02008301} \\
\cline{1-4}
$U/I$ ($\%$) & $-0.5\pm1.8$ & $-3.3\pm2.6$ & $-0.9\pm1.5$ \\ 
$Q/I$ ($\%$) & $-1.8\pm1.8$ & $3.5\pm2.6$ & $-1.0\pm1.5$ \\
\cline{1-4}
PA ($^\circ$) & $-82\pm28$ & $-22\pm15$ & $-69\pm32$ \\
PD ($\%$) & $1.8\pm1.8$ & $4.9\pm2.6$ & $1.4\pm1.5$ \\
\cline{1-4}
MDP$_{99}$ ($\%$) & 5.4 & 7.8 & 4.5 \\
\cline{1-4}
\colhead{} & \multicolumn{3}{c}{03003101} \\
\cline{1-4}
$U/I$ ($\%$) & $4.4\pm4.1$ & $-6.9\pm4.2$ & $1.3\pm3.0$ \\ 
$Q/I$ ($\%$) & $0.6\pm4.1$ & $6.1\pm4.2$ & $2.1\pm3.0$ \\
\cline{1-4}
PA ($^\circ$) & $41\pm27$ & $-24\pm13$ & $16\pm35$ \\
PD ($\%$) & $4.4\pm4.1$ & $9.2\pm4.2$ & $2.5\pm3.0$ \\
\cline{1-4}
MDP$_{99}$ ($\%$) & 12.6 & 12.6 & 9.2 \\
\cline{1-4}
\colhead{} & \multicolumn{3}{c}{03010001} \\
\cline{1-4}
$U/I$ ($\%$) & $-4.1\pm4.8$ & $-8.4\pm5.3$ & $-5.1\pm3.7$ \\ 
$Q/I$ ($\%$) & $0.6\pm4.8$ & $-11.1\pm5.3$ & $-2.1\pm3.7$ \\
\cline{1-4}
PA ($^\circ$) & $-41\pm33$ & $-71\pm11$ & $-56\pm19$ \\
PD ($\%$) & $4.2\pm4.8$ & $13.9\pm5.3$ & $5.6\pm3.7$ \\
\cline{1-4}
MDP$_{99}$ ($\%$) & 14.6 & 16.2 & 11.1 \\
\cline{1-4}
\colhead{} & \multicolumn{3}{c}{03010101} \\
\cline{1-4}
$U/I$ ($\%$) & $-11.1\pm4.0$ & $-11.1\pm3.5$ & $-11.1\pm2.7$ \\ 
$Q/I$ ($\%$) & $-2.5\pm4.0$ & $0.7\pm3.5$ & $-1.3\pm2.7$ \\
\cline{1-4}
PA ($^\circ$) & $-51\pm10$ & $-43\pm9$ & $-48\pm7$ \\
PD ($\%$) & $11.4\pm4.0$ & $11.2\pm3.5$ & $11.2\pm2.7$ \\
\cline{1-4}
MDP$_{99}$ ($\%$) & 12.0 & 10.5 & 8.1 \\
\cline{1-4}
\colhead{} & \multicolumn{3}{c}{05250601} \\
\cline{1-4}
$U/I$ ($\%$) & $-1.9\pm2.3$ & $0.2\pm2.1$ & $-1.3\pm1.6$ \\ 
$Q/I$ ($\%$) & $5.1\pm2.3$ & $3.2\pm2.1$ & $4.5\pm1.6$ \\
\cline{1-4}
PA ($^\circ$) & $-10\pm12$ & $1\pm19$ & $-8\pm10$ \\
PD ($\%$) & $5.4\pm2.3$ & $3.2\pm2.1$ & $4.6\pm1.6$ \\
\cline{1-4}
MDP$_{99}$ ($\%$) & 6.9 & 6.4 & 4.8 \\
\cline{1-4}
\enddata
\caption{Polarization properties of dip time intervals. Errors are reported at 68\% CL.}
\end{deluxetable*}

\begin{deluxetable*}{lccc}[!htb]
\tabletypesize{\normalsize}
\tablewidth{0pt}
\label{tab:Polarization_properties_of_off-dip_time_intervals_2}
\tablehead{
\colhead{} & \multicolumn{3}{c}{off-dip} \\
\cline{2-4}
\colhead{} & \colhead{2--3.5\,keV} & \colhead{3.5--8\,keV} & \colhead{2--8\,keV}}
\startdata
\colhead{} & \multicolumn{3}{c}{02008201} \\
\cline{1-4}
$U/I$ ($\%$) & $-1.1\pm0.4$ & $-2.0\pm0.6$ & $-1.2\pm0.4$ \\ 
$Q/I$ ($\%$) & $1.6\pm0.4$ & $2.6\pm0.6$ & $1.7\pm0.4$ \\
\cline{1-4}
PA ($^\circ$) & $-17\pm6$ & $-19\pm6$ & $-17\pm5$ \\
PD ($\%$) & $1.9\pm0.4$ & $3.3\pm0.6$ & $2.1\pm0.4$ \\
\cline{1-4}
MDP$_{99}$ ($\%$) & 1.3 & 2.0 & 1.1 \\
\cline{1-4}
\colhead{} & \multicolumn{3}{c}{02008301} \\
\cline{1-4}
$U/I$ ($\%$) & $-1.4\pm0.3$ & $-2.5\pm0.5$ & $-1.5\pm0.3$ \\ 
$Q/I$ ($\%$) & $1.3\pm0.3$ & $3.0\pm0.5$ & $1.5\pm0.3$ \\
\cline{1-4}
PA ($^\circ$) & $-24\pm5$ & $-20\pm4$ & $-23\pm4$ \\
PD ($\%$) & $1.9\pm0.3$ & $3.9\pm0.5$ & $2.1\pm0.3$ \\
\cline{1-4}
MDP$_{99}$ ($\%$) & 0.9 & 1.5 & 0.8 \\
\cline{1-4}
\colhead{} & \multicolumn{3}{c}{03003101} \\
\cline{1-4}
$U/I$ ($\%$) & $-2.1\pm0.5$ & $-1.9\pm0.6$ & $-2.0\pm0.4$ \\ 
$Q/I$ ($\%$) & $2.0\pm0.5$ & $2.8\pm0.6$ & $2.2\pm0.4$ \\
\cline{1-4}
PA ($^\circ$) & $-23\pm5$ & $-17\pm5$ & $-22\pm4$ \\
PD ($\%$) & $2.9\pm0.5$ & $3.3\pm0.6$ & $3.0\pm0.4$ \\
\cline{1-4}
MDP$_{99}$ ($\%$) & 1.6 & 1.8 & 1.2 \\
\cline{1-4}
\colhead{} & \multicolumn{3}{c}{03010001} \\
\cline{1-4}
$U/I$ ($\%$) & $-3.5\pm0.5$ & $-4.2\pm0.6$ & $-3.7\pm0.4$ \\ 
$Q/I$ ($\%$) & $2.6\pm0.5$ & $2.2\pm0.6$ & $2.5\pm0.4$ \\
\cline{1-4}
PA ($^\circ$) & $-27\pm3$ & $-31\pm3$ & $-28\pm2$ \\
PD ($\%$) & $4.4\pm0.5$ & $4.7\pm0.6$ & $4.5\pm0.4$ \\
\cline{1-4}
MDP$_{99}$ ($\%$) & 1.5 & 1.7 & 1.2 \\
\cline{1-4}
\colhead{} & \multicolumn{3}{c}{03010101} \\
\cline{1-4}
$U/I$ ($\%$) & $-3.8\pm0.6$ & $-3.8\pm0.6$ & $-3.8\pm0.4$ \\ 
$Q/I$ ($\%$) & $1.4\pm0.6$ & $1.9\pm0.6$ & $1.5\pm0.4$ \\
\cline{1-4}
PA ($^\circ$) & $-35\pm4$ & $-32\pm4$ & $-34\pm3$ \\
PD ($\%$) & $4.1\pm0.6$ & $4.3\pm0.6$ & $4.1\pm0.4$ \\
\cline{1-4}
MDP$_{99}$ ($\%$) & 1.7 & 2.0 & 1.3 \\
\cline{1-4}
\colhead{} & \multicolumn{3}{c}{05250601} \\
\cline{1-4}
$U/I$ ($\%$) & $-2.9\pm0.4$ & $-2.1\pm0.4$ & $-2.7\pm0.3$ \\ 
$Q/I$ ($\%$) & $1.8\pm0.4$ & $2.4\pm0.4$ & $1.9\pm0.3$ \\
\cline{1-4}
PA ($^\circ$) & $-29\pm3$ & $-20\pm4$ & $-27\pm3$ \\
PD ($\%$) & $3.4\pm0.4$ & $3.2\pm0.4$ & $3.3\pm0.3$ \\
\cline{1-4}
MDP$_{99}$ ($\%$) & 1.2 & 1.3 & 0.9 \\
\cline{1-4}
\enddata
\caption{Polarization properties of off-dip time intervals. Errors are reported at 68\% CL.}
\end{deluxetable*}

\section{Polarization Properties and Light Curves in 2023, 2024, and 2026} \label{app:appendix_pol_and_lc_2023_2024}

\begin{figure*}[!hbt]
\centering
\begin{tabular}{cc}
\includegraphics[width=0.9\columnwidth]{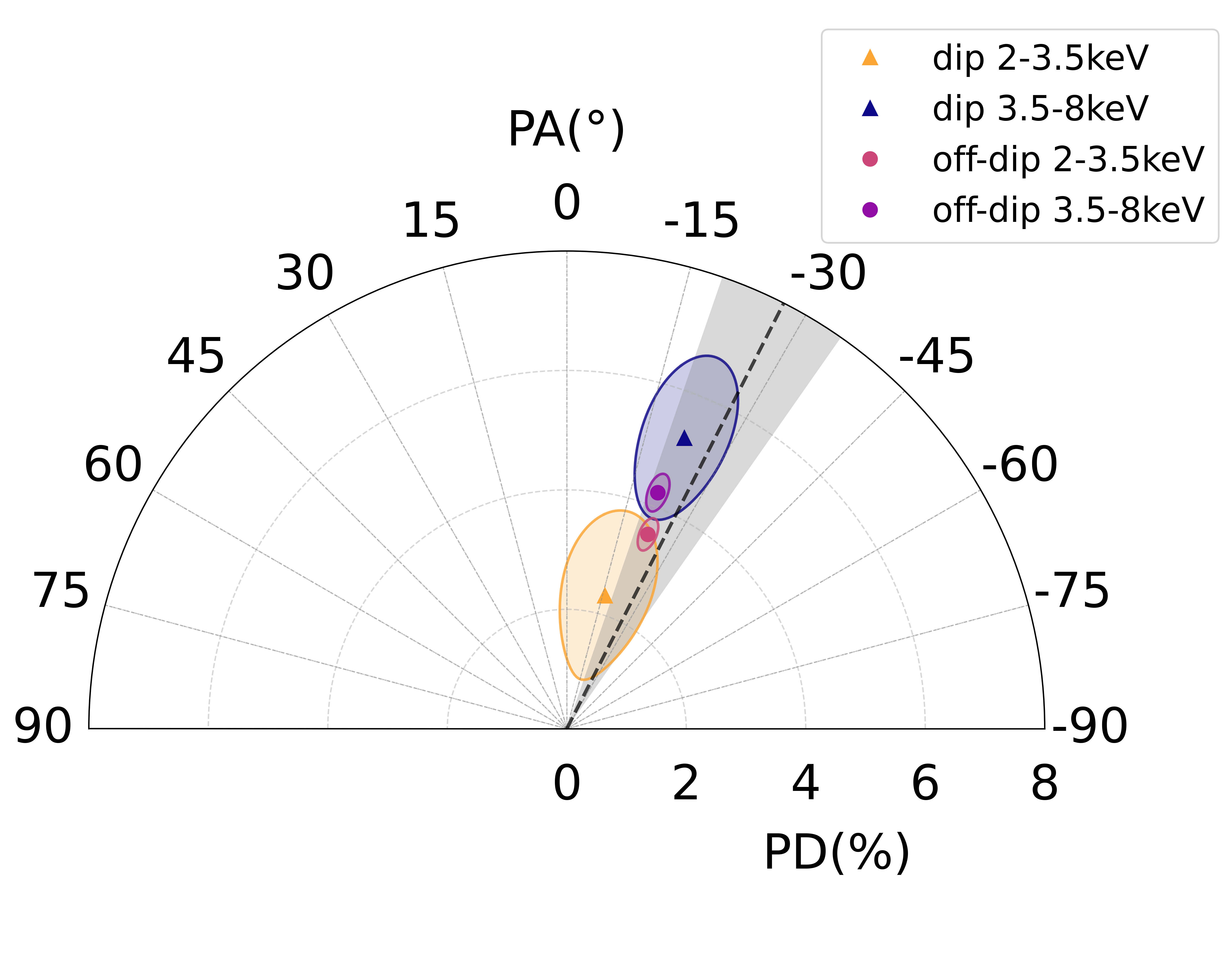} &
\includegraphics[width=0.9\columnwidth]{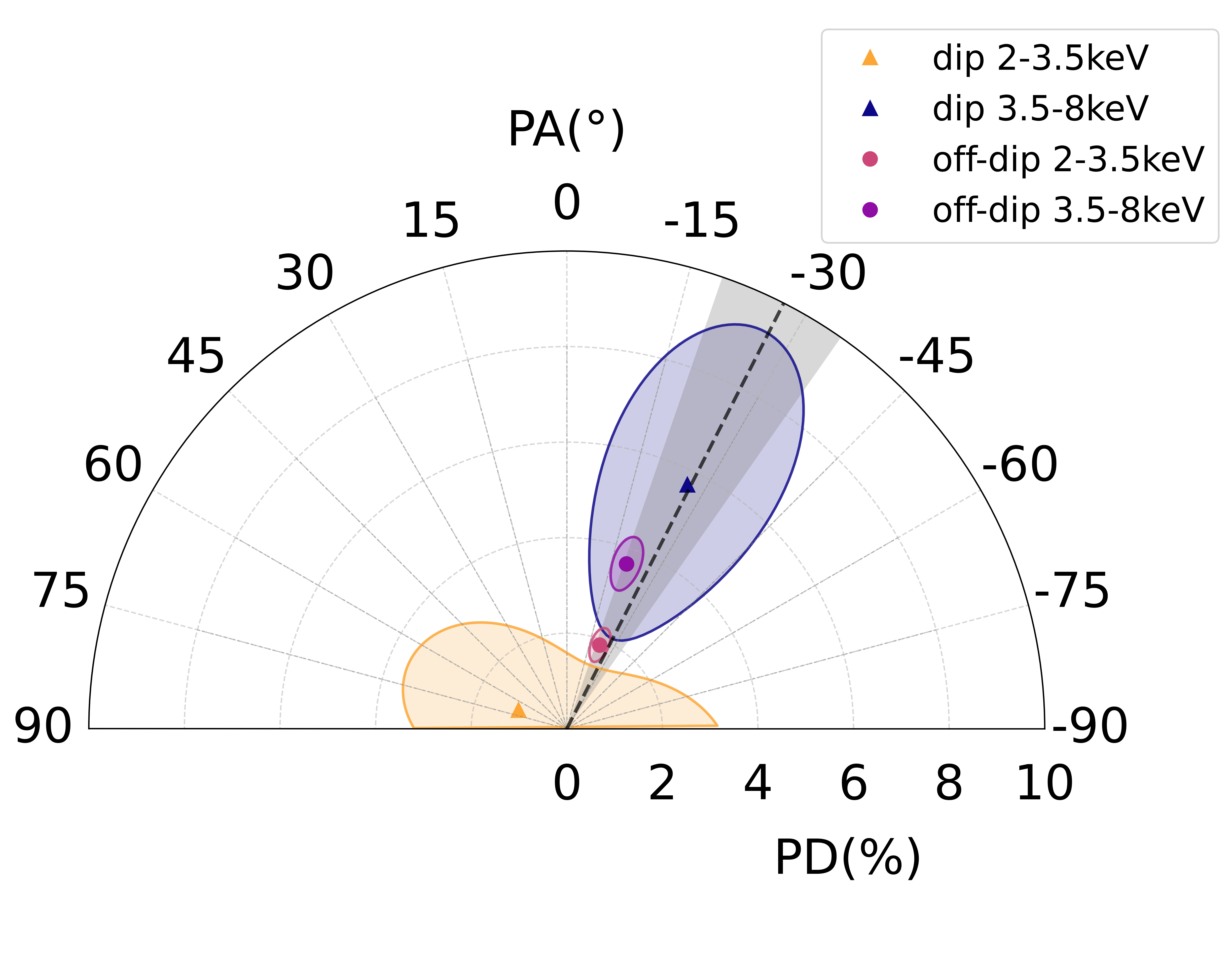} \\
\includegraphics[width=0.9\columnwidth]{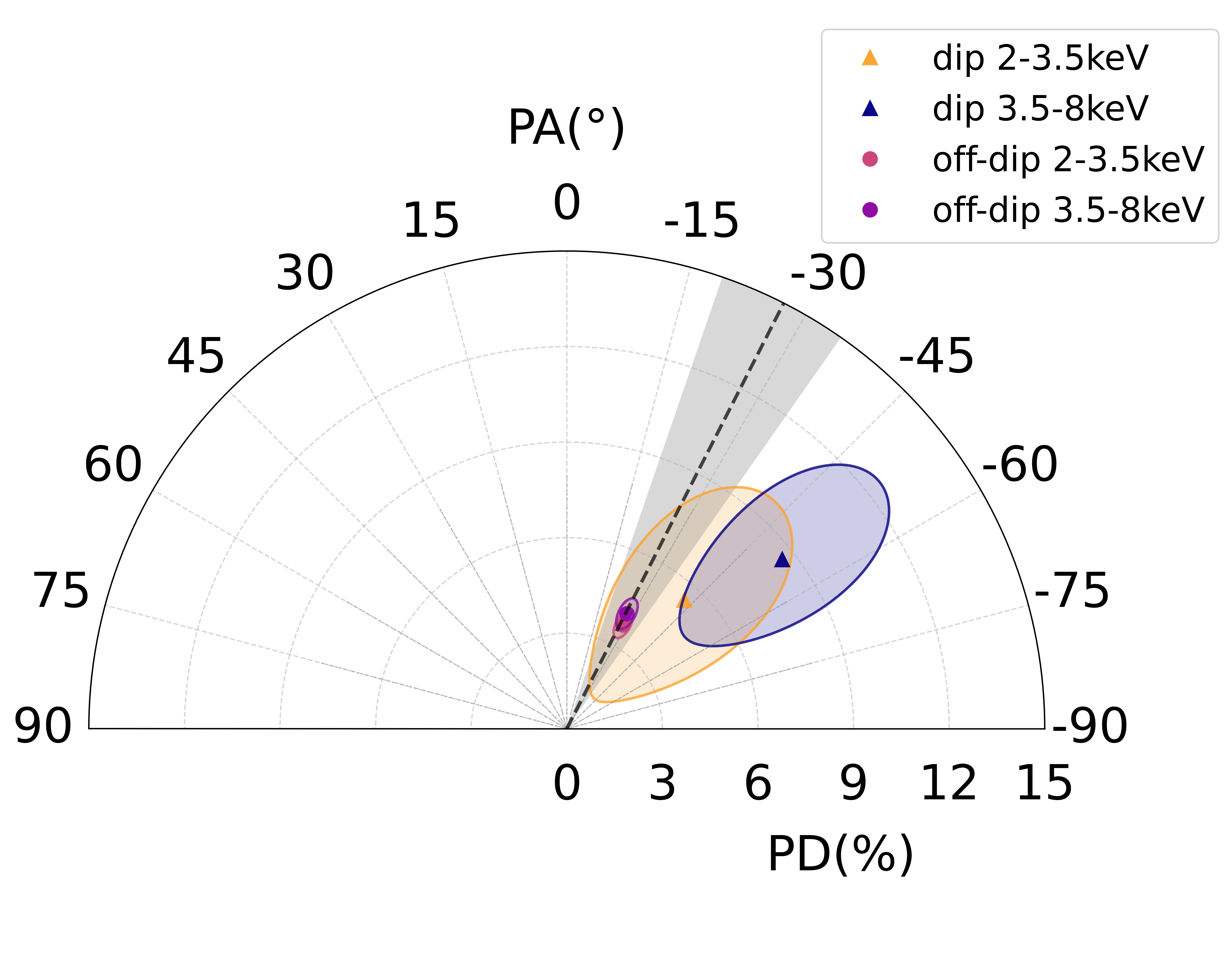} &
\includegraphics[width=0.9\columnwidth]{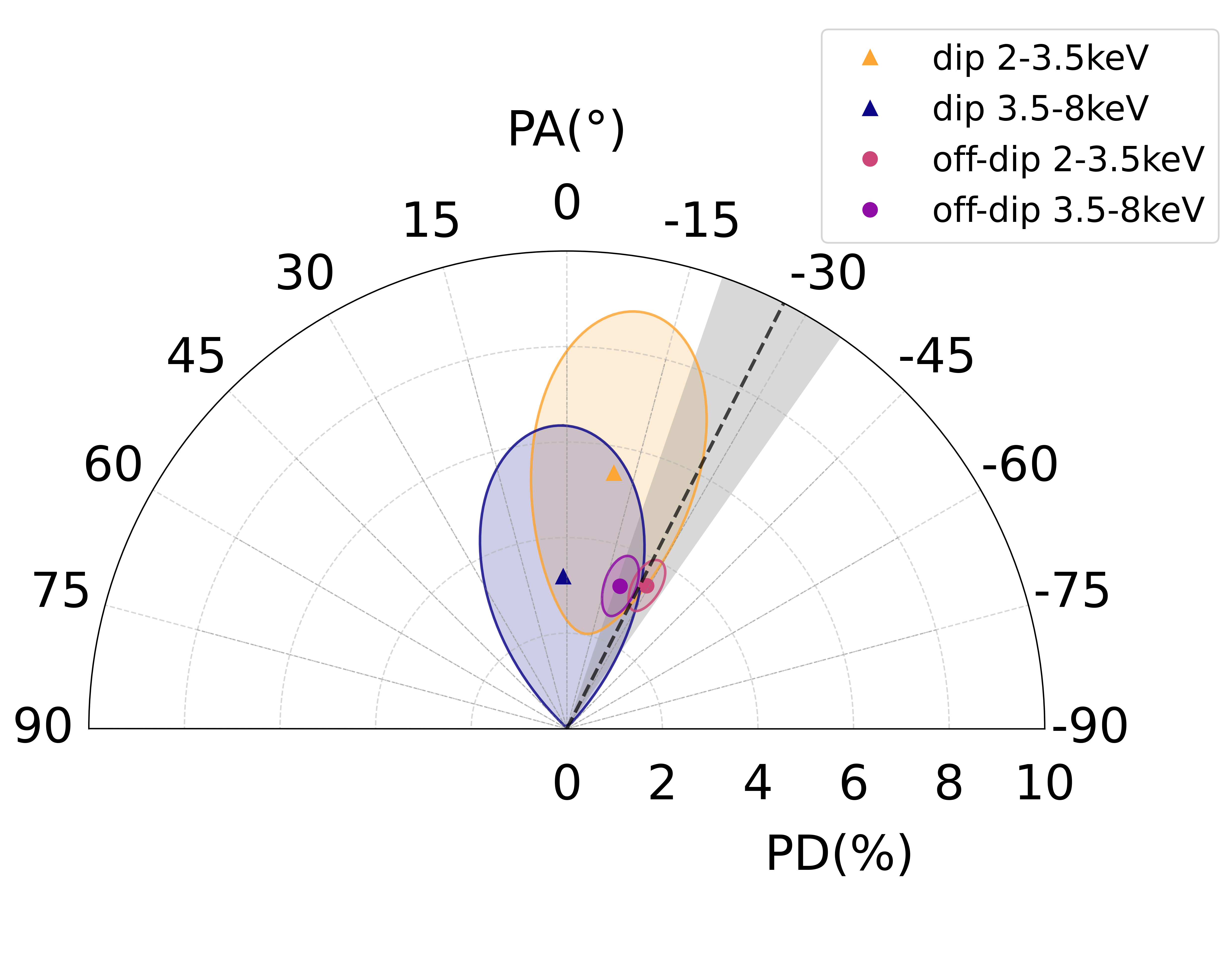}
\end{tabular}
\caption{Polarization comparison between dip and off-dip states. The data used in each panel are as follows: (a) combines observations 01002901 and 01250101; (b) combines observations 02008201 and 02008301; (c) incorporates 03003101, 03010001, and 03010101; (d) is based on observation 05250601. The gray shaded region denotes the radio jet direction of -27$^{\circ}\pm$8$^{\circ}$ \citep{2021Sci...371.1046M}. Allowed regions are reported at 68\% CL.}
\label{fig:dip_off-dip_other}
\end{figure*}

In the \ixpe observations of Cyg~X-1 in 2023, 2024, and 2026, we also found similar dips characterized by an increase in hardness with a decrease in the flux. The dips were too shallow to yield statistically significant polarization data, even after combining observation IDs in similar states, as shown in Figure~\ref{fig:dip_off-dip_other}. Among the remaining observation IDs (02008401, 02008501, 02008601, 03002201, 03002599b) for which polarization properties and light curves are not shown, no dip was found.

\begin{figure*}[!hbt]
\centering
\begin{tabular}{ccc}
\includegraphics[width=0.3\columnwidth]{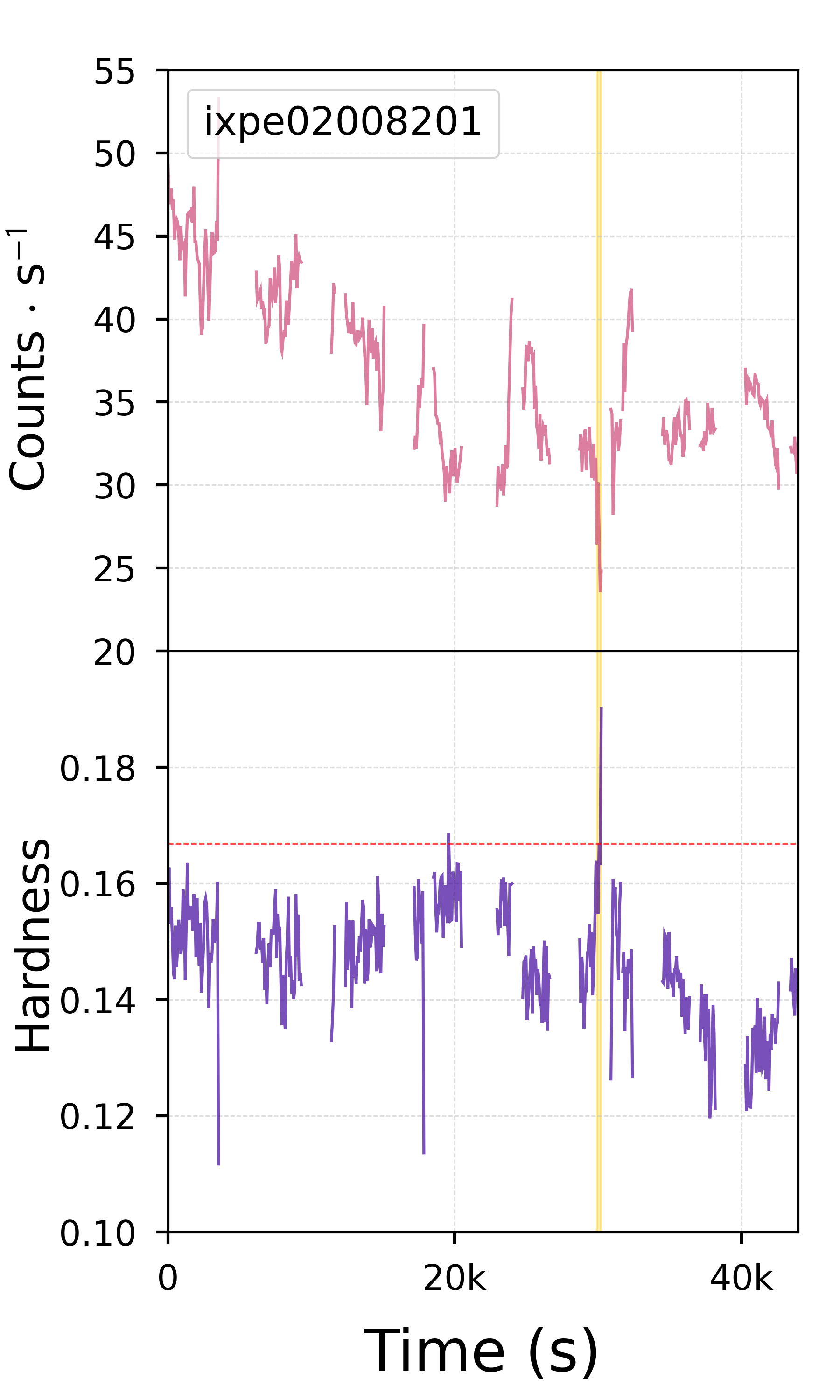} &
\includegraphics[width=0.3\columnwidth]{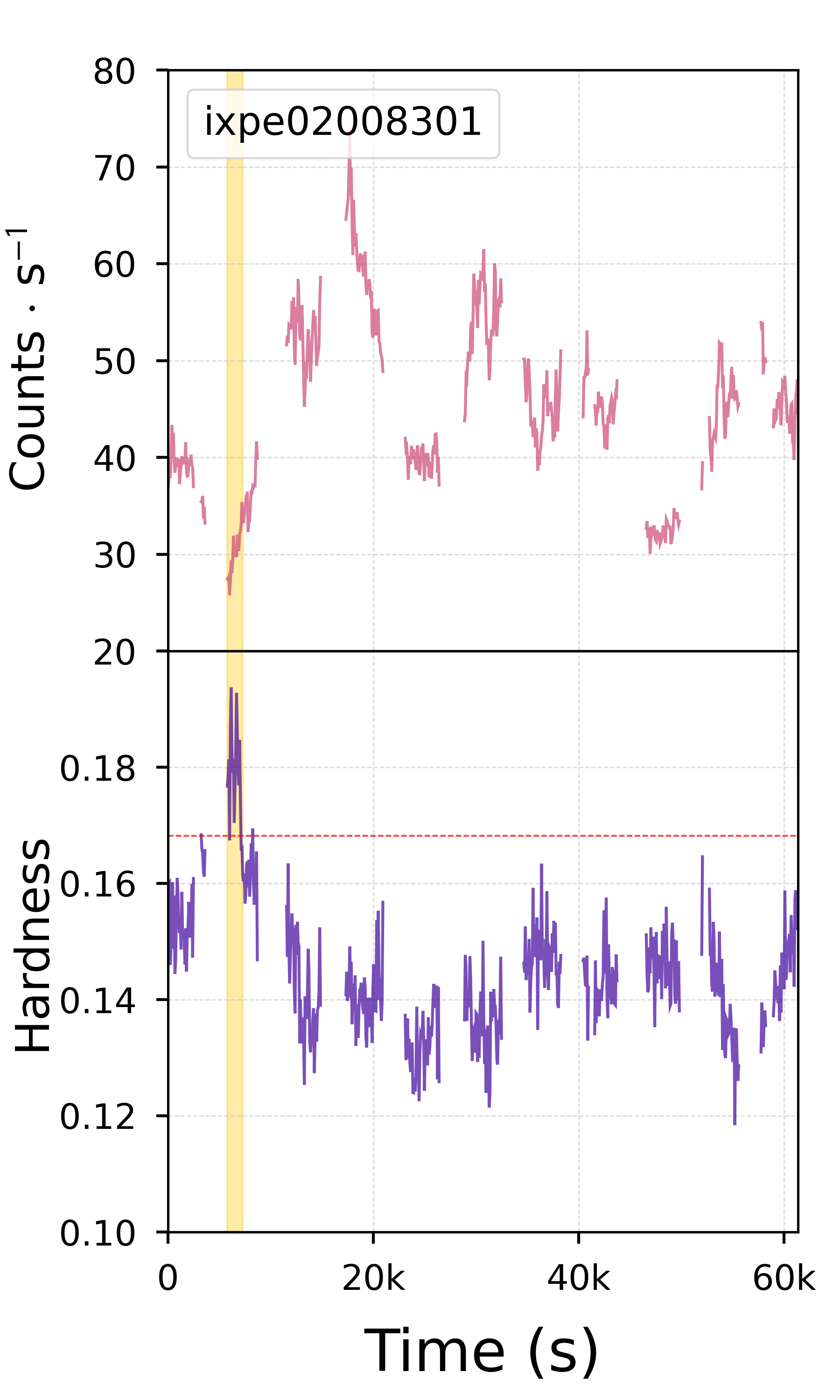} &
\includegraphics[width=0.3\columnwidth]{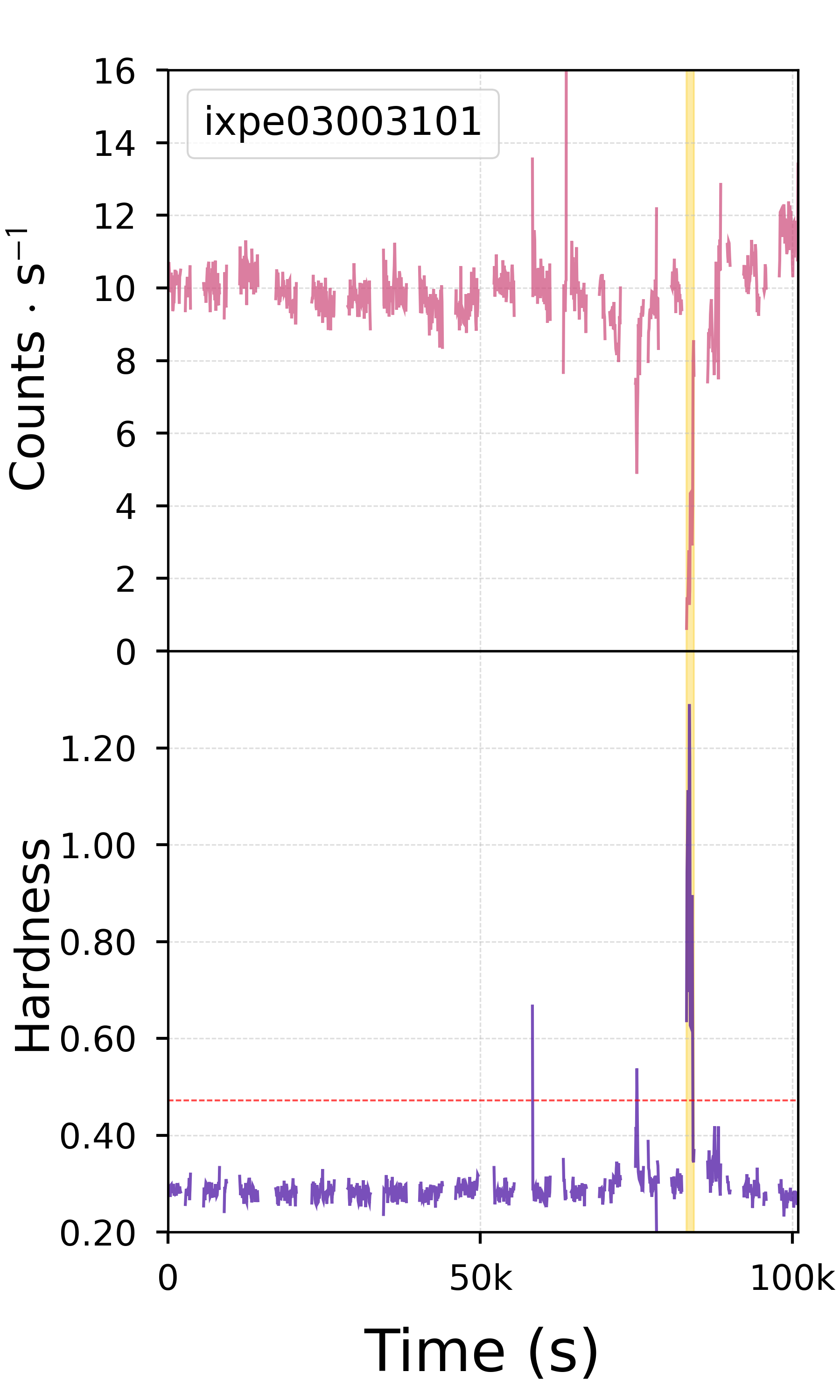} \\
\includegraphics[width=0.3\columnwidth]{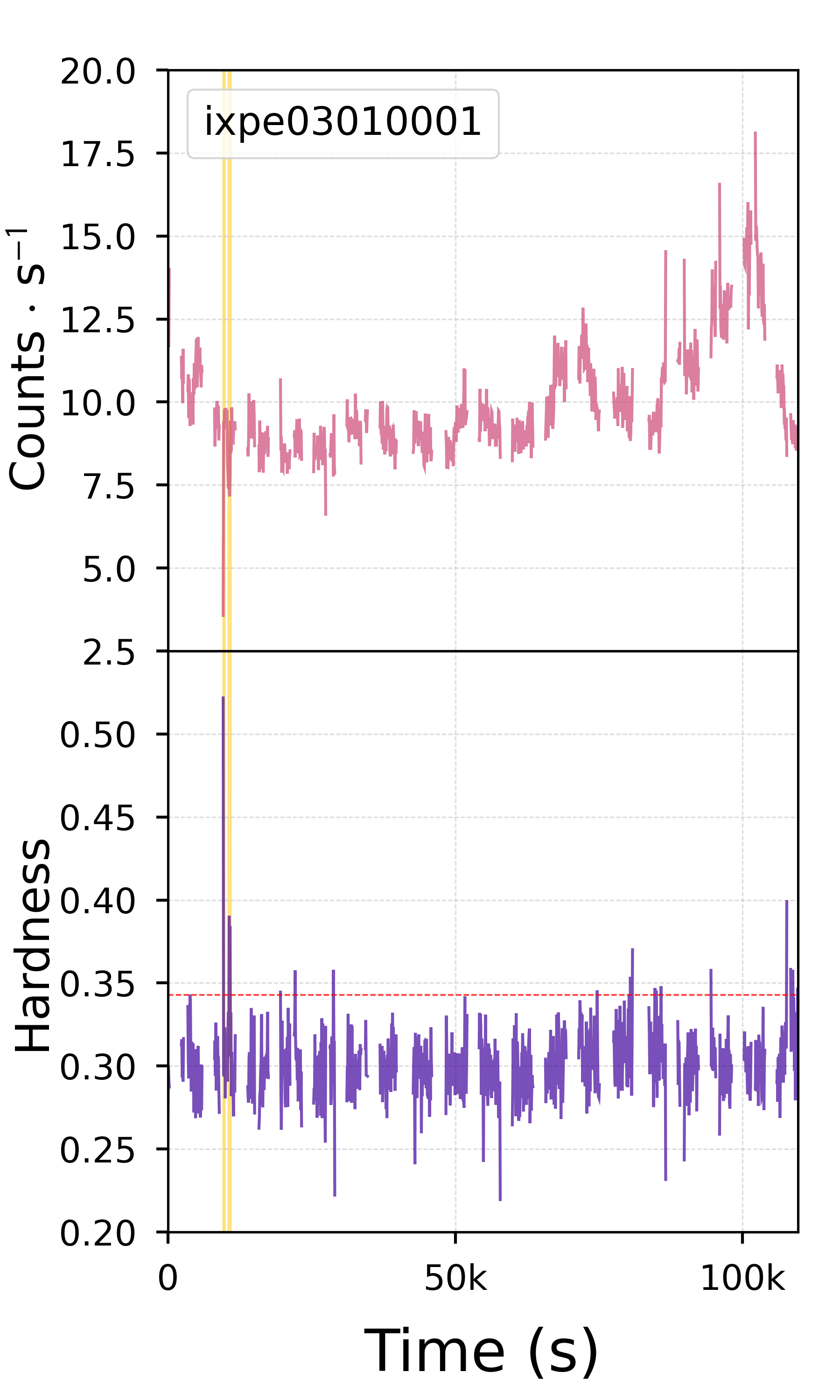} &
\includegraphics[width=0.3\columnwidth]{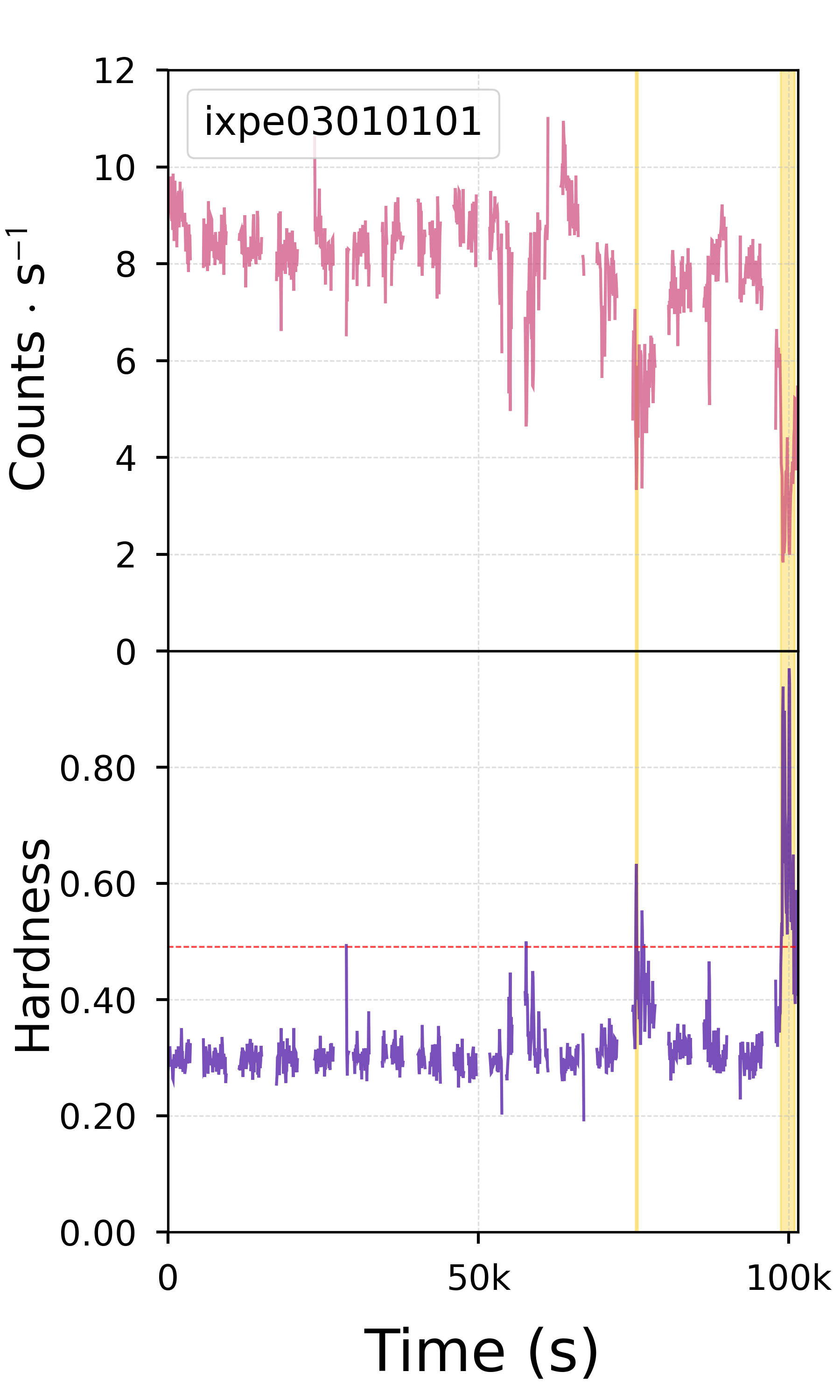} &
\includegraphics[width=0.3\columnwidth]{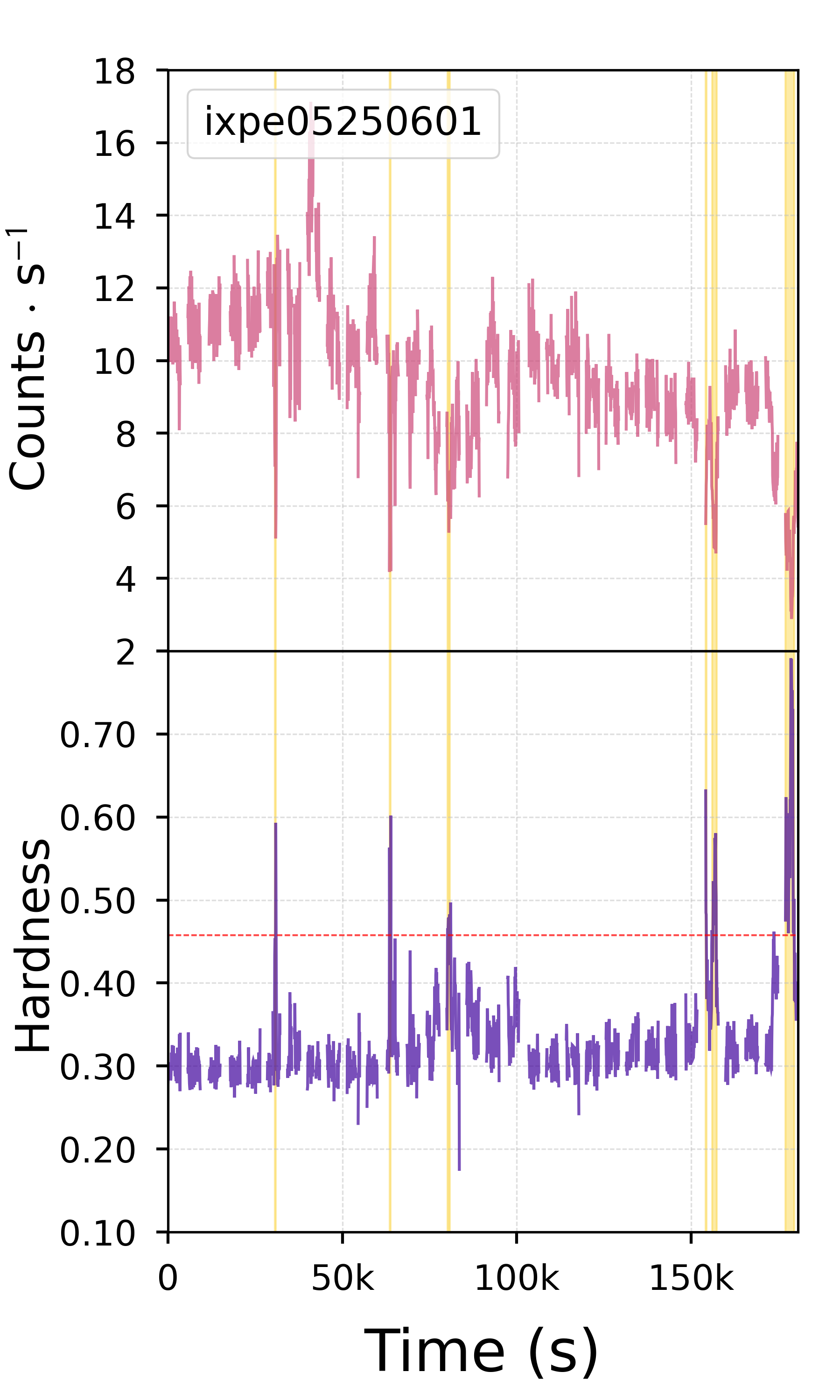} \\
\end{tabular}
\caption{Light curves and hardness ratios of Cyg~X-1 from \ixpe observations in 2023, 2024 and 2026, showing dips.}
\label{fig:IXPE_lc_hardness_other}
\end{figure*}

Figure~\ref{fig:IXPE_lc_hardness_other} shows the light curves and hardness ratios for all observations in which dips were detected in 2023, 2024, and 2026, from which the dip intervals were extracted. Table~\ref{tab:Polarization_properties_of_dip_time_intervals_2} and Table~\ref{tab:Polarization_properties_of_off-dip_time_intervals_2} present the polarization properties of all observations in which dips were detected in 2023, 2024, and 2026.

\section{Independent Consistency Check of the Overlap Dip Spectral Model using \nicer Data} \label{app:nicer_consistency}

\begin{figure*}[!hbt]
\centering
\includegraphics[width=0.49\linewidth]{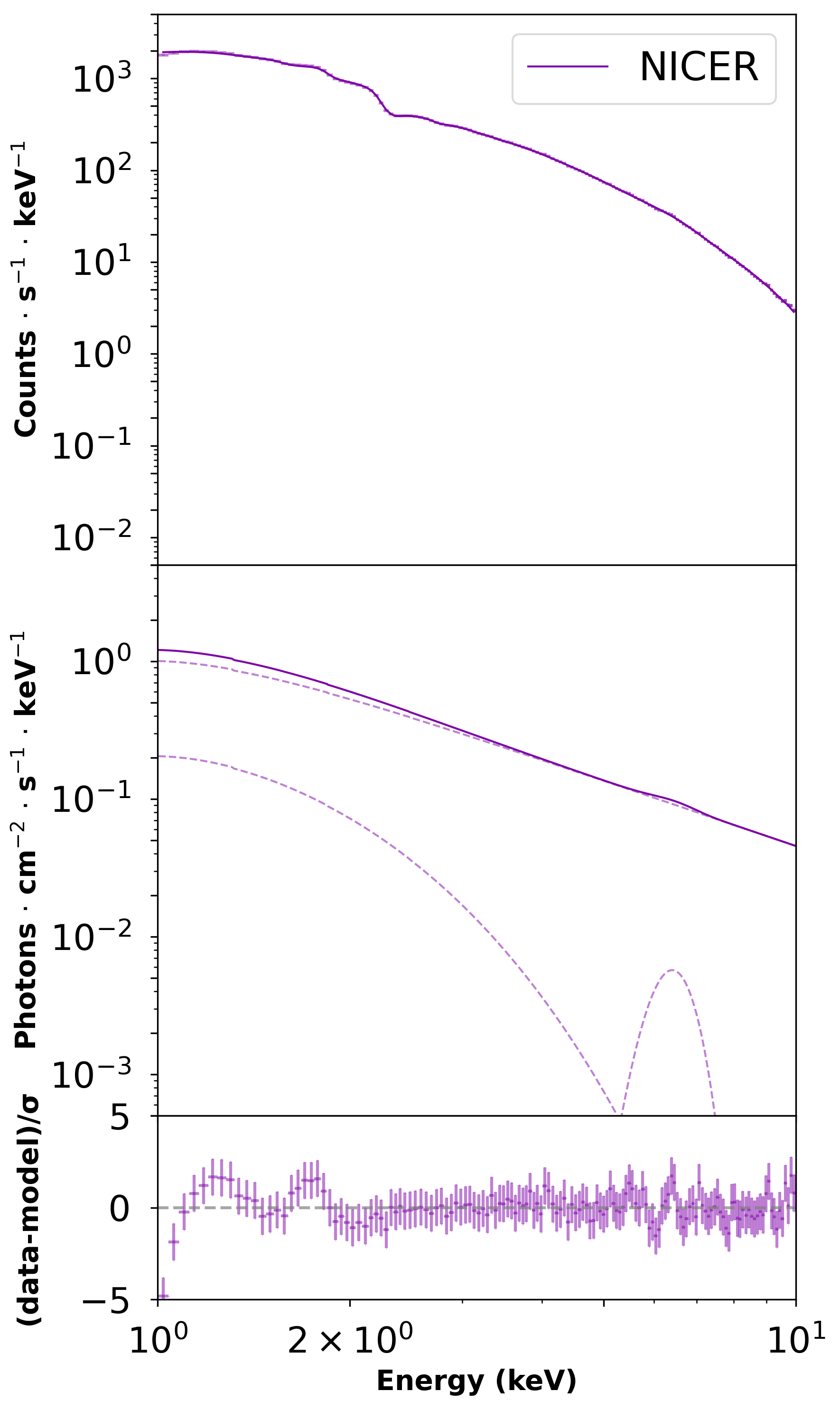}
\includegraphics[width=0.49\linewidth]{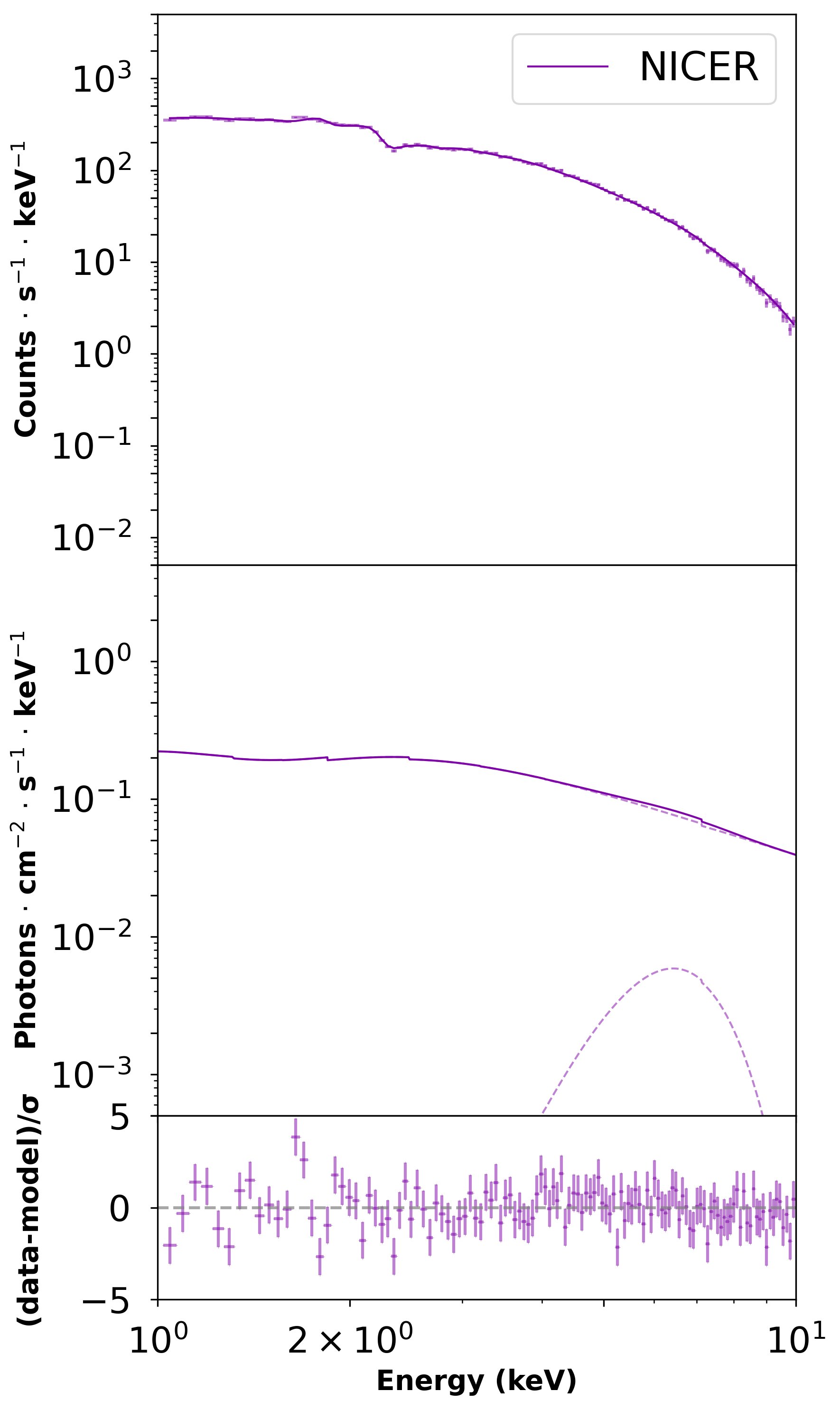}
\caption{The spectra of the off-dip (left) and dip (right) time intervals using only \nicer datasets.}
\label{fig:Spectrum_dip_offdip_nicer}
\end{figure*}

To verify that the modifications due to \texttt{MBPO} do not introduce systematics, we performed a separate spectral fit to the \nicer data without the \texttt{MBPO} model. The results in Table~\ref{tab:Best-fit_Parameters_nicer} and Figure~\ref{fig:Spectrum_dip_offdip_nicer} show that the best-fit parameters for the \texttt{TBPCF} and \texttt{DISKBB} components agree with those obtained for the full broadband spectral model, see Table~\ref{tab:Best-fit_Parameters}.

\begin{deluxetable*}{lccc}[!htb]
\tabletypesize{\scriptsize}
\label{tab:Best-fit_Parameters_nicer}
\tablewidth{0pt}
\tablehead{
\colhead{Component} & 
\colhead{Parameter (unit)} & 
\colhead{Off-dip} & 
\colhead{Overlap dip}
}
\startdata
TBABS & $N_{\text{H}}$ ($10^{22}$ cm$^{-2}$) & $[0.316]$ & $[0.316]$ \\[0.5em]
TBPCF & $N_{\text{H}}$ ($10^{22}$ cm$^{-2}$) & $[5.00]$ & $4.96^{+0.18}_{-0.18}$ \\
 & $pcf$ & $<0.095$ & $0.734^{+0.016}_{-0.019}$ \\[0.5em]
DISKBB & $T_{in}$ (keV) & $0.566^{+0.033}_{-0.021}$ & $[0.566]$ \\
 & $norm$ ($10^{2}$) & $5.5^{+3.5}_{-0.8}$ & $<1.1$ \\[0.5em]
POWERLAW & $Gamma$ & $1.580^{+0.035}_{-0.021}$ & $1.553^{+0.062}_{-0.058}$ \\
 & $norm$ & $1.740^{+0.096}_{-0.050}$ & $1.443^{+0.121}_{-0.169}$ \\[0.5em]
GAUSSIAN & $LineE$ (keV) & $6.4^{+0.1}_{-0.1}$ & $[6.4]$ \\
 & $Sigma$ & $0.48^{+0.17}_{-0.17}$ & $1.11^{+0.68}_{-0.72}$ \\
 & $norm$ ($10^{-2}$) & $0.70^{+0.28}_{-0.21}$ & $1.72^{+2.13}_{-1.24}$ \\[0.5em]
\hline
$\chi^{2}$/d.o.f. & & 99/147 & 139/125 \\
\enddata
\caption{Best-fit parameters for \nicer data only. $N_{\text{H}}$ is fixed at 0.316, as in Table~\ref{tab:Best-fit_Parameters}.
Errors are reported at 90\% CL.}
\end{deluxetable*}

%% the collaboration and author truncation commands are used. It has to


\end{document}